\documentclass[twocolumn]{aastex7}

\usepackage{amsmath,amssymb,amsfonts}
\usepackage{bm}

\usepackage{float}
\usepackage{framed}
\usepackage{soul}

\usepackage{afterpage}

\newcommand{\alfven}{Alfvén }
\newcommand{\db}{$\delta B_0/B_0$}
\newcommand{\awesom}{\textbf{\texttt{aweSOM}} }

\begin{document}

\title{Statistics of Current and Vorticity Structures in Relativistic Turbulence}

%% The \author command is the same as before except it now takes an optional
%% argument which is the 16 digit ORCID. The syntax is:
%% \author[xxxx-xxxx-xxxx-xxxx]{Author Name}
%%
%% This will hyperlink the author name to the author's ORCID page. 

% \correspondingauthor{Zachary Davis}
% \email{zkdavis@hawaii.edu}

\author[0000-0002-0959-9991]{Zachary Davis}
\affiliation{Institute for Astronomy, University of Hawaii, Manoa, 2680 Woodlawn Dr., Honolulu, HI
96822, USA}
\email{zkdavis@hawaii.edu}

\author[0000-0001-8822-8031]{Luca Comisso}
\affiliation{Department of Astronomy and Department of Physics, Columbia University, 538 West 120th Street, New York, NY 10027, USA} 
\email{luca.comisso@columbia.edu}

\author[0000-0002-2160-7288]{Colby Haggerty}
\affiliation{Institute for Astronomy, University of Hawaii, Manoa, 2680 Woodlawn Dr., Honolulu, HI
96822, USA}
\email{colbyh@hawaii.edu}

\author[0000-0002-3226-4575]{Joonas Nättilä}
\affiliation{Department of Physics, University of Helsinki, P.O. Box 64, University of Helsinki, FI-00014, Finland}
\email{joonas.nattila@helsinki.fi}

\begin{abstract}
Coherent structures created through turbulent cascades play a key role in energy dissipation and particle acceleration. In this work, we investigate both current and vorticity sheets in 3D particle-in-cell simulations of decaying relativistic turbulence in pair plasma by training a self-organizing map to recognize these structures. We subsequently carry out an extensive statistical analysis to reveal their geometric and structural properties. This analysis is systematically applied across a range of magnetizations ($\sigma$) and fluctuating-to-mean magnetic field strengths (\db) to assess how these parameters influence the resulting structures. We find that the structures' geometric properties form power-law distributions in their probability density functions (PDFs), with the exception of the structure width, which generally exhibits an exponential distribution peaking around 2 electron skin depths. The measurements show weak dependence on $\sigma$ but a strong dependence on \db. Finally, we investigate the spatial relationship between current sheets and vorticity sheets. We find that most current sheets are directly associated with at least one vorticity sheet neighbor and are often situated between two vorticity sheets. These findings provide a detailed statistical framework for understanding the formation and organization of coherent structures in relativistic magnetized turbulence, allowing for their incorporation into updated theoretical models for structure-based energy dissipation and particle acceleration processes crucial for interpreting high-energy astrophysical observations.
\end{abstract}

%% https://astrothesaurus.org

\section{Introduction} \label{sec:intro}

Understanding particle acceleration responsible for broad-spectrum emission from high-energy sources has been a central effort of the astrophysics community for many decades. Typically, this effort focuses on collisionless shocks \citep{Bell1978I,Bell1978II,Drury1983,Blandford1987} or fast magnetic reconnection \citep{Zenitani2001,giannios2013,Sironi2014} as explanations for the often-needed non-thermal particle distributions required to explain broad-spectrum emission. However, the large separation between astrophysical system scales and plasma kinetic scales, combined with the almost perfectly conducting nature of these plasmas, makes turbulence inevitable. Such turbulence can arisefrom the nonlinear stages of shocks \citep{Bell2004,Caprioli2014,Haggerty2020,Caprioli2020}, magnetic reconnection \citep{lazarian2012}, large-scale driving of the system \citep{Elmegreen2004}, or even in the absence of large-scale events through plasma instabilities \citep{Borse2021}. 
Regardless of its formation, recent works have shown turbulence to be an efficient accelerator of non-thermal particles in a diverse range of regimes including compressible magnetohydrodynamic (MHD) turbulence \citep{Gootkin2025}, non-relativistic turbulence \citep{Comisso2022}, and the focus of this paper: \textit{relativistic} turbulence, where the \alfven velocity $v_A$ is approximately $c$ \citep{comisso2018,comisso2019,zhdankin2020,Joonas2021,vega2022}.

Turbulence has been extensively applied to model many astrophysical sources including understanding particle acceleration in the solar wind \citep{Cranmer2007} and cosmic-ray acceleration in the interstellar medium (ISM) \citep{Bustard2021}. In \textit{relativistic} turbulence, particle acceleration via turbulence also provides a natural mechanism for converting large amounts of magnetic energy into particle energy \citep{comisso2018}, which is needed to explain the large energy outputs in high-energy emission. For example, relativistic turbulence can help resolve the \textit{sigma problem} \citep{Rees1974,Kennel1984} by allowing for efficient conversion of magnetic energy into particle energy in models of pulsar wind nebulae (PWNe) \citep{Lyutikov2019}, or in astrophysical jets, which, if launched via the Blandford-Znajek mechanism \citep{Blandford1977}, are expected to be Poynting-flux dominated, including blazar jets \citep{marscher2014,Davis2022,Zhang2023,Mehlhaff2025} and gamma-ray bursts (GRBs) \citep{Bykov1996}. Additionally, relativistic turbulent acceleration has been proposed as a mechanism for producing ultra-high-energy cosmic rays (UHECRs) \citep{Comisso2024} and TeV neutrinos \citep{Fiorillo2025}.

Turbulence is often described as a process by which energy initially injected at large scales forms eddies that turn over and break apart, forming smaller eddies in a cascade that eventually ends when the energy can be efficiently dissipated. An analytical understanding is usually developed using scaling theories, the most influential of which being \citet{kolmogorov1941}. The principal goal of cascade theories is to quantify how energy injected at large scales is transferred to smaller scales through nonlinear interactions, ultimately reaching the eventual dissipation scale. In hydrodynamical turbulence, the dissipation scale corresponds to vorticity filaments \citep{Vincent1991}. These are one-dimensional structures characterized by large vorticity that are distributed intermittently throughout the fluid. In MHD, several extensions of the phenomenological cascade have been proposed to account for the effects of the magnetic field \citep{iroshnikov1964,goldreich1995,Boldyrev2005}. Additionally, in magnetized turbulence, vorticity filaments are expected to be replaced by large regions of current or \textit{current sheets} where electromagnetic dissipation is expected to occur \citep{comisso2018,comisso2019}. Current sheets in relativistic turbulence are often reconnecting and prominent sites of magnetic energy dissipation and particle acceleration \citep{comisso2018,comisso2019,zhdankin2020,joonas2022,dong2022}. Although turbulent acceleration is often viewed as a slow Type II Fermi acceleration \citep{fermi1949} where particles slowly gain energy by scattering off magnetic fluctuations, in the relativistic turbulence regime (where $\frac{v^2}{c^2} \approx \frac{v}{c} \approx 1$), particle acceleration can be very efficient \citep{comisso2018,comisso2019,zhdankin2020}. This stochastic acceleration is further enhanced by a rapid energy gain when particles are in or near current sheets \citep{comisso2018,comisso2019}. Furthermore, large regions of current are expected to be in close proximity to regions of vorticity \citep{Matthaeus1982} where they may influence dissipation in the fluid via the pressure tensor \citep{Yang2017,Yang2017NC,Parashar2016}.

Despite the importance of coherent structures in particle acceleration, traditional theories such as quasi-linear theory (QLT) \citep{bernstein1966} largely overlook their role. QLT treats turbulent fields as small-amplitude perturbations to a mean magnetic field and represents the turbulence as a superposition of uncorrelated wave modes, allowing the particle response to be treated analytically \citep[see, e.g.,][]{demidem2020}. Unfortunately, QLT faces several limitations due to the restriction of small-amplitude turbulent fluctuations and inconsistencies with fully kinetic particle-in-cell (PIC) simulations, including the inability to account for anisotropy in the particle distribution \citep{comisso2019,comisso2020,Comisso2021}, recreate the observed particle distribution without additional terms in the advection coefficient \citep{zhdankin2020,Joonas2021,Davis2022}, or account for acceleration in and around current sheets \citep{comisso2018,comisso2019}. The shortcomings of QLT in relativistic turbulence have led \citet{lemoine2021} to put forward an alternative approach that follows particle momentum in a set of frames where the electric field vanishes and acceleration arises from interactions with coherent structures. In light of these recent advances in particle acceleration in turbulence, there is a clear need for a detailed accounting of the statistics of coherent structures, one which can be used to develop a more complete model of turbulent particle acceleration.

In this work, we aim to advance the statistical understanding of coherent structures in turbulence that are relevant for energy dissipation and particle acceleration. Characterizing the statistical distribution of coherent structures is key to building a simplified model of turbulent acceleration that both agrees with kinetic simulations and can be used efficiently enough to model the emission of astrophysical sources. In this study, we focus primarily on current sheets and vorticity sheets. Current sheets are often subject to magnetic reconnection, making them ideal sites for magnetic energy dissipation. Moreover, cascade models for MHD turbulence often describe current sheets as the dissipative structures at the end of a cascade \citep{she1994,dubrulle1994,muller2003}, possibly tying the statistical fluctuations of the inertial range to the intermittency and distributions of current sheets \citep{Davis2024}. Vorticity sheets, often occurring alongside current sheets, are also important sites of energy dissipation \citep{Yang2017} and may trace regions of shear-flow reconnection, where turbulent heating is enhanced \citep{Haggerty2025}. In this work, we set out to understand the statistical properties of current and vorticity sheets and their dependence on parameters understood to be key to particle heating and acceleration in relativistic turbulence. Previous works have investigated the statistical features of current sheets \citep{Zhdankin2013,zhdankin2016}, analyzed current sheets with machine learning techniques \citep{bussov2021,serrano2024,Ha2025}, segmented current sheets and vorticity sheets through wavelet analysis \citep{yoshimatsu2009}, and tied dissipative structures to the statistics of the inertial range \citep{Davis2024}. Our work significantly advances these previous works by analyzing both current sheets and vorticity sheets, studying the statistical relationship between the two, and systematically investigating how magnetization and the ratio of fluctuation to mean magnetic field (parameters key to particle heating and acceleration) control the statistical understanding of their properties and organization.

To characterize current and vorticity sheets, we begin by outlining the PIC simulation setup (Section \ref{sec:sim-setup}) and subsequently describe our method for identifying current and vorticity sheets (Section \ref{sec:struct_ident}). Next, we describe the algorithm for measuring the sheet properties (Section \ref{sec:measurements}) before going through the results (Section \ref{sec:results}). Finally, we discuss our results and their possible implications before concluding (Section \ref{sec:discussion}).

\section{Simulation Setup}
\label{sec:sim-setup}
We analyze fully kinetic particle-in-cell (PIC) simulations of relativistic plasma turbulence following the numerical setup of \citet{comisso2018,comisso2019}. The simulations are performed in a triply periodic cubic domain of side length $L$. The plasma consists of a uniform electron-positron pair population of total density $n_0$, sampled from a Maxwell-Jüttner distribution with dimensionless temperature $\theta_0 = k_B T_0/m c^2 = 0.3$, where $T_0$ is the initial temperature, $m$ the electron mass, $k_B$ the Boltzmann constant, and $c$ the speed of light. A uniform background magnetic field $\mathbf{B}_0 = B_0 \hat{\mathbf{z}}$ is imposed, and turbulence is seeded by large-scale transverse magnetic fluctuations of root-mean-square amplitude $\delta B_0 = \langle \delta B^2 \rangle^{1/2}$, with a spectrum peaking at $k_p = 6\pi/L$, defining the \textit{coherence scale}, $l_0 = 2\pi/k_p$. Different mean-field strengths are explored, with ratios $\delta B_0 / B_0 \in \{0.5, 1, 2\}$.

The relative strength of the initial magnetic fluctuations is quantified by the magnetization parameter $\sigma = \delta B_0^2/4\pi h_0$, where $h_0 = n_0 w_0 m c^2$ is the plasma enthalpy density and $w_0 = K_3(\theta_0^{-1}) / K_2(\theta_0^{-1}) \approx 1.88$ is the initial enthalpy per particle, with $K_n(z)$ denoting the modified Bessel function of the second kind of order $n$. Including the mean field, the total magnetization is $\sigma + \sigma_{B_0} = (\delta B_0^2 + B_0^2)/4\pi h_0$. Simulations span magnetizations $\sigma \in \{2.5, 5, 10, 20, 40\}$, corresponding to the relativistic turbulence regime with \alfven velocity fluctuations $v_A = c\sqrt{\sigma/(1 + \sigma)} \sim c$. 

The computational domain is discretized into $1024^3$ cells, with an average of four computational particles per cell. The spatial resolution is $\Delta x = d_{e0}/3$, where $d_{e0} = c/\omega_{p0}$ is the initial plasma skin depth and $\omega_{p0}$ the relativistic plasma frequency. Previous studies of relativistic turbulence have verified convergence with these parameters \citep{comisso2018,comisso2019}. Simulations are evolved until turbulence is fully developed, and plasma properties are analyzed at $t \sim 3 l_0/c$, when the turbulent cascade is well established. Results are discussed in terms of normalized units: magnetic field $b = B/B_0$, current density $j = J/en_0 c$, and fluid bulk velocity $v = V/c$ obtained by averaging the velocities of individual particles.

\section{Structure Identification Methodology}
\label{sec:struct_ident}
To identify and characterize current and vorticity structures within the turbulent plasma, we employ the GPU-accelerated data clustering and segmentation software \awesom \citep{Ha2025}. 
\awesom is a machine learning library that implements self-organizing maps (SOMs) to perform unsupervised clustering of user-given data. 
SOM is a machine learning techniques that maps high-dimensional data to a lower-dimensional grid of nodes, where these nodes group data with similar properties \citep{Kohonen1982}. Using an SOM clustering allows us to select current and vorticity sheets without a rigid, hand-selected cutoff. Although \awesom is capable of combining multiple \textit{features} to include nonlinear trends when clustering, here we choose to use one feature for a given structure type: for current sheets we use the magnitude of the current density $j$, and for vorticity sheets we use the magnitude of the vorticity $\omega = \lvert \nabla \times \mathbf{v} \rvert$. When using \awesom\, we normalize the data by a method we refer to as RMScap. We first find the feature's root-mean-square (RMS; $F_{\mathrm{rms}}$). The data are then divided by $F_{\mathrm{rms}}$ times a given threshold ($T_{\mathrm{rms}}$). Finally, any data greater than 1 are capped at one:
\begin{equation}
    \label{eq:rmscap}
    F_i = \begin{cases} 
    1 & \text{if } \dfrac{F_i}{F_{\mathrm{rms}} T_{\mathrm{rms}}} > 1 \\[0.5em]
    \dfrac{F_i}{F_{\mathrm{rms}} T_{\mathrm{rms}}} & \text{otherwise}
    \end{cases}
\end{equation}
Using the normalization scheme in Equation \ref{eq:rmscap} allows us to directly compare the results of the SOM clustering to RMS cutoffs that have been used in previous studies to segment coherent structures \citep{Zhdankin2013,wan2016}.
    
\begin{figure}[H]
    \centering
    \includegraphics[width=\columnwidth]{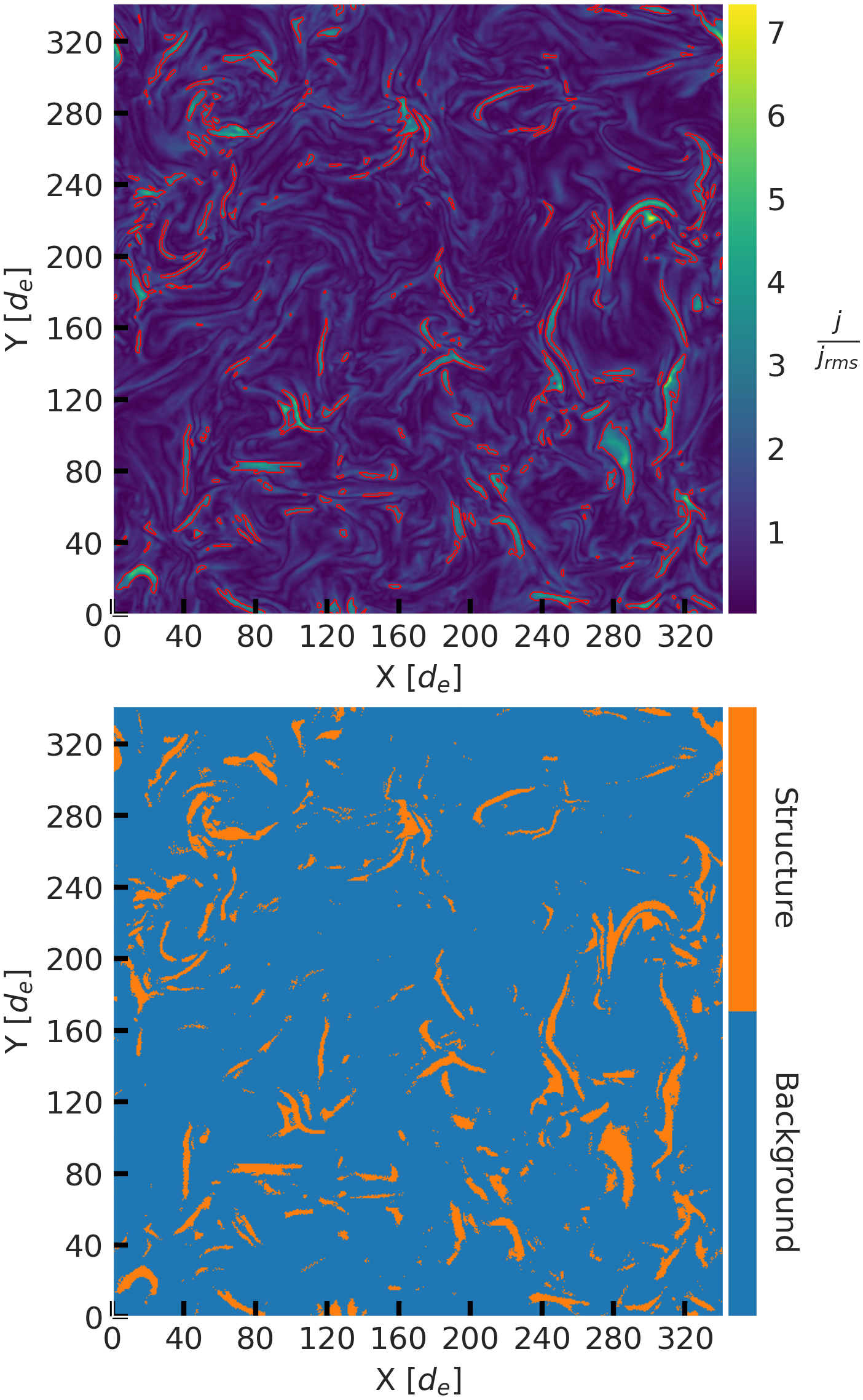} 
    \caption{\awesom\ clustering for $\sigma = 10$, \db{} $ = 1$, and $T_{\mathrm{rms}} = 2$. Top: current density field ($j/j_{\mathrm{rms}}$) in a slice of the simulation with axes in units of $d_e$. Red contour lines highlight regions exceeding the RMS threshold. Bottom: binary clustering result that segments the slice into current sheets (orange) and background (blue).}
    \label{fig:2d_clustering}
\end{figure}

For each feature, we train a model using \awesom\, where we need to define the initial learning rate $\alpha_0$, the number of training steps $N_{\mathrm{train}}$, the map aspect ratio $H$, and the merging threshold $m_{\mathrm{th}}$. We adopt the suggestions from \citet{Ha2025} for choosing these values and use $\alpha_0 = 0.1$, $H = 0.6$, and a slightly smaller value of $m_{\mathrm{th}} = 0.2$. The parameter $m_{\mathrm{th}}$ is related to grouping similar clusters at the end of training and requires the cluster to be tighter. However, values around $m_{\mathrm{th}} = 0.25$, as suggested in \citet{Ha2025}, produce negligible differences in clustering and thus do not significantly affect the results. For $N_{\mathrm{train}}$, \citet{Ha2025} found convergence at $10\% L^3$ or $10\%$ of the cells in the simulation, but in this work we adjusted this to $N_{\mathrm{train}} = 100\% L^3$ to guarantee convergence for every case. 
We train the model on all seven simulations in Table \ref{tab:current_sheet_sum} before applying the model. Once trained, data can be quickly mapped to the trained cluster nodes to reveal which category the data belong to. In our work, due to using a single feature per model, \awesom separates the data into clusters. We identify the cluster related to the coherent structure by choosing the cluster with a filling fraction similar to that of cells with $F > T_{\mathrm{rms}} F_{\mathrm{rms}}$. We provide an example of this clustering in Figure \ref{fig:2d_clustering}.

\section{Isolating and Measuring Individual Structures}
\label{sec:measurements}
\begin{figure*}[!htb]
\centering 
\includegraphics[width=0.49\textwidth]{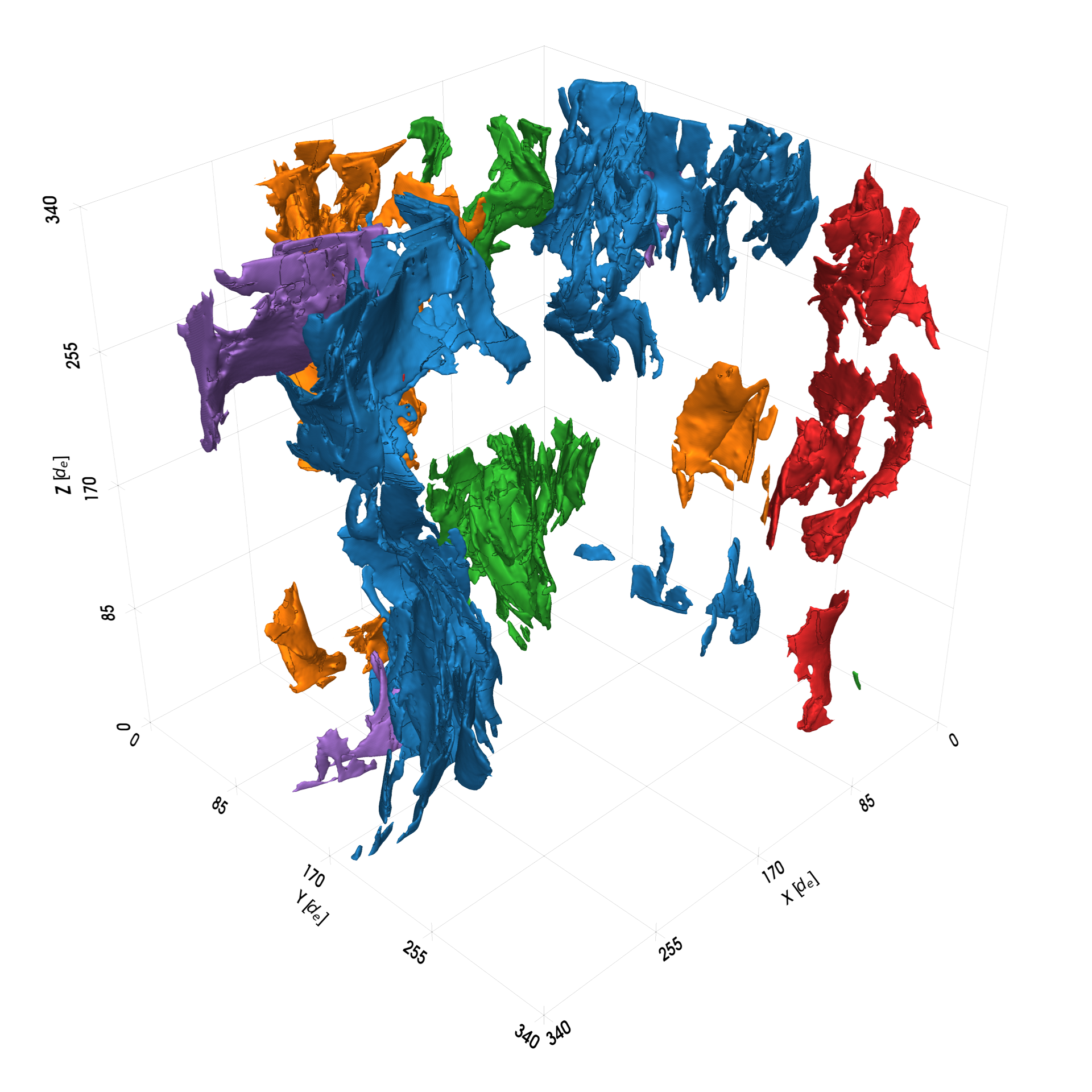}\hfill% 
\includegraphics[width=0.49\textwidth]{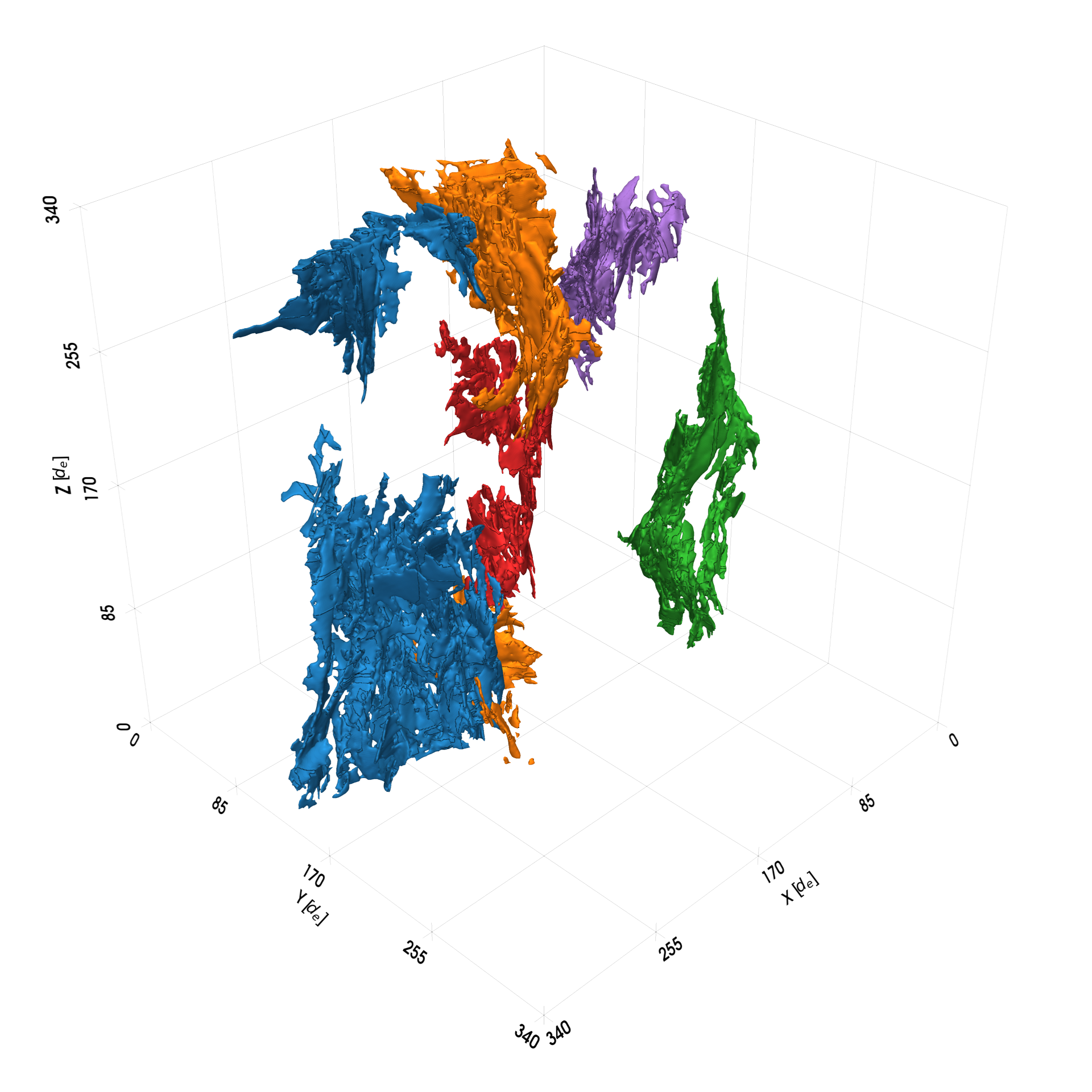}\hfill% 
\includegraphics[width=0.49\textwidth]{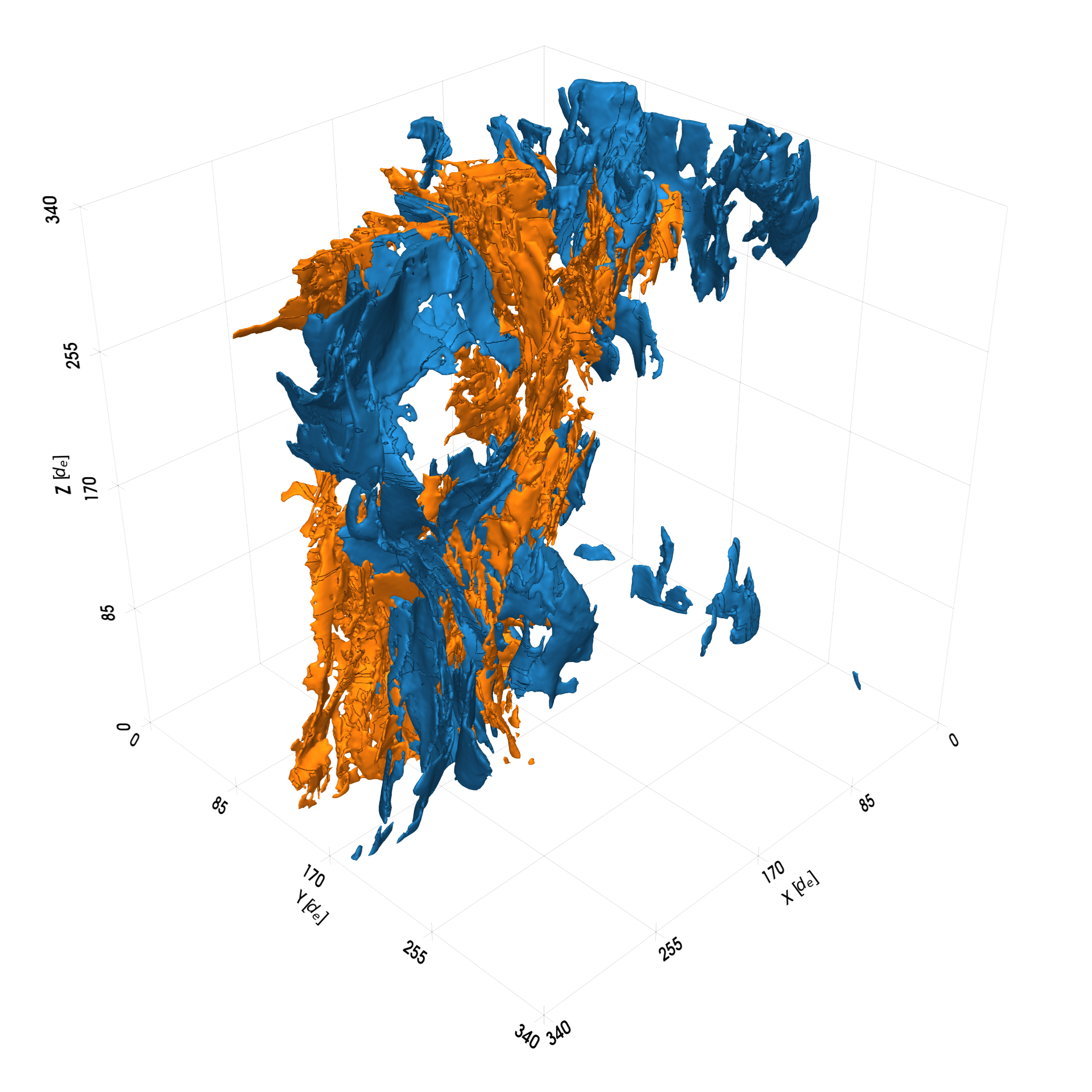} 
\caption{Three-dimensional visualization of the largest structures identified in our simulations for $T_{\text{rms}} = 2$, $\sigma =10$ and \db$ =1$. Top Left: the 5 largest current sheets. Top Right: the 5 largest vorticity sheets. Bottom: Select current sheets (Blue) and vorticity sheets (Orange) shown together for comparison.} 
\label{fig:10_largest} 
\end{figure*}

\begin{figure*}[!htb]
    \centering
    \includegraphics[width=0.98\textwidth]{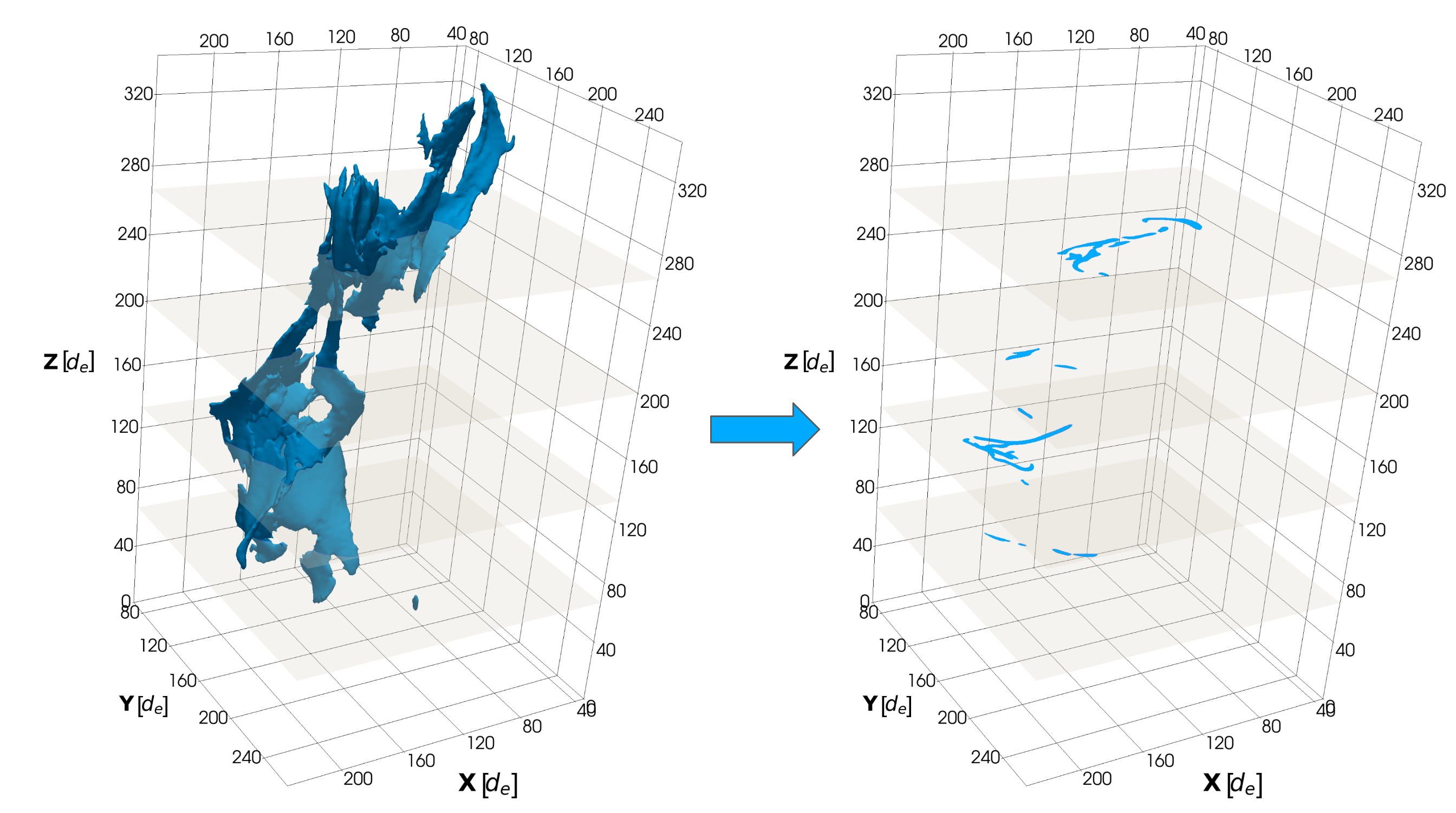}
    \caption{An illustration showing how we segment one structure, slice it, and then perform measurements on the segments. Left: a single isolated structure. Right: the same structure sliced perpendicular to the $z$-axis, showing multiple segments within each slice.}
    \label{fig:structure_2_seg_illustration}
\end{figure*}

Once clustered by \awesom, the structures still need to be processed so that unique, isolated structures can be separated from each other. To do this, we separate individual structures by finding regions of continuously connected points. This is done using the library \textit{cc3d} \citet{cc3d} , where points sharing a face are considered connected, including across periodic boundaries. We filter out structures that have a volume less than $\approx 90 \, d_e^3$( or structures who occupy less than $\approx 0.0002\%$ of the simulations volume) to improve computation and reduce errors typical of small structures. Once separated, the simulations can contain thousands of individual structures. Examples of these structures are shown in Figure \ref{fig:10_largest}.
Here, we define the length along the mean field, $l_\parallel$, as the total length along the $z$-axis spanned by the structure. Since each structure is elongated along $z$, we can simplify further measurements by slicing the structures along the $z$-axis every 2 grid cells. We choose this value to correspond roughly to the inertial length, thus allowing for computational savings with minimal loss of unique data. Most slices of structures consist of several disconnected regions in the slice but remain connected through the $z$-axis, which we refer to as \textit{segments} (see Figure \ref{fig:structure_2_seg_illustration}).
For each segment with an area greater than $1 \ d_{e}^2$ in a slice, we fit a spline to the segment. The spline fitting is a complex procedure that takes several steps and is described in Appendix \ref{sec:ap_spline_fitting}. Once a spline is fitted, we use it to measure the arc length of the spline, which we take as the value of the size of the structure perpendicular to the mean magnetic field, $l_\perp$. 

\begin{figure}[htbp]
    \centering
    \includegraphics[width=\columnwidth]{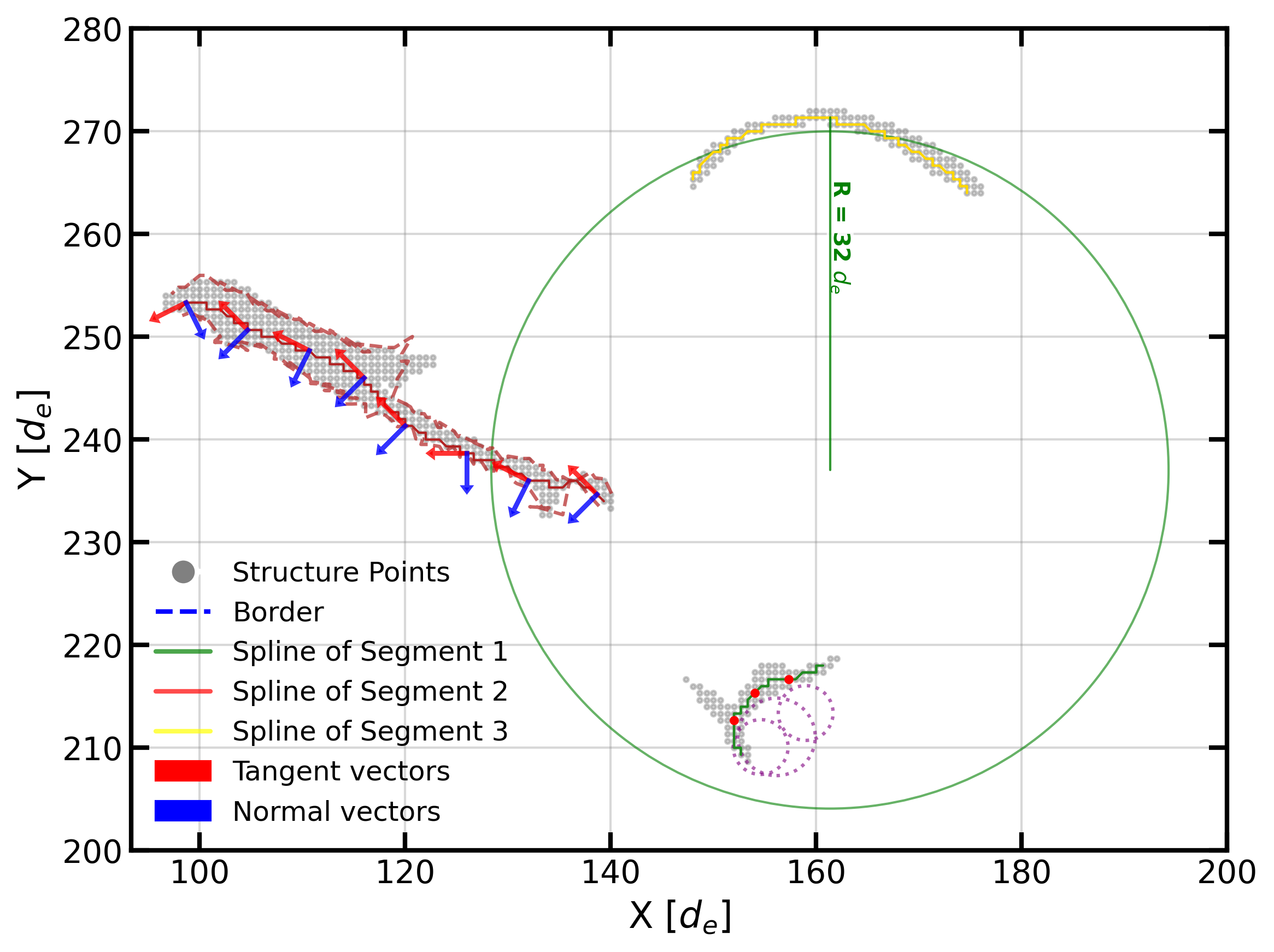}
    \caption{An example of a slice with the various measurements highlighted. The spline fit is shown in green (segment 1), red (segment 2), and yellow (segment 3). Segment 2 shows tangent vectors (red arrows) and normal vectors (blue arrows). The boundary computed is indicated by the red dashed line around segment 2. Segment 1 shows several small circles to illustrate individual measurements of $\kappa$. Segment 3 shows the radius of curvature implied from $\kappa_3$ with the large green circle.}
    \label{fig:example_slice_measurements}
\end{figure}

For every two points on the curve, we compute the tangent ($\mathbf{T}$) and normal vectors ($\mathbf{N}$) at that point. Similarly to the $z$-axis spacing, we skip points within an electron inertial length to reduce computational cost. The normal vectors at a point are used to calculate the structure width at that point. Specifically, for a given point, we traverse along the normal vector until we are no longer on the segment. We repeat this for the opposite direction, and the length of traversal in both directions is summed to define the structure width $w$ at this point. We further measure a local curvature parameter $\kappa = \left| {d\mathbf{T}}/ds \right|$, where $ds$ is the distance along the spline. The parameter $\kappa$ is an extremely local measure that displays small fluctuations in the local curvature. $\kappa$ can probe strong curvature that may exist inside structures, particularly bent field lines in the exhaust of current sheets and field structure around plasmoids. For a curvature that better represents the larger-scale curvature often directly observed in the structures, and more relevant for the global formation and stability of the structure, we implement a three-point curvature measure $\kappa_3$. To calculate $\kappa_3$, we first select the start, middle, and end points and check that they are not collinear. These three points uniquely define a circle that passes through them, which can be found by solving the three-point circle equation. An example of the measurements is shown in Figure \ref{fig:example_slice_measurements}.

When examining all recorded values of $w$, $\kappa$, and $\kappa_3$, we find that we typically oversample and have many repeated measurements along the spline. For this reason, we average these measurements over a given segment so that every segment has a single value for $w$, $l_\perp$, $\kappa$, and $\kappa_3$, while every structure has a unique value for $l_\parallel$. Additionally, for each segment we examine the aspect ratio $\alpha = l_\perp/w$, which due to the segment average of $w$ can result in values less than 1. For each of the measurements, we compile probability density functions (PDFs) presented in Section \ref{sec:results}.

\section{Results}
\label{sec:results}
Due to the large amount of data analyzed, we choose to first present the results for current sheets and vorticity sheets separately, and then examine the trends and relationships between them. We then consider simulation-wide statistics that include the number of structures ($N_s$), filling fraction ($f$), and co-dimension ($C_0$). The co-dimension is measured using the same box-counting algorithm on all structures as described in \citet{Davis2024}. All measurements are repeated for two values of $T_{\mathrm{rms}} = 2, 3$. The broad results for each simulation are summarized in Table~\ref{tab:current_sheet_sum}.
\begin{deluxetable}{c@{\hspace{2pt}}c@{\hspace{2pt}}c@{\hspace{2pt}}c@{\hspace{2pt}}c@{\hspace{2pt}}c@{\hspace{2pt}}c@{\hspace{2pt}}c}
\tablewidth{\columnwidth}
% \tablewidth{0pt}
\tabletypesize{\small}
\tablecolumns{8}
\tablecaption{A results summary from each simulation. The columns from left to right show the feature used to train the model, the $T_{\mathrm{rms}}$ value used to train the model, the magnetization ($\sigma$) and magnetic fluctuation strength (\db) of the simulation, the filling fraction of structures ($f$), the number of structures ($N_s$), the number of segments ($N_{\mathrm{seg}}$), and the co-dimension ($C_0$).
\label{tab:current_sheet_sum}}
\tablehead{
\colhead{Feature} & \colhead{$T_{\text{rms}}$} & \colhead{$\sigma$} & \colhead{\db} & \colhead{$f$} & \colhead{$N_s$} & \colhead{$N_{\text{seg}}$} & \colhead{$C_0$}
}
\startdata
$j$ & 2 & 5 & 1 & 0.068 & 1692 & 38201 & 0.74 $\pm$ 0.03 \\
\hline
$j$ & 2 & 10 & 2 & 0.071 & 932 & 24984 & 0.73 $\pm$ 0.01 \\
\hline
$j$ & 2 & 10 & 1 & 0.068 & 1651 & 36304 & 0.75 $\pm$ 0.02 \\
\hline
$j$ & 2 & 10 & 0.5 & 0.071 & 1568 & 45823 & 0.72 $\pm$ 0.03 \\
\hline
$j$ & 2 & 20 & 1 & 0.068 & 1525 & 34051 & 0.76 $\pm$ 0.02 \\
\hline
$j$ & 2 & 40 & 1 & 0.068 & 1330 & 31037 & 0.75 $\pm$ 0.02 \\
\hline
$j$ & 3 & 2.5 & 1 & 0.016 & 707 & 12040 & 1.1 $\pm$ 0.007 \\
\hline
$j$ & 3 & 5 & 1 & 0.015 & 786 & 11666 & 1.1 $\pm$ 0.008 \\
\hline
$j$ & 3 & 10 & 2 & 0.02 & 621 & 11660 & 1.1 $\pm$ 0.007 \\
\hline
$j$ & 3 & 10 & 1 & 0.014 & 772 & 10823 & 1.1 $\pm$ 0.01 \\
\hline
$j$ & 3 & 10 & 0.5 & 0.012 & 732 & 11475 & 1.2 $\pm$ 0.01 \\
\hline
$j$ & 3 & 20 & 1 & 0.014 & 702 & 10363 & 1.1 $\pm$ 0.01 \\
\hline
$j$ & 3 & 40 & 1 & 0.014 & 649 & 10073 & 1.1 $\pm$ 0.01 \\
\hline
$\omega$ & 1.5 & 2.5 & 1 & 0.23 & 1210 & 69298 & 0.21 $\pm$ 0.03 \\
\hline
$\omega$ & 1.5 & 5 & 1 & 0.23 & 1508 & 81889 & 0.19 $\pm$ 0.03 \\
\hline
$\omega$ & 1.5 & 10 & 2 & 0.21 & 1605 & 68147 & 0.24 $\pm$ 0.03 \\
\hline
$\omega$ & 1.5 & 10 & 1 & 0.22 & 1977 & 97030 & 0.17 $\pm$ 0.04 \\
\hline
$\omega$ & 1.5 & 10 & 0.5 & 0.21 & 1080 & 81647 & 0.2 $\pm$ 0.03 \\
\hline
$\omega$ & 1.5 & 20 & 1 & 0.22 & 2535 & 106853 & 0.16 $\pm$ 0.04 \\
\hline
$\omega$ & 1.5 & 40 & 1 & 0.22 & 2851 & 106868 & 0.16 $\pm$ 0.04 \\
\hline
$\omega$ & 2 & 2.5 & 1 & 0.076 & 2137 & 53878 & 0.59 $\pm$ 0.04 \\
\hline
$\omega$ & 2 & 5 & 1 & 0.073 & 3143 & 57275 & 0.59 $\pm$ 0.04 \\
\hline
$\omega$ & 2 & 10 & 2 & 0.077 & 2617 & 52988 & 0.6 $\pm$ 0.04 \\
\hline
$\omega$ & 2 & 10 & 1 & 0.072 & 4325 & 58735 & 0.59 $\pm$ 0.05 \\
\hline
$\omega$ & 2 & 10 & 0.5 & 0.084 & 2354 & 64711 & 0.51 $\pm$ 0.04 \\
\hline
$\omega$ & 2 & 20 & 1 & 0.072 & 4733 & 59654 & 0.59 $\pm$ 0.05 \\
\hline
$\omega$ & 2 & 40 & 1 & 0.071 & 4527 & 58572 & 0.59 $\pm$ 0.05 \\
\hline
\enddata
\end{deluxetable}
\subsection{Current Sheets}
\label{sec:current-sheets}

\begin{figure}[htbp]
    \centering 
    \includegraphics[width=\columnwidth,height=0.88\textheight,keepaspectratio]{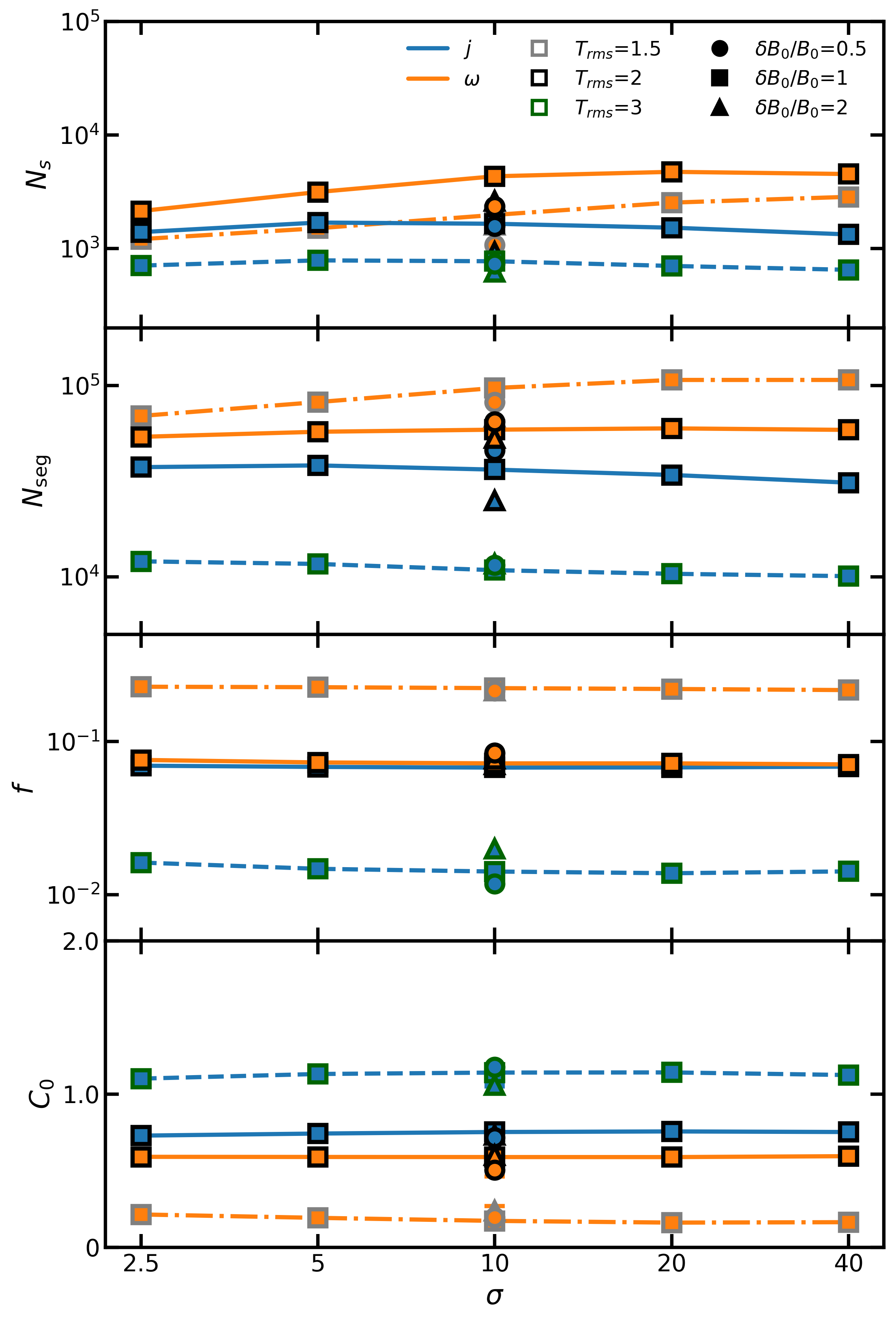}
    \caption{A results summary from each simulation. Orange shows results for $\omega$, and blue shows results for $j$. For $\sigma = 10$, different values of \db\ are indicated by different marker shapes. $T_{\mathrm{rms}}$ values of 1.5, 2, and 3 are indicated by dot-dashed, solid, and dashed line styles and grey, black, and dark green border colors, respectively.}
    \label{fig:cs_simulation_trends}
\end{figure}

\begin{figure}[htbp]
    \centering
    \includegraphics[width=\columnwidth]{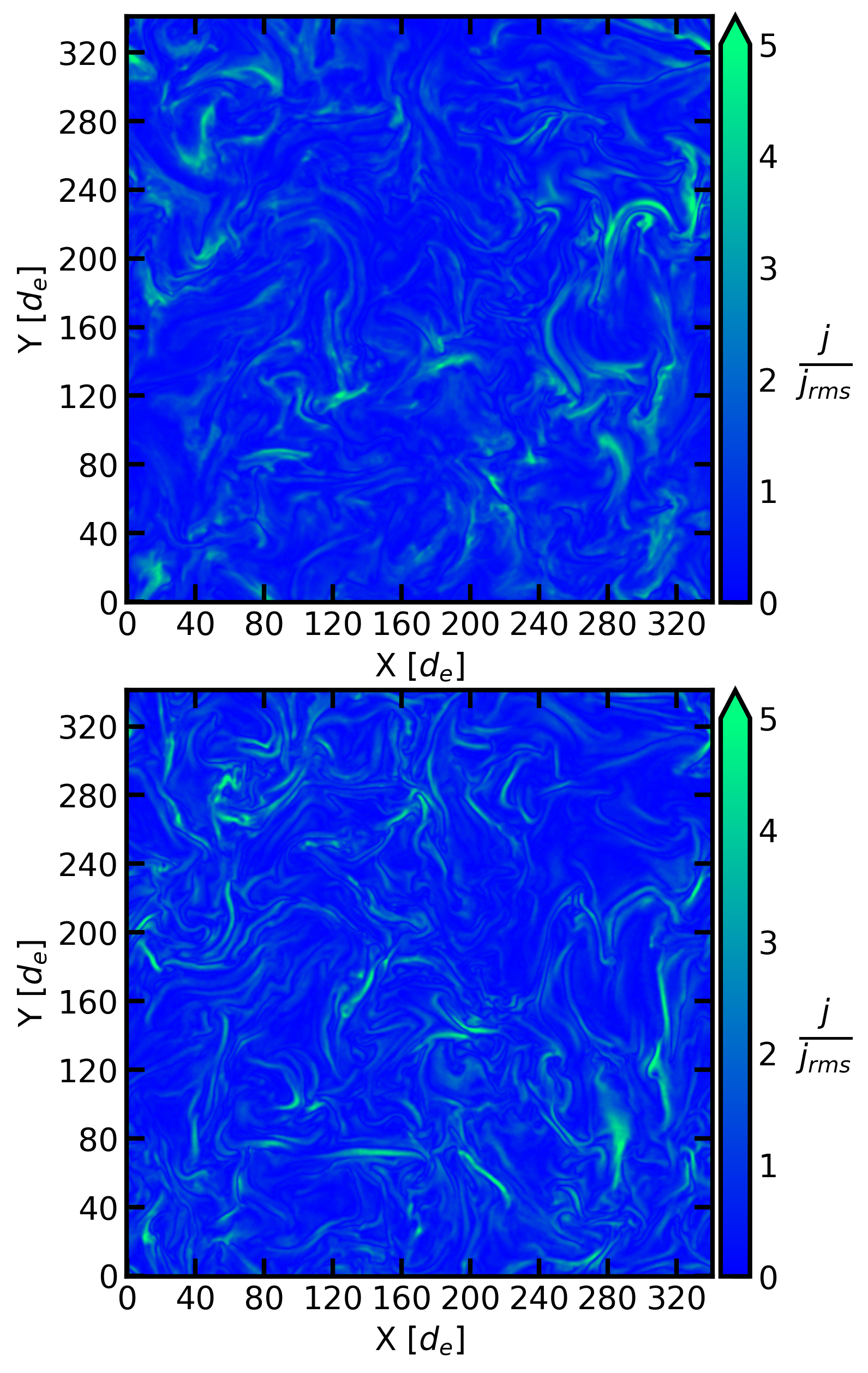}
    \caption{2D slices of the simulations showing the current density for $\sigma = 2.5$, \db$=1$ (top) and for $\sigma=40$, \db$=1$ (bottom). }
    \label{fig:slice_compare_j_sigma_vert}
\end{figure}

With respect to $\sigma$, the current sheets are very similar with comparable values of $N_s$, $f$, and $C_0$. These trends are shown in Figure \ref{fig:cs_simulation_trends}. This indicates that there is very little change in the development of current sheets once $\sigma \gg 1$. This point is further illustrated in Figure~\ref{fig:slice_compare_j_sigma_vert}, where we compare slices from simulations with $\sigma = 2.5$ and $\sigma = 40$. Despite the different magnetizations, the current density structures appear qualitatively very similar. 

\begin{figure}[htbp]
    \centering
    \includegraphics[width=\columnwidth]{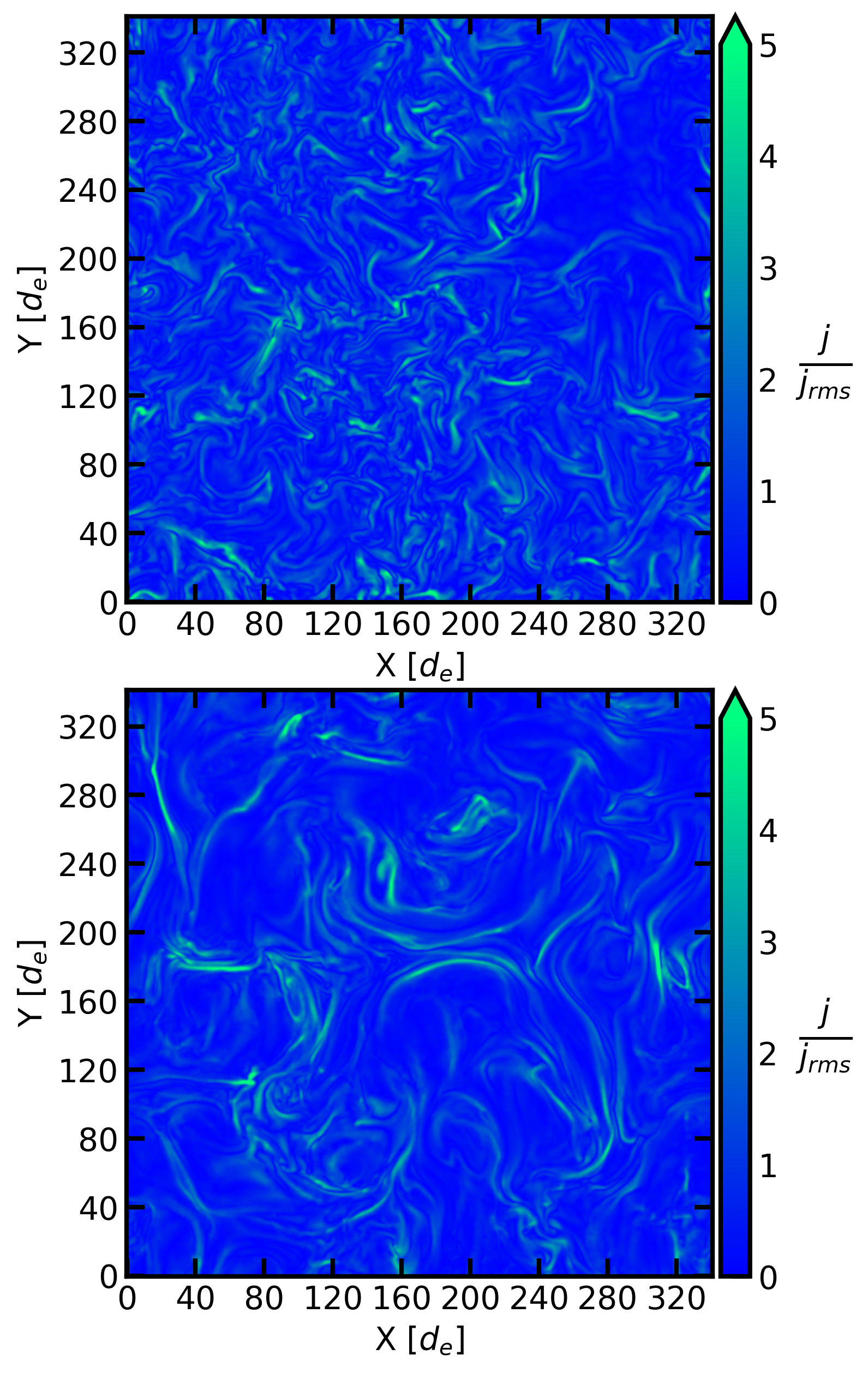}
    \caption{2D slices of the simulations showing the current density for $\sigma = 10$, \db$=0.5$ (top) and for $\sigma=10$, \db$=2$ (bottom).}
    \label{fig:cs_slice_db_compare}
\end{figure}

When analyzing \db, only $N_s$ increases monotonically, while $f$ and $C_0$ do not show a specific trend (see Figure \ref{fig:cs_simulation_trends}). Despite this, for changes in \db, Figure \ref{fig:cs_slice_db_compare} shows a distinct development, as current sheets appear less coherent as \db\ decreases.
This indicates that, for the development and morphology of turbulent current sheets well in the relativistic regime, \db\ plays a more significant role.

\begin{figure}[htbp]
    \centering
    \includegraphics[width=\textwidth,height=0.88\textheight,keepaspectratio]{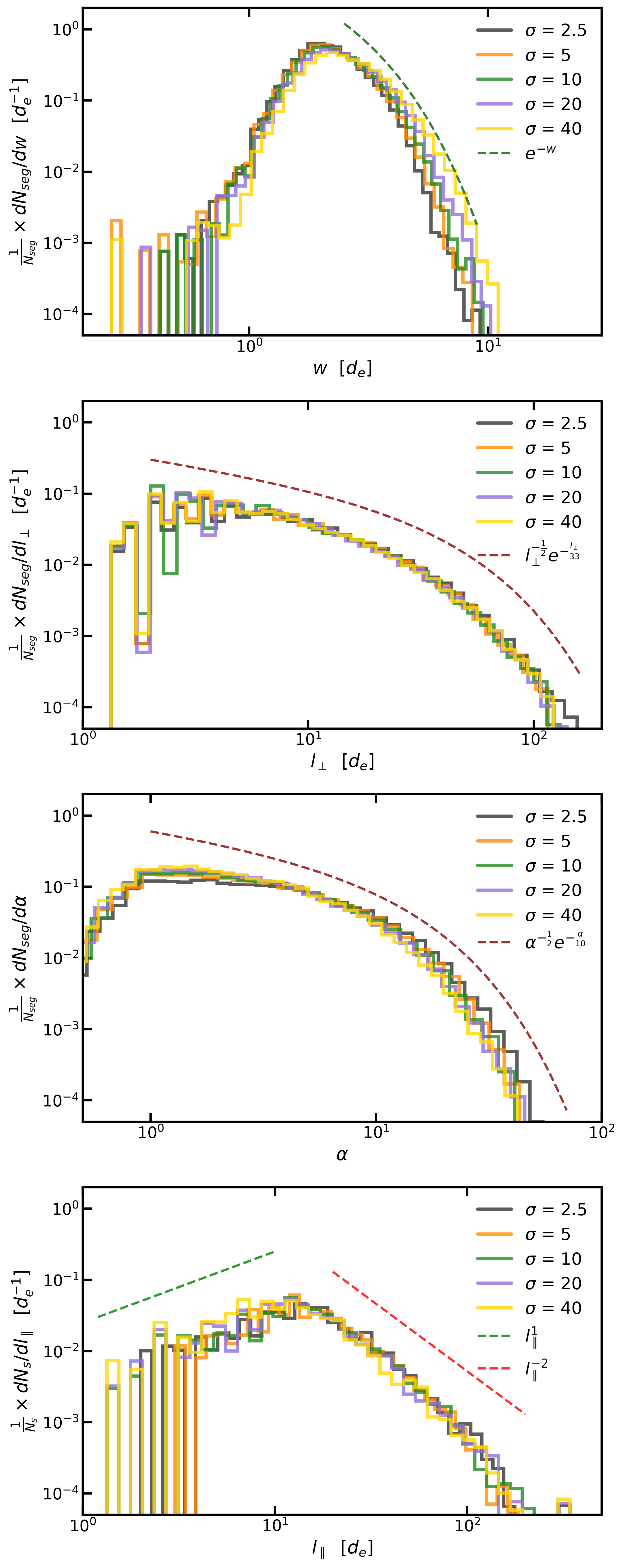}
    \caption{PDFs of current sheet measurements for different values of $\sigma$ (see legend), with illustrative fits shown as dashed lines (see legend). From top to bottom, the panels show the structure width ($w$), perpendicular length ($l_\perp$), aspect ratio ($\alpha$), and length along the mean field ($l_\parallel$).}
    \label{fig:cs_sigma_primary}
\end{figure}

\begin{figure}[htbp]
    \centering
    \includegraphics[width=\textwidth,height=0.44\textheight,keepaspectratio]{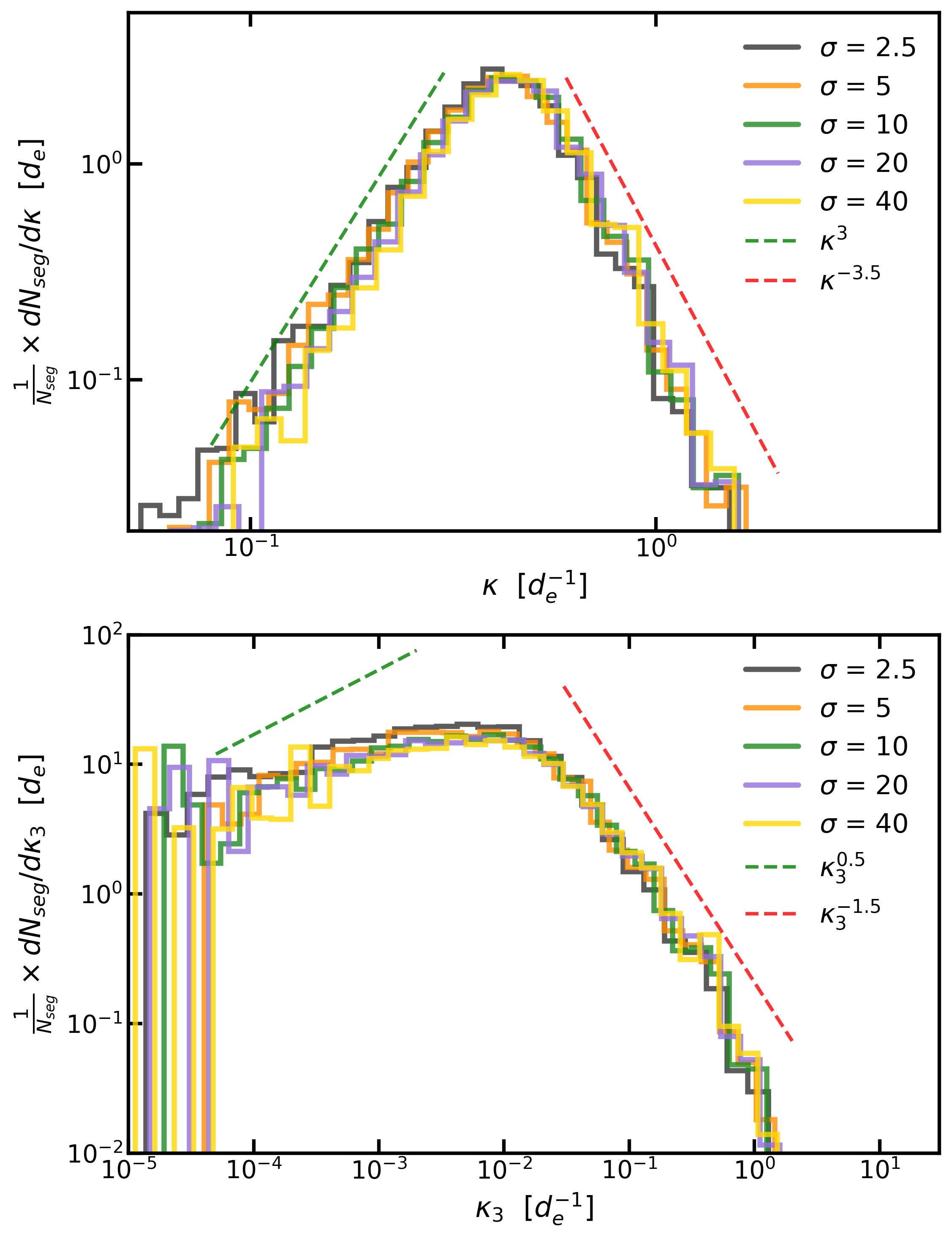}
    \caption{PDFs of current sheet measurements for different values of $\sigma$ (see legend), with illustrative fits shown as dashed lines (see legend). The top panel shows the local curvature $\kappa$, while the bottom panel shows the three-point curvature $\kappa_3$.}
    \label{fig:cs_sigma_curvature}
\end{figure}

\begin{figure}[htbp]
    \centering
    \includegraphics[width=\textwidth,height=0.88\textheight,keepaspectratio]{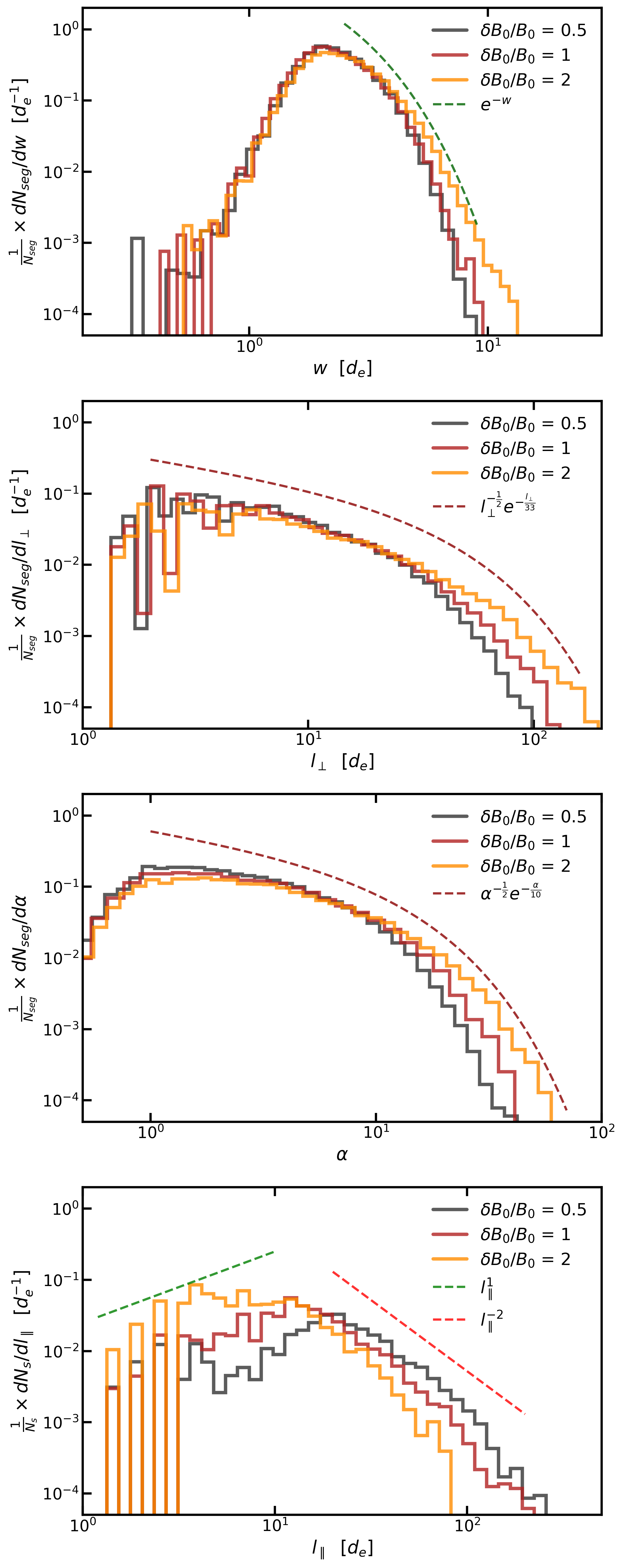}
    \caption{PDFs of current sheet measurements for different values of \db\ (see legend), with illustrative fits shown as dashed lines (see legend). From top to bottom, the panels show the structure width ($w$), perpendicular length ($l_\perp$), aspect ratio ($\alpha$), and length along the mean field ($l_\parallel$).}
    \label{fig:cs_db_primary}
\end{figure}

\begin{figure}[htbp]
    \centering
    \includegraphics[width=\textwidth,height=0.44\textheight,keepaspectratio]{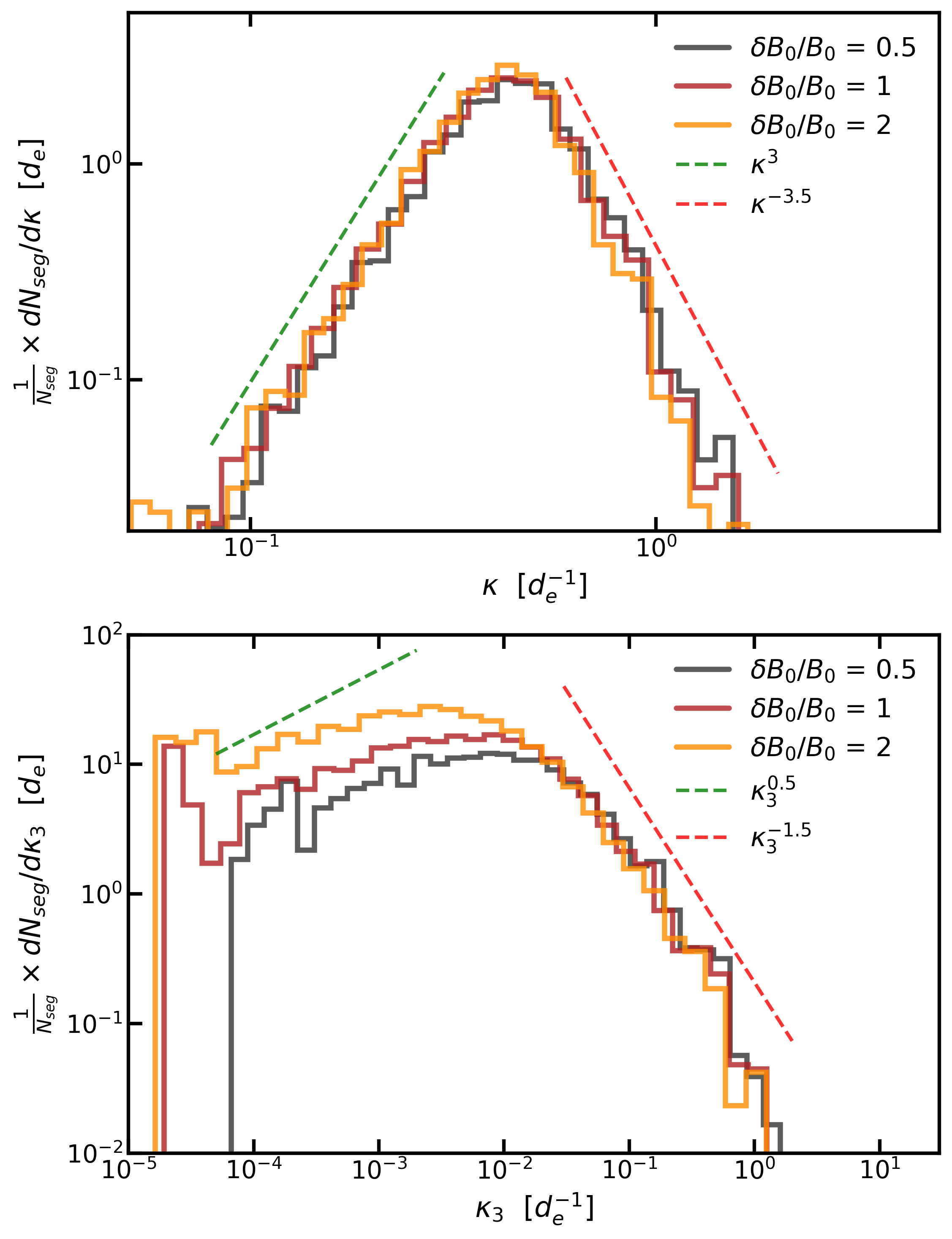}
    \caption{PDFs of current sheet measurements for different values of \db (see legend), with illustrative fits shown as dashed lines (see legend). The top panel shows the local curvature $\kappa$, while the bottom panel shows the three-point curvature $\kappa_3$.}
    \label{fig:cs_db_curvature}
\end{figure}

\label{sec:current_sheets_hists}
For each measurement described in Section \ref{sec:measurements}, we plot the resulting PDFs for all values of $\sigma$ in Figures \ref{fig:cs_sigma_primary} and \ref{fig:cs_sigma_curvature}. PDFs for all values of \db\ are shown in Figures \ref{fig:cs_db_primary} and \ref{fig:cs_db_curvature}. In these plots, we display only the fiducial results for $T_{\mathrm{rms}} = 2$ for both current and vorticity sheets. PDFs for other values of $T_{\mathrm{rms}}$ are provided in Appendix \ref{ap:rms_hists} and show qualitatively similar features to the fiducial case.

The width $w$, shown at the top of Figure \ref{fig:cs_sigma_primary} for $\sigma$ and at the top of Figure \ref{fig:cs_db_primary} for \db, shows a strong peak around $2 \ d_e$ with a slight possible power-law extension as $w$ increases until an exponential decay. A width value that peaks around $2 \ d_e$ is similar to the results found in \citet{serrano2024} for a 2D turbulence setup. In order to model and compare the distribution's dependence, we fit $w$ to the exponential,
\begin{equation}
    \frac{1}{N_{\mathrm{seg}}} \times \frac{dN_{\mathrm{seg}}}{dw} = N_0 e^{-\beta w},
    \label{eq:fit_w_exponential}
\end{equation}
where $N_0$ normalizes the distribution to 1, and $\beta$ is the decay rate. $w$ is fit to Equation \ref{eq:fit_w_exponential} in the range of $2 \leq w \leq 10$. For easy reference, we display the $\beta = 1$ case for all $w$ plots. A visual trend can be seen in which increasing $\sigma$ increases the maximum $w$. Conversely, for \db, increasing values of \db\ lead to an increase in the maximum $w$. Further discussion of the trends and fits can be found in Section \ref{sec:trends}, and all fits are summarized in tables in Appendix \ref{ap:trends_tables}.

Both $l_\perp$ and $\alpha$ are well described by power laws (possibly broken power laws) with exponential decay. In their respective subplots in Figure \ref{fig:cs_sigma_primary}, we choose to fit them with a power law with exponential decay,
\begin{equation}
     \label{eq:pwl_exp_fit}
     \frac{1}{N_{\mathrm{seg}}} \times \frac{dN_{\mathrm{seg}}}{dl_{\perp}} = N_0 l_\perp^{p}e^{-\beta l_\perp},
\end{equation}
where $p$ is the power-law index. For $\alpha$, we use the same functional form as in Equation \ref{eq:pwl_exp_fit}. The quantity $l_\perp$ was fit over the range $2 \leq l_\perp \leq 200$, while the aspect ratio $\alpha$ is fit for $1 \leq \alpha \leq 60$. In all figures of $l_\perp$, we show a reference fit with $p = 1/2$ and $\beta = 1/33$. For $\alpha$, the reference fit has $p = 1/2$ and $\beta = 1/10$. These results show little variation with $\sigma$, as seen in Figure \ref{fig:cs_sigma_primary}, but in Figure \ref{fig:cs_db_primary} they both have their maximum extent truncated with decreasing \db. It is worth noting here that $\beta$ for $l_\perp$ is approximately a third of the coherence length of the simulation. Though some current sheets do reach the coherence length, they are not common.

The rest of the measurements, $l_\parallel$, $\kappa$, and $\kappa_3$, are fit with a broken power law,
\begin{equation}
    \label{eq:broken_pwl_fit}
    \frac{1}{N_{\mathrm{seg}}} \times \frac{dN_{\mathrm{seg}}}{dl_{\parallel}} = N_0 \times \begin{cases} 
    l_\parallel^{p_1} & \text{if } l_\parallel < \lambda \\[0.5em]
   l_\parallel^{p_2} & \text{otherwise} 
    \end{cases},
\end{equation}
where $p_1$ and $p_2$ are the power-law indices before and after the peak, and $\lambda$ is the power-law break. Here we are not focused on the break itself, as it is closely related to the mean, and we only show reference power laws before and after the break (all values of $\lambda$ can be found in Appendix \ref{ap:rms_hists}). The quantity $l_\parallel$ is unique since it only has a single measurement per structure by definition, which makes its histogram less resolved. Nevertheless, the histogram is well resolved, particularly after the peak. For $l_\parallel$, we show reference values $p_1 = 1$ and $p_2 = -2$, with the fit applied over the range $5 \leq l_\parallel \leq 150$. The results look nearly identical across $\sigma$ but generally decrease for \db. The local curvature $\kappa$ is sharply peaked around $\kappa \approx 0.5$, with reference values $p_1 = 3$ and $p_2 = -3.5$. The quantity $\kappa$ is fit over the range $10^{-1} \leq \kappa \leq 2$. The behavior of $\kappa$ remains similar across $\sigma$ with a slight possible trend toward larger values for decreasing \db. The three-point curvature $\kappa_3$, our three-point measurement that samples the larger-scale curvature, has a much flatter peak with values before the peak being flatter with larger values of \db. Beyond this, $\kappa_3$ has little dependence on \db\ or $\sigma$. We use reference values of $p_1 = 0.5$ and $p_2 = 1.5$ for the power laws of $\kappa_3$, with the fit performed over the range $10^{-4} \leq \kappa_3 \leq 1$.

\subsection{Vorticity Sheets}
\label{sec:vorticity-sheets}

\begin{figure}[htbp]
    \centering
    \includegraphics[width=\columnwidth]{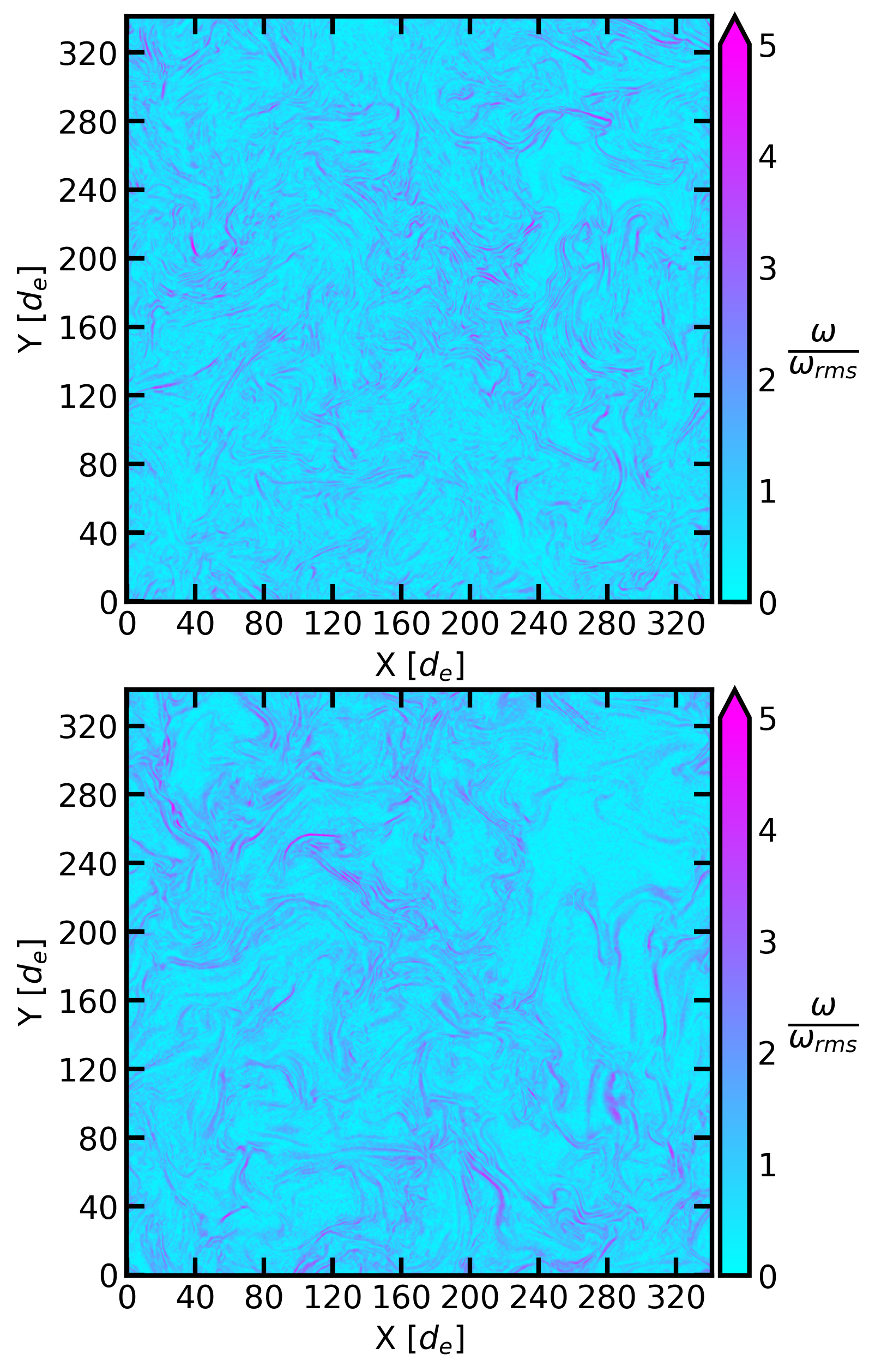}
      \caption{2D slices of the simulations showing the vorticity for $\sigma = 2.5$, \db$=1$ (top) and for $\sigma=40$, \db$=1$ (bottom).}
    \label{fig:vs_slice_sigma_compare}
\end{figure}

\begin{figure}[htbp]
    \centering
    \includegraphics[width=\columnwidth]{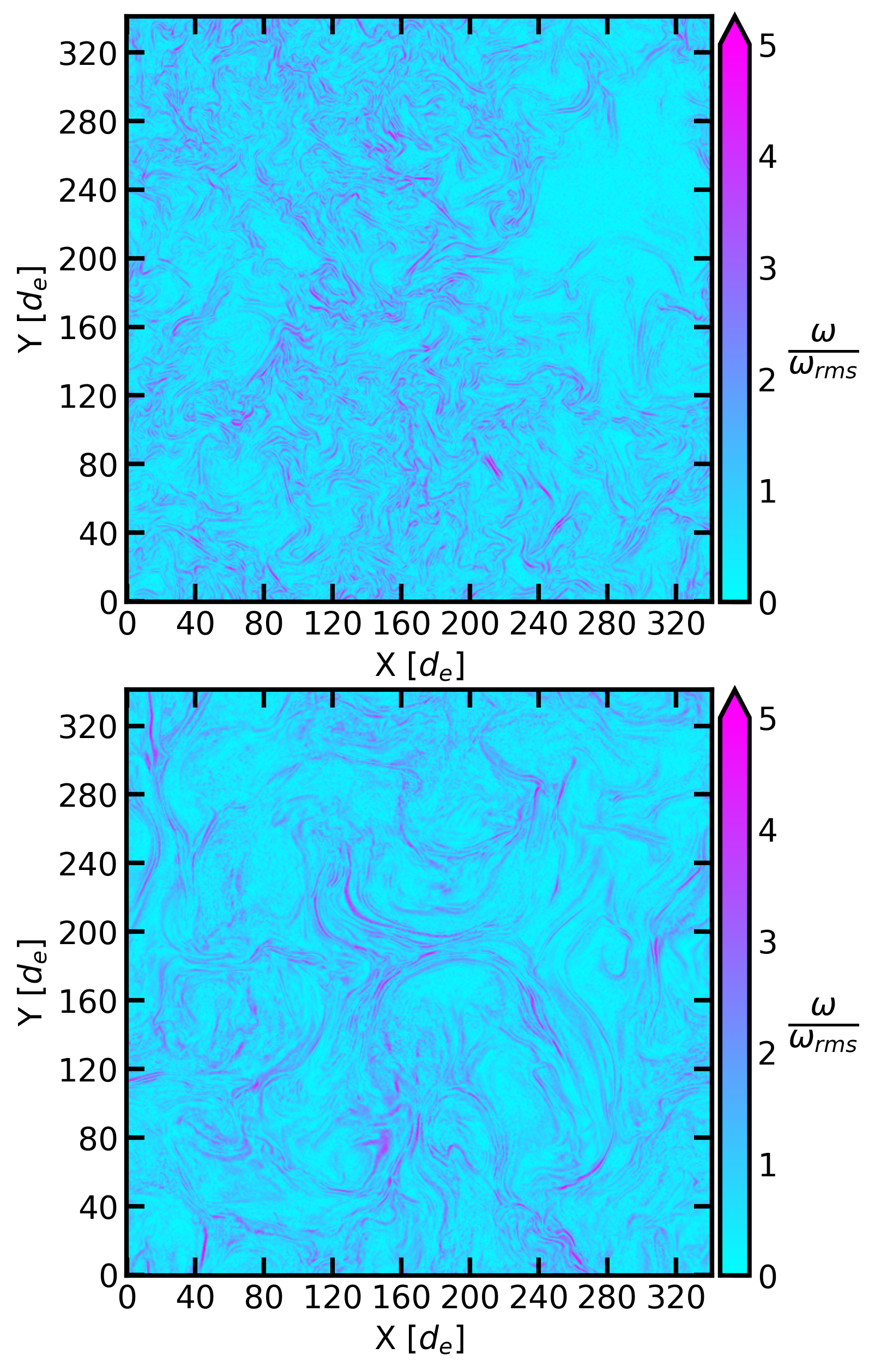}
    \caption{2D slices of the simulations showing the vorticity for $\sigma = 10$, \db$=0.5$ (top) and for $\sigma=10$, \db$=2$ (bottom).}
    \label{fig:vs_slice_db_compare}
\end{figure}

\begin{figure}[htbp]
    \centering
    \includegraphics[width=\textwidth,height=0.88\textheight,keepaspectratio]{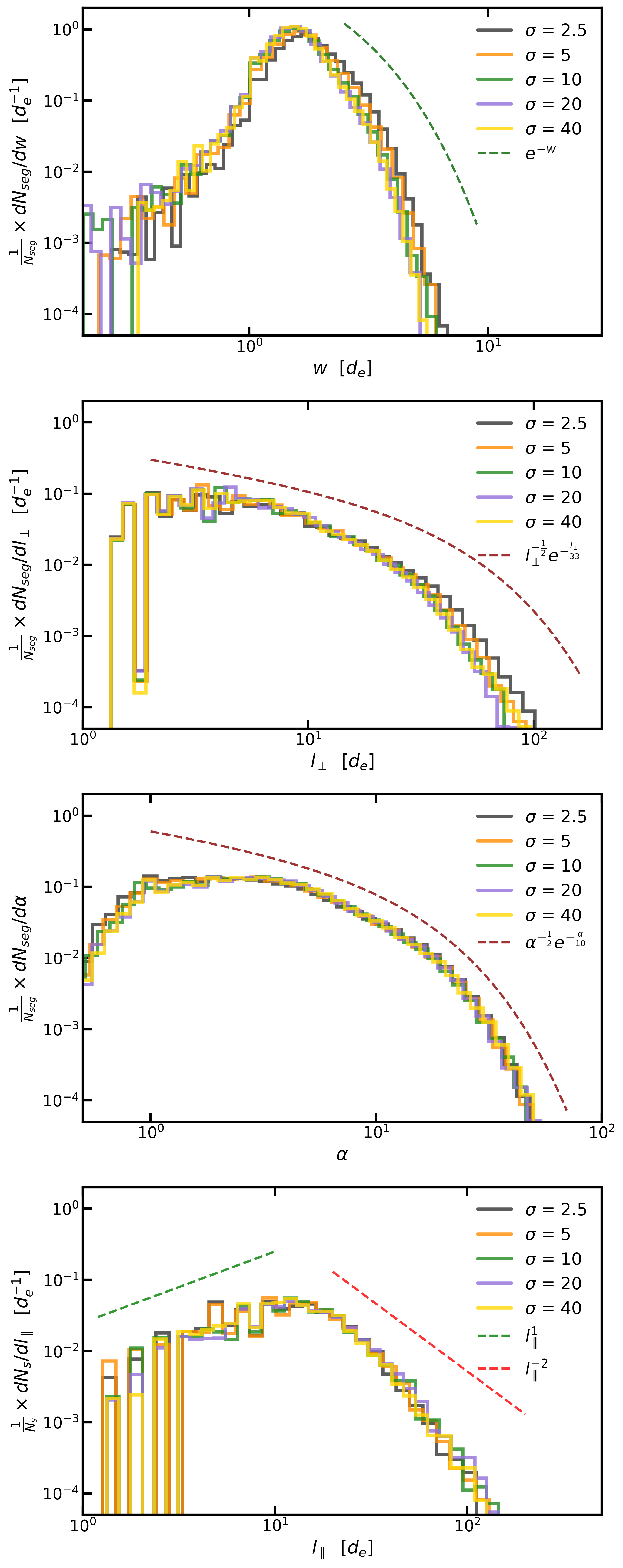}
    \caption{PDFs of vorticity sheet measurements for different values of $\sigma$ (see legend), with illustrative fits shown as dashed lines (see legend). From top to bottom, the panels show the structure width ($w$), perpendicular length ($l_\perp$), aspect ratio ($\alpha$), and length along the mean field ($l_\parallel$).}
    \label{fig:vs_sigma_primary}
\end{figure}

\begin{figure}[htbp]
    \centering
    \includegraphics[width=\textwidth,height=0.44\textheight,keepaspectratio]{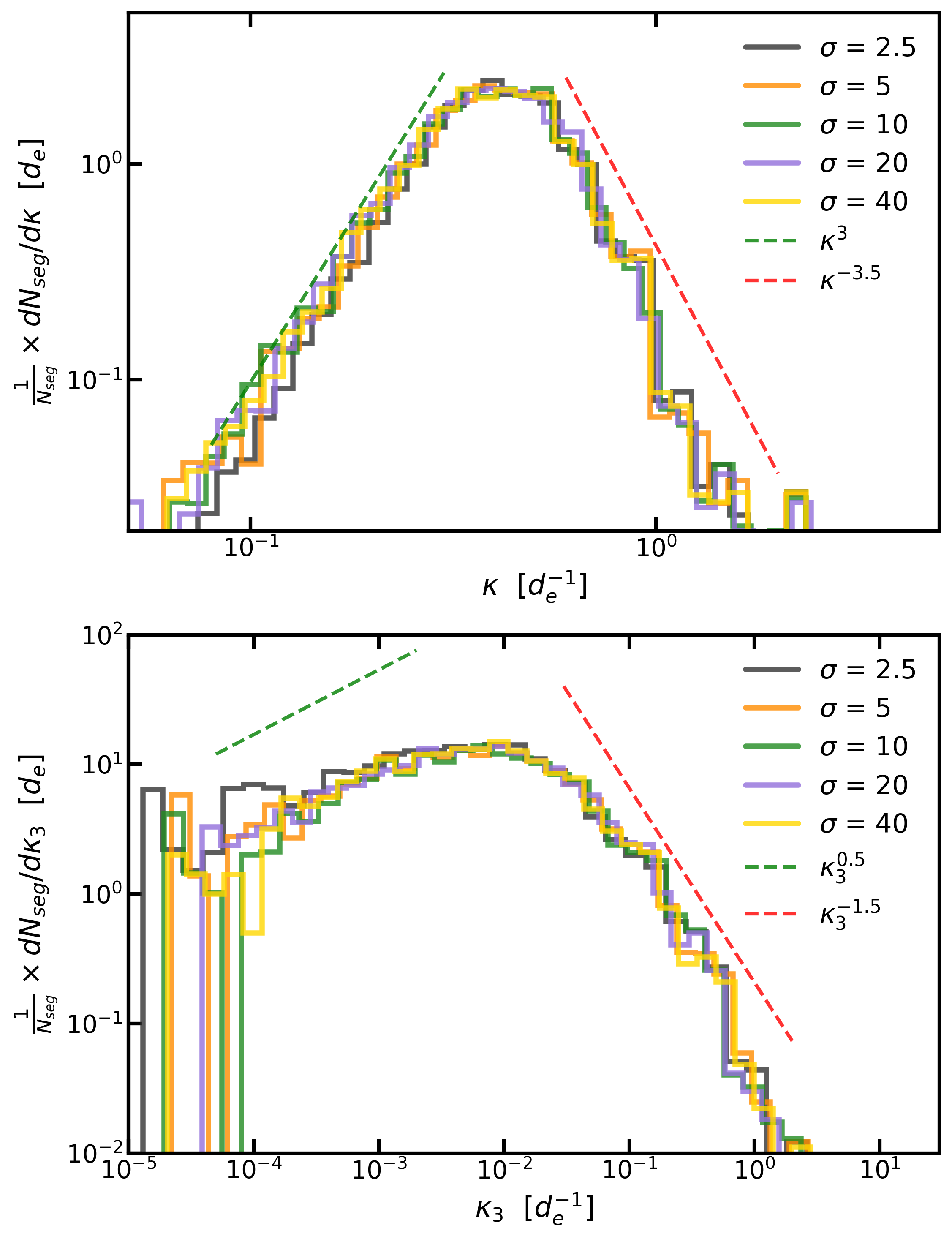}
    \caption{PDFs of vorticity sheet measurements for different values of $\sigma$ (see legend), with illustrative fits shown as dashed lines (see legend). The top panel shows the local curvature $\kappa$, while the bottom panel shows the three-point curvature $\kappa_3$.}
    \label{fig:vs_sigma_curvature}
\end{figure}

\begin{figure}[htbp]
    \centering
    \includegraphics[width=\textwidth,height=0.88\textheight,keepaspectratio]{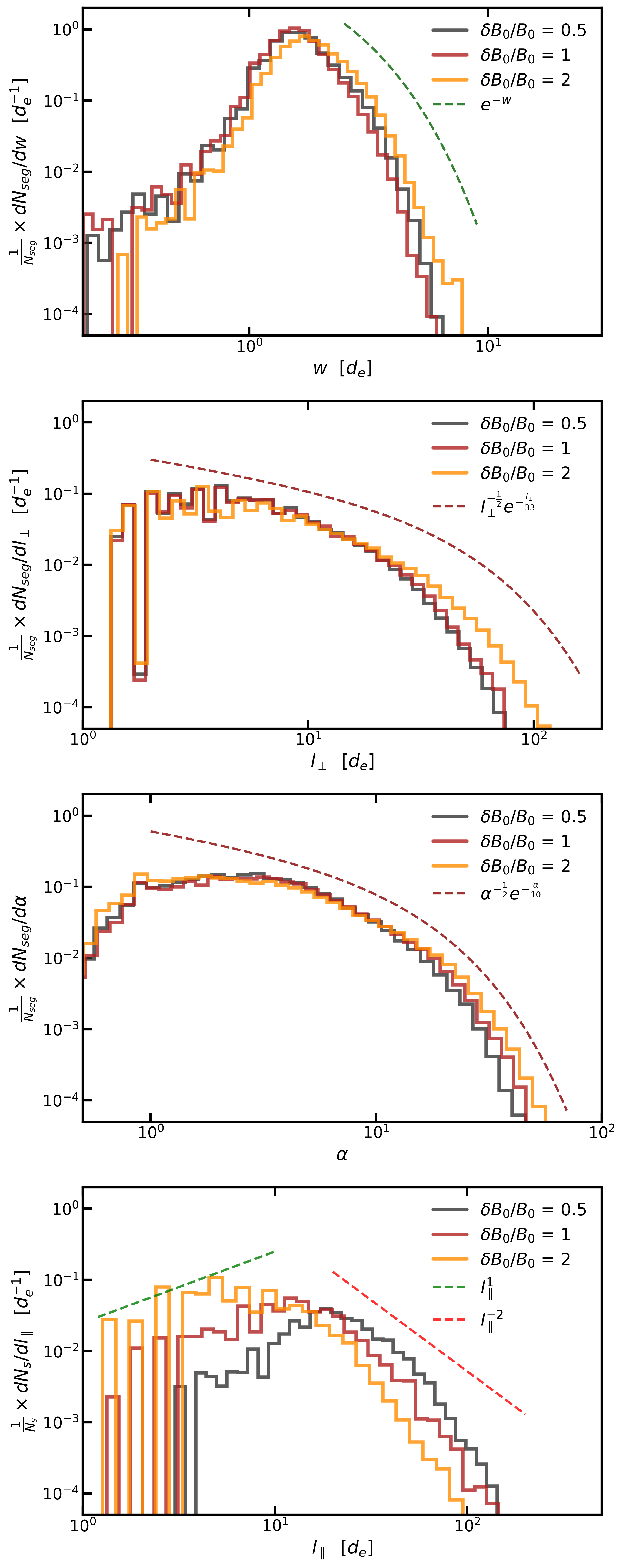}
    \caption{PDFs of vorticity sheet measurements for different values of \db\ (see legend), with illustrative fits shown as dashed lines (see legend). From top to bottom, the panels show the structure width ($w$), perpendicular length ($l_\perp$), aspect ratio ($\alpha$), and length along the mean field ($l_\parallel$).}
    \label{fig:vs_db_primary}
\end{figure}

\begin{figure}[htbp]
    \centering
    \includegraphics[width=\textwidth,height=0.44\textheight,keepaspectratio]{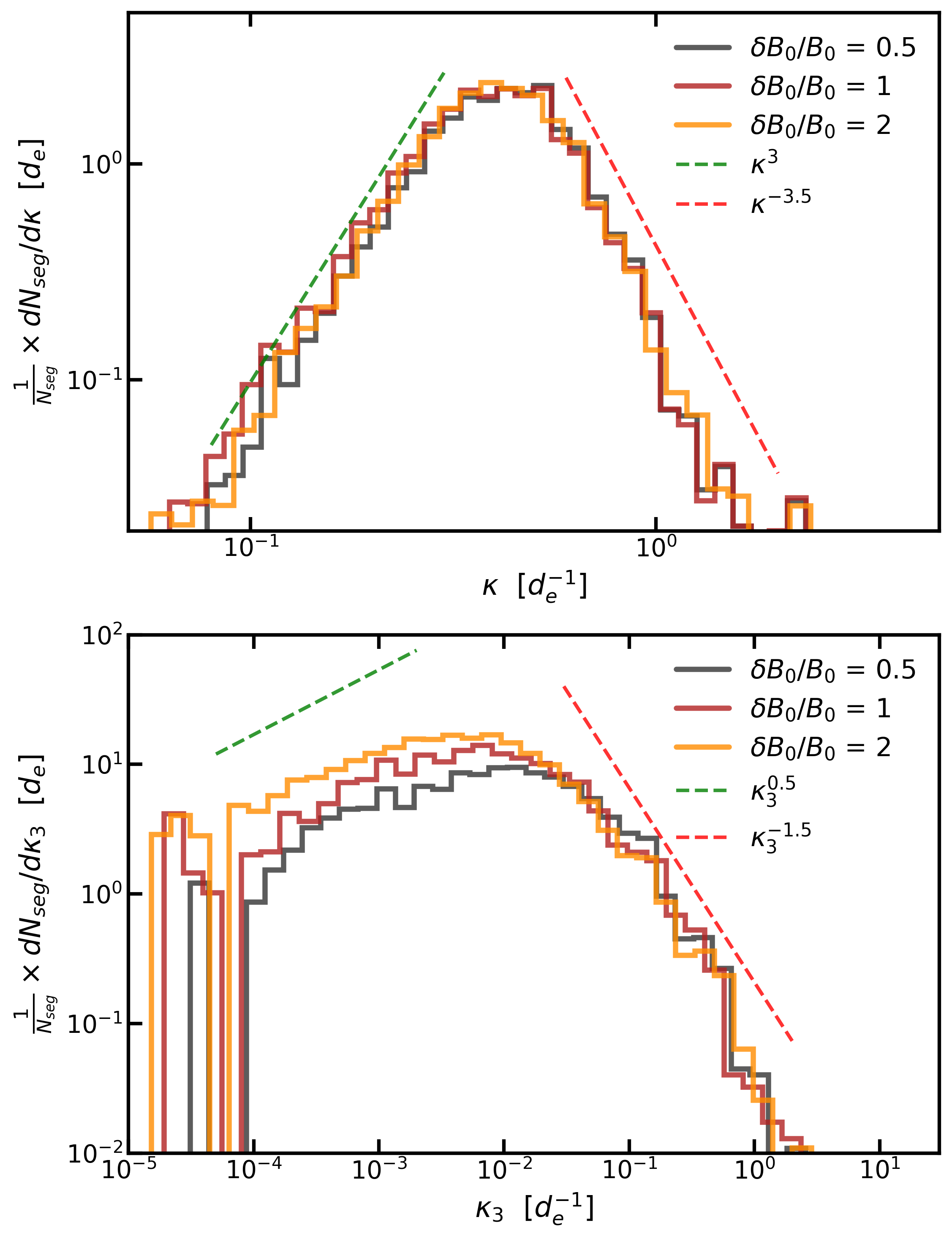}
    \caption{PDFs of vorticity sheet measurements for different values of \db (see legend), with illustrative fits shown as dashed lines (see legend). The top panel shows the local curvature $\kappa$, while the bottom panel shows the three-point curvature $\kappa_3$.}
    \label{fig:vs_db_curvature}
\end{figure}

The regions of high vorticity in the plasma exhibit trends similar to those of the current sheets, particularly the weak dependence on $\sigma$ for $N_s$, $f$, $N_{\mathrm{seg}}$, and $C_0$. The values of $N_s$, $f$, $N_{\mathrm{seg}}$, and $C_0$ are listed in Table \ref{tab:current_sheet_sum} and displayed in Figure \ref{fig:cs_simulation_trends}. Like current sheets, vorticity sheets show little variation in these quantities with $\sigma$. Larger trends are seen with changes in \db, although the effect of \db\ is less visually evident than for current sheets. Direct comparisons are shown for $\sigma$ in Figure \ref{fig:vs_slice_sigma_compare} and for \db\ in Figure \ref{fig:vs_slice_db_compare}. The values themselves are similar when using the same value of $T_{\mathrm{rms}}$. Though $T_{\mathrm{rms}} = 2$ is the fiducial value for the vorticity sheets as well, in our analysis we found vorticity to be more sensitive to this threshold. Increasing it significantly leads to very sparse regions of vorticity, whereas lowering it slightly to $T_{\mathrm{rms}} = 1.5$ results in it taking up most of the volume. For this work, we produced results for the fiducial case $T_{\mathrm{rms}} = 2$ and one additional case with $T_{\mathrm{rms}} = 1.5$.   
We calculate each measurement described in Section \ref{sec:measurements} and present them as PDFs with varying $\sigma$ in Figures \ref{fig:vs_sigma_primary} and \ref{fig:vs_sigma_curvature}. The corresponding dependence on \db\ is shown in Figures \ref{fig:vs_db_primary} and \ref{fig:vs_db_curvature}. Each PDF is fit using the same functional forms, fitting ranges, and reference curves as in Section \ref{sec:current_sheets_hists} for the current sheets, allowing for direct comparison. All fit results are provided in Appendix \ref{ap:trends_tables}.  

The vorticity sheet width $w$, shown in the top panels of Figures \ref{fig:vs_sigma_primary} and \ref{fig:vs_db_primary}, is generally larger than that of the current sheets because the latter exhibit a sharper exponential cutoff. Trends with $\sigma$ are more difficult to observe with the width of the vorticity sheets. With \db, $w$ tends to increase, similar to the current sheets, although in this case the trend is not clearly monotonic. 
Vorticity sheet measurements of $l_\perp$, shown in the second panels of Figures \ref{fig:vs_sigma_primary} and \ref{fig:vs_db_primary}, exhibit a similar shape to their current-sheet counterparts but may develop a more pronounced power-law break for small values of $l_\perp$. Additionally, vorticity sheets typically have less extreme maximal values of $l_\perp$. The quantity $l_\perp$ shows little to no change with $\sigma$ but does tend to increase with \db.

For the vorticity sheets, the aspect ratio $\alpha$, shown in the third panels of Figures \ref{fig:vs_sigma_primary} and \ref{fig:vs_db_primary}, follows a clear broken power law. The distribution appears initially flat before breaking and transitioning into a power law with an exponential cutoff. This stands in contrast to current sheets where there was significantly less evidence for a broken power law. Otherwise, the behavior of $\alpha$ is broadly consistent between vorticity and current sheets. No significant trend is observed with $\sigma$, but the maximal extent of $\alpha$ tends to decrease with decreasing \db. 
The quantity $l_\parallel$ in vorticity sheets is consistent with that in current sheets, following a broken power law with no dependence on $\sigma$ but with a maximal extent that decreases with \db. The main difference between them is that vorticity sheets have a harder slope after the peak. In turn, this makes them typically shorter along the $z$-direction than current sheets. 

The local curvature $\kappa$ for the vorticity sheets has no distinguishable differences from its counterpart in current sheets. In both cases, the PDF forms a sharply peaked broken power law around $\kappa \approx 0.5 \ d_e^{-1}$. The trends in $\kappa$ are similarly difficult to distinguish and have small, if any, dependence on $\sigma$ and \db. 

Measurements of $\kappa_3$ for vorticity sheets likewise maintain the smoothly broken power-law shape observed in current sheets. The main difference is that the pre-peak probability is lower, indicating a lower overall probability of vorticity sheets exhibiting large-radius curvatures, particularly for large values of \db, compared with current sheets. Otherwise, the PDFs trends for $\kappa_3$ remain similar to those of the current sheets; they show little noticeable change with $\sigma$, but the sheets generally become more curved with decreasing \db. 

\subsection{Parameter Dependence}
\label{sec:trends}
To gain further insight into the results from the PDFs in Sections \ref{sec:current-sheets} and \ref{sec:vorticity-sheets}, we compare the statistical means across the parameter range. We also examine the parameter dependence of specific fit parameters before looking briefly at specific interdependencies of parameters. 

For each PDF, we compute the mean and use the standard deviation ($\sigma_{\mathrm{std}}$) to evaluate the standard error $\sigma_{\mathrm{se}} = \sigma_{\mathrm{std}}/\sqrt{N_{m}}$. Here, $N_m$ is the number of measurements included in the mean. The corresponding error bars are included in all figures of the means. Due to the large number of measurements, the error bars may be too small to distinguish, but complete results are summarized in the tables in Appendix \ref{ap:mean_tables}. Each plot of the mean shows values for both $\omega$ (orange) and $j$ (blue), as well as results for $T_{\mathrm{rms}} = 1.5$ (circle, dotted line), 2 (square, solid line), and 3 (triangle, dashed line). 
\begin{figure}[htbp]
    \centering
    
    \includegraphics[width=\columnwidth]{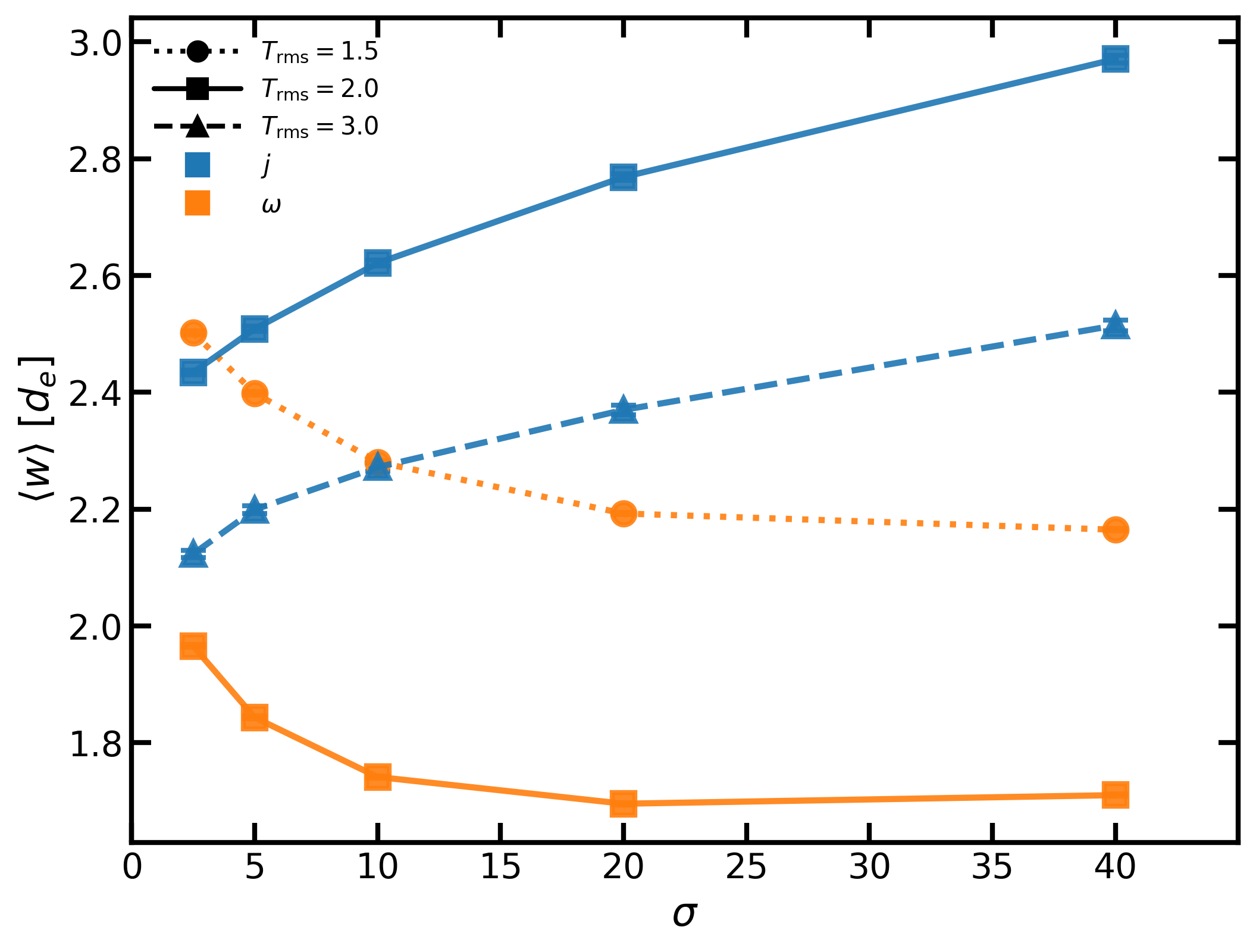}
    % \caption{Caption for the first figure}
    
    \includegraphics[width=\columnwidth]{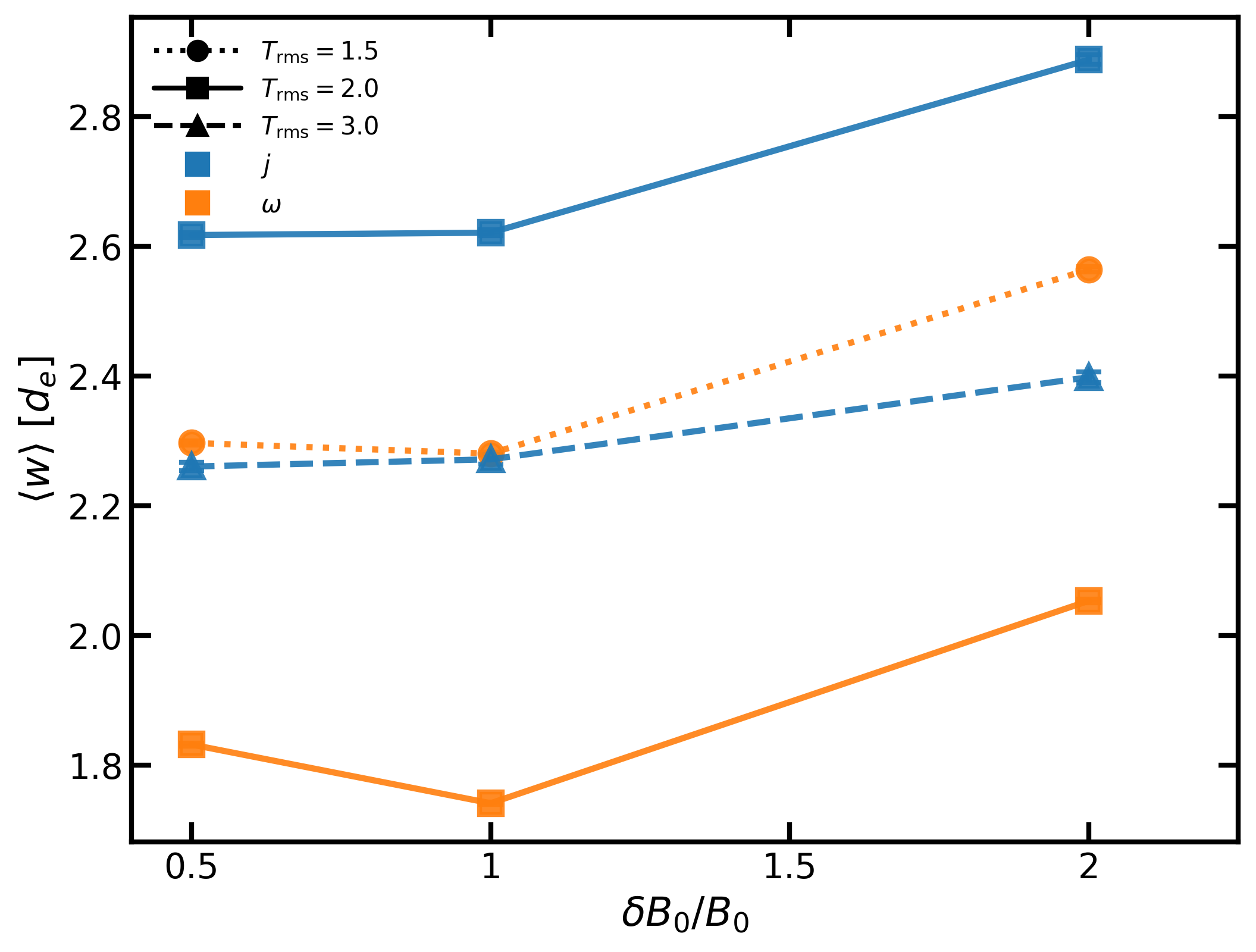}
    \caption{Mean values of $w$ for current sheets (blue) and vorticity sheets (orange) for different values of $\sigma$ (top) and \db\ (bottom). Linestyles indicate different values of $T_{rms}$: dashed for $T_{\text{rms}}=3$, solid for $T_{rms}=2$, and dotted for $T_{\text{rms}} = 1.5$.}
    \label{fig:width_mean}
\end{figure}
The mean values of $w$, shown in Figure \ref{fig:width_mean}, have consistent trends that show a slight widening of current sheets with $\sigma$ while vorticity sheets tend to become thinner. Both vorticity and current sheets approach an asymptotic behavior at large $\sigma$. With decreasing \db, both current and vorticity sheets initially decrease in size but plateau at \db $= 1$. 
\begin{figure}[htbp]
    \centering
    
    \includegraphics[width=\columnwidth]{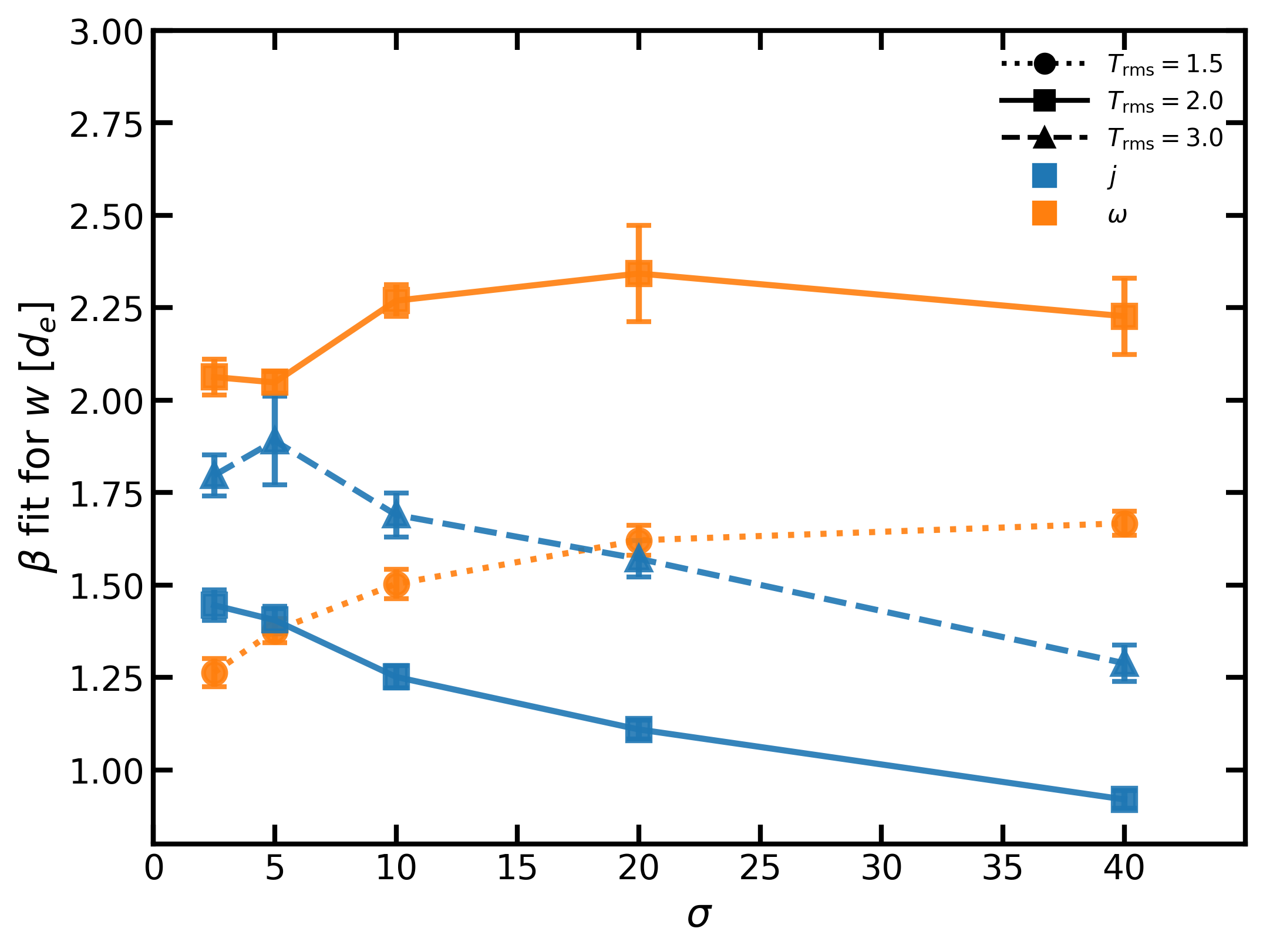}
    % \caption{Caption for the first figure}
    
    \includegraphics[width=\columnwidth]{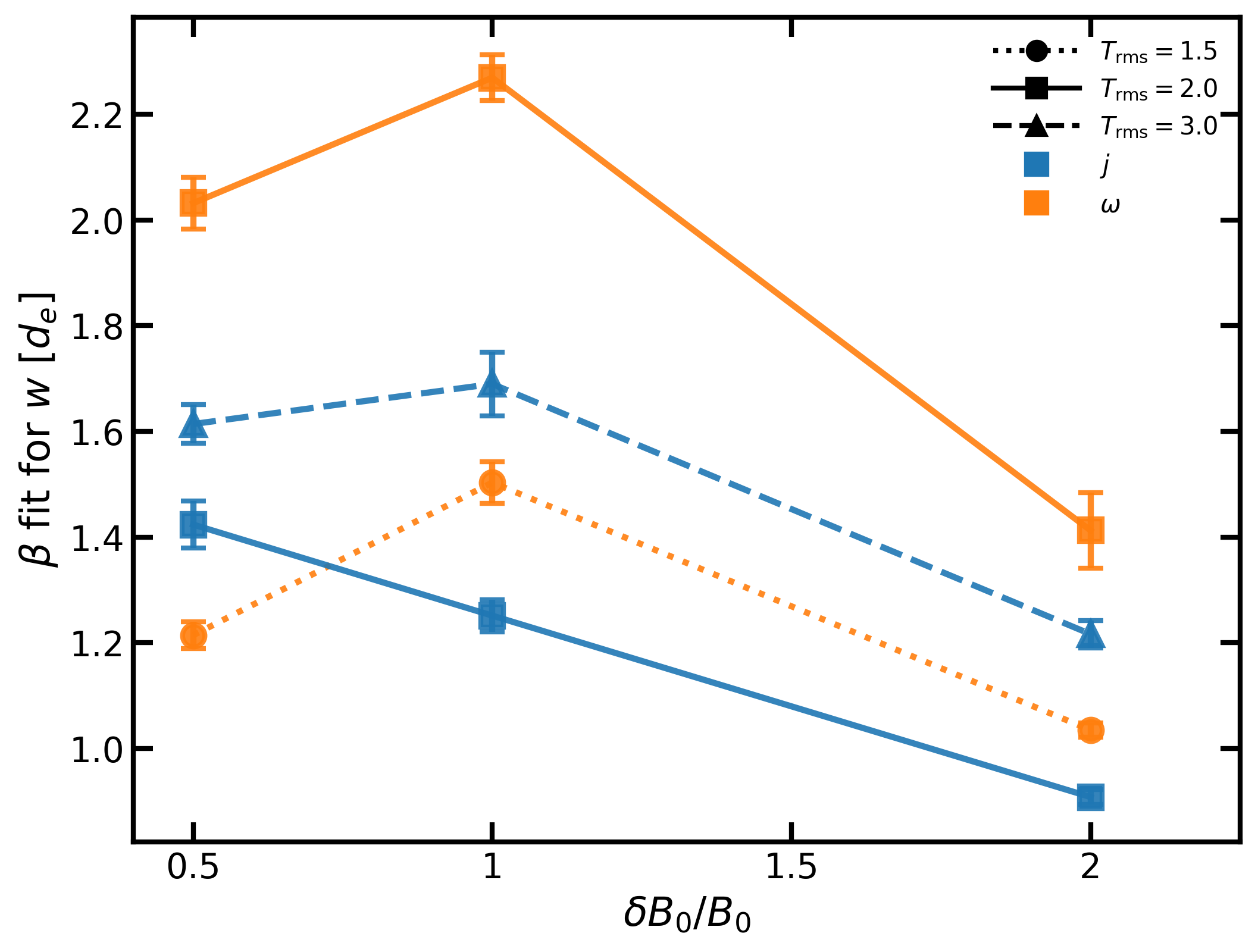}
    \caption{Values of the exponential fit parameters $\beta$ for the measurements $w$. Current sheets are shown in blue and vorticity sheets in orange for different values of $\sigma$ (top) and \db\ (bottom). Linestyles indicate different values of $T_{\text{rms}}$: dashed for $T_{\text{rms}}=3$, solid for $T_{\text{rms}}=2$, and dotted for $T_{\text{rms}} = 1.5$.}
    \label{fig:beta_for_width_seg_mean}
\end{figure}
The majority of the variation in the mean value of $w$ arises from changes in the exponential tail, as seen in Figure \ref{fig:beta_for_width_seg_mean}, where the fit for $\beta$ shows a similar dependence on $\sigma$ and \db. 
\begin{figure}[htbp]
    \centering
    
    \includegraphics[width=\columnwidth]{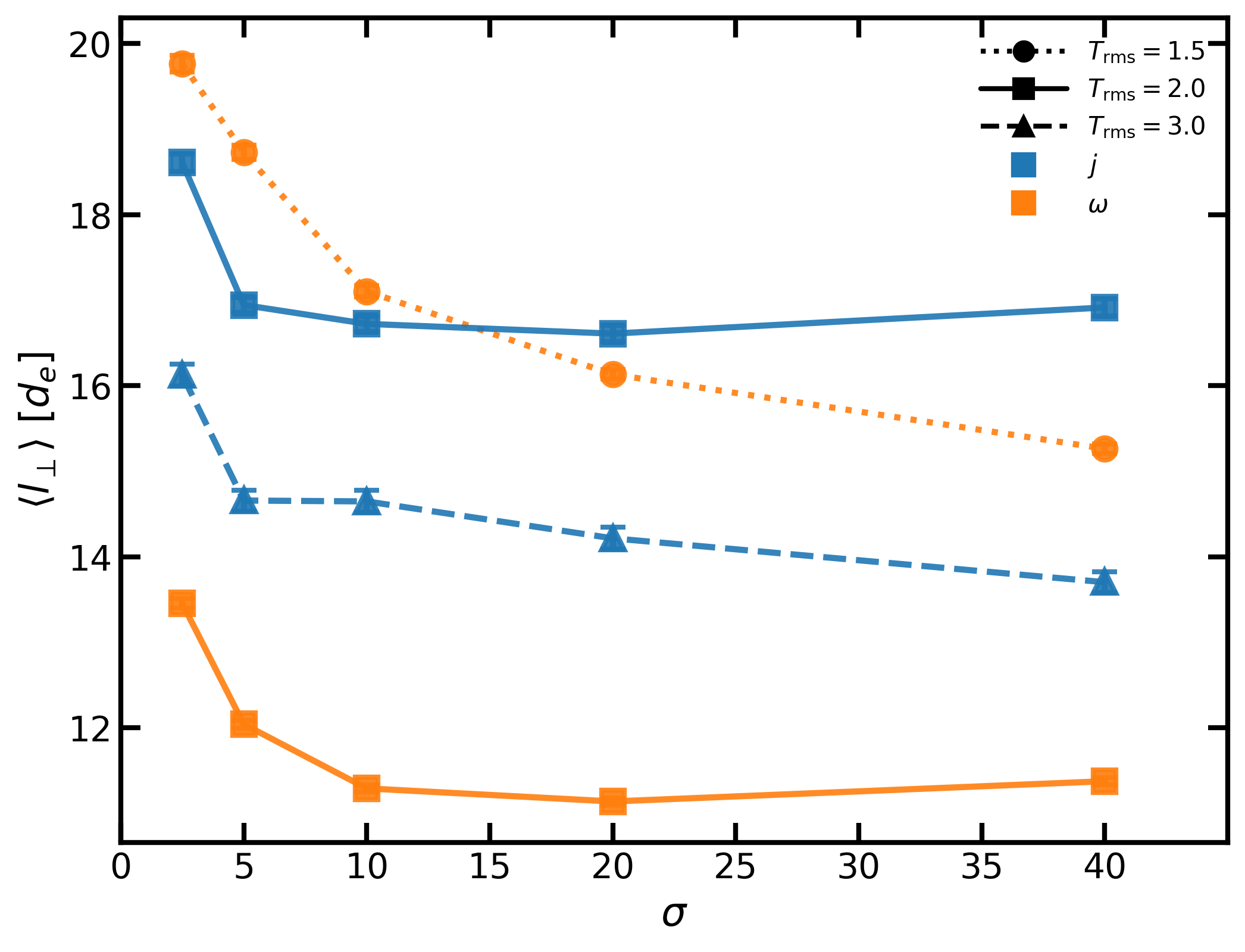}
    
    \includegraphics[width=\columnwidth]{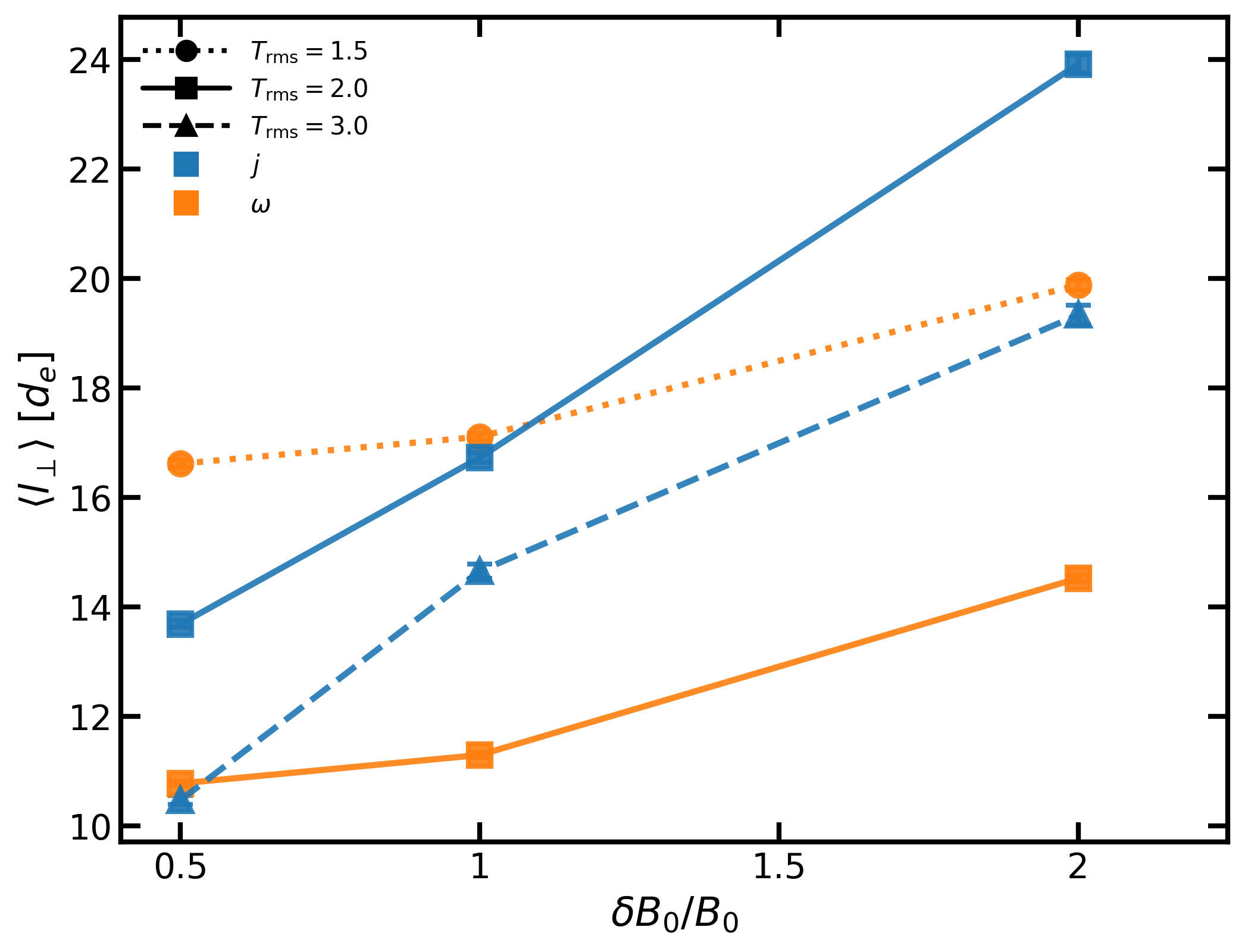}
    \caption{Mean values of $l_\perp$ for current sheets (blue) and vorticity sheets (orange) for different values of $\sigma$ (top) and \db\ (bottom). Linestyles indicate different values of $T_{\text{rms}}$: dashed for $T_{\text{rms}}=3$, solid for $T_{\text{rms}}=2$, and dotted for $T_{\text{rms}} = 1.5$.}
    \label{fig:arc_length_mean}
\end{figure}
The mean values of $l_\perp$ for both current and vorticity sheets are generally independent of $\sigma$, aside from an initial decrease between $\sigma = 2.5$ and $\sigma = 5$. With increasing \db, the mean $l_\perp$ increases rapidly for current sheets, while for vorticity sheets increase with \db\ is more modest. 
\begin{figure}
    \centering
    \includegraphics[width=\columnwidth]{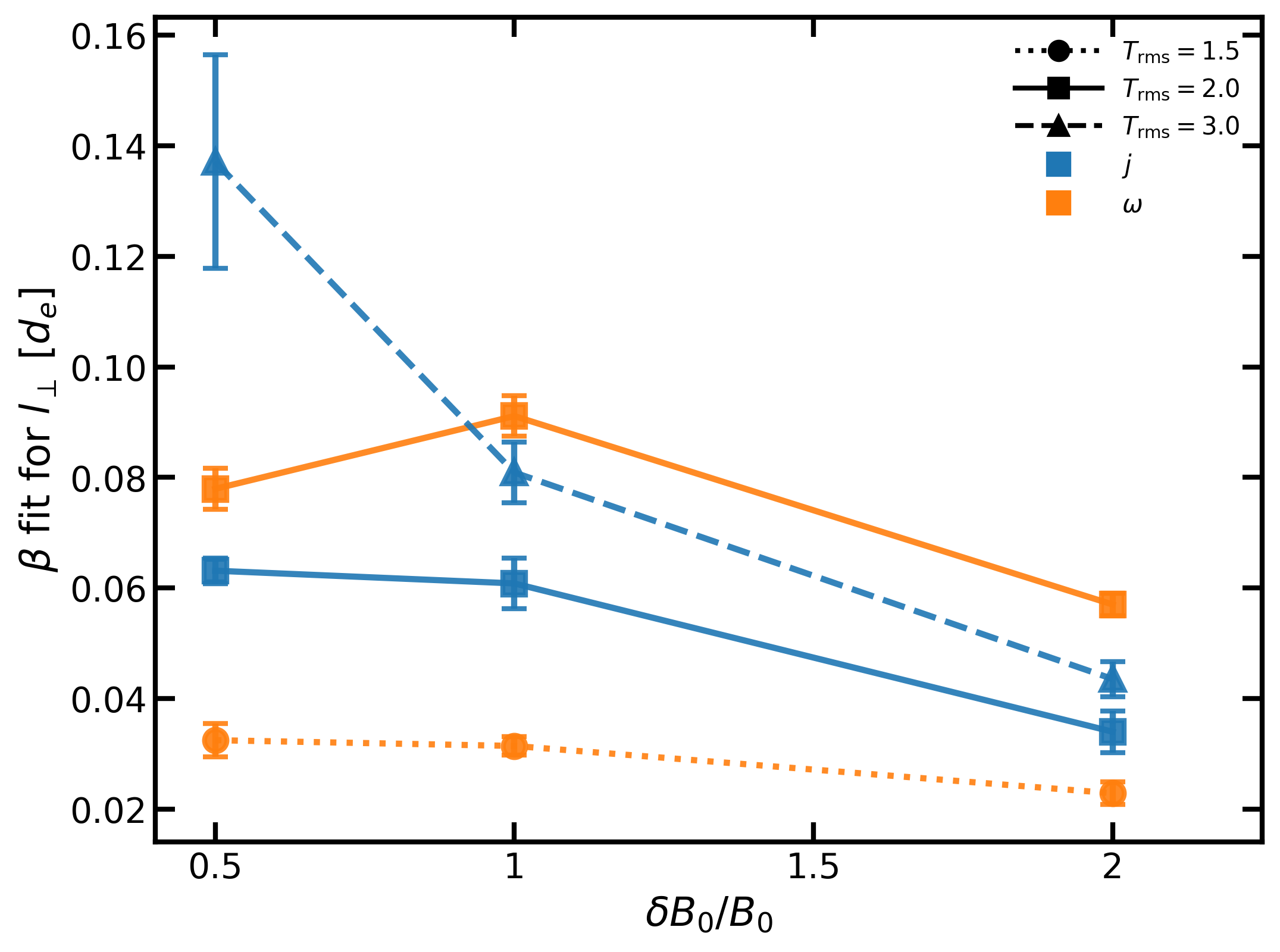}
    \caption{Values of the exponential fit parameters $\beta$ for the measurements $l_\perp$. Current sheets are shown in blue and vorticity sheets in orange for different values of $\sigma$ (top) and \db\ (bottom). Linestyles indicate different values of $T_{\text{rms}}$. Dashed shows $T_{\text{rms}}=3$, solid $T_{\text{rms}}=2$, and dotted $T_{\text{rms}} = 1.5$.}
    \label{fig:beta_for_arc_length_vs_db}
\end{figure}
The dependence of the mean $l_\perp$ on \db\ arises mainly from the modest change in the strength of the exponential cutoff shown in Figure \ref{fig:beta_for_arc_length_vs_db}. The power-law portion of the fit remains relatively constant. 
\begin{figure}[htbp]
    \centering
    
    \includegraphics[width=\columnwidth]{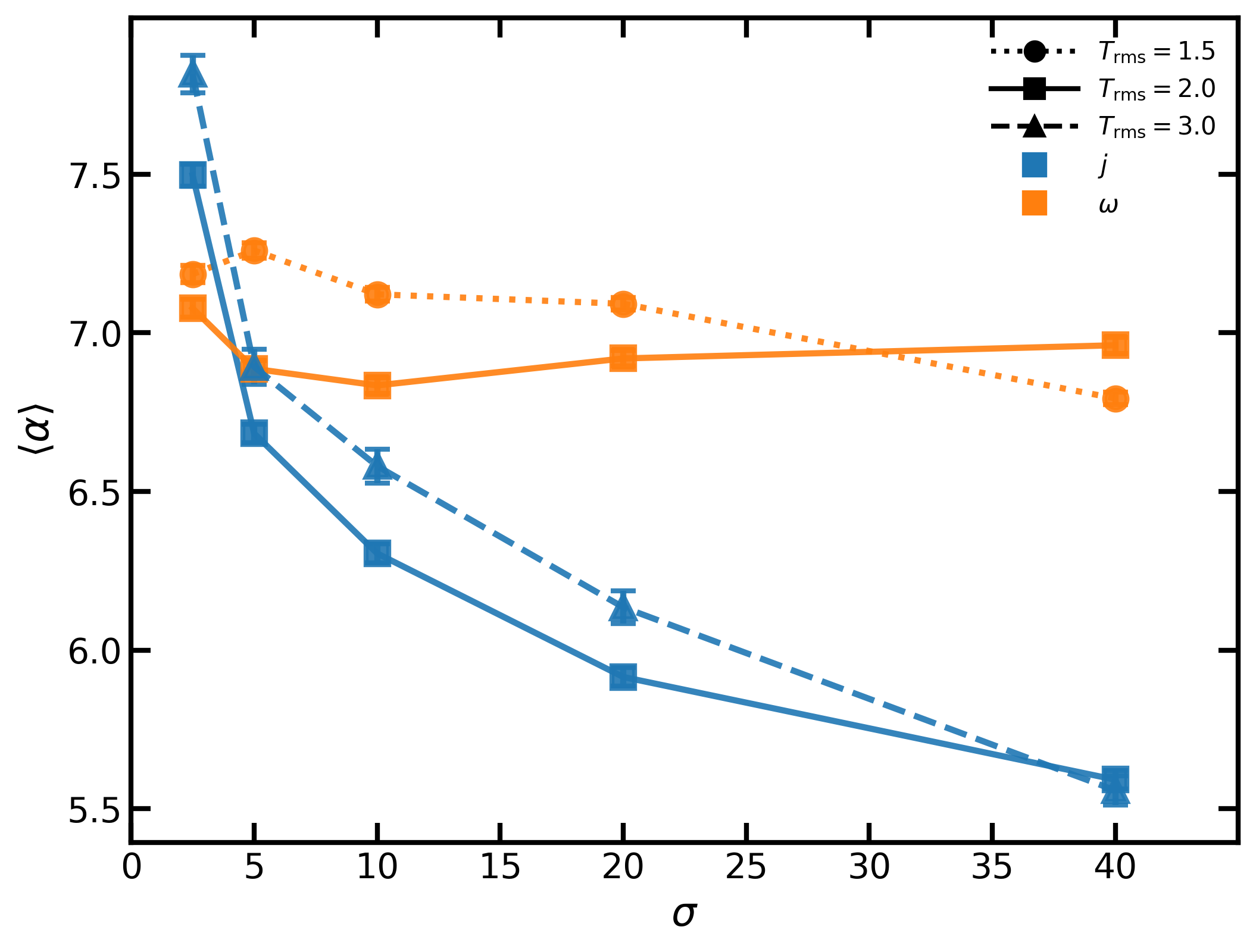}
    
    \includegraphics[width=\columnwidth]{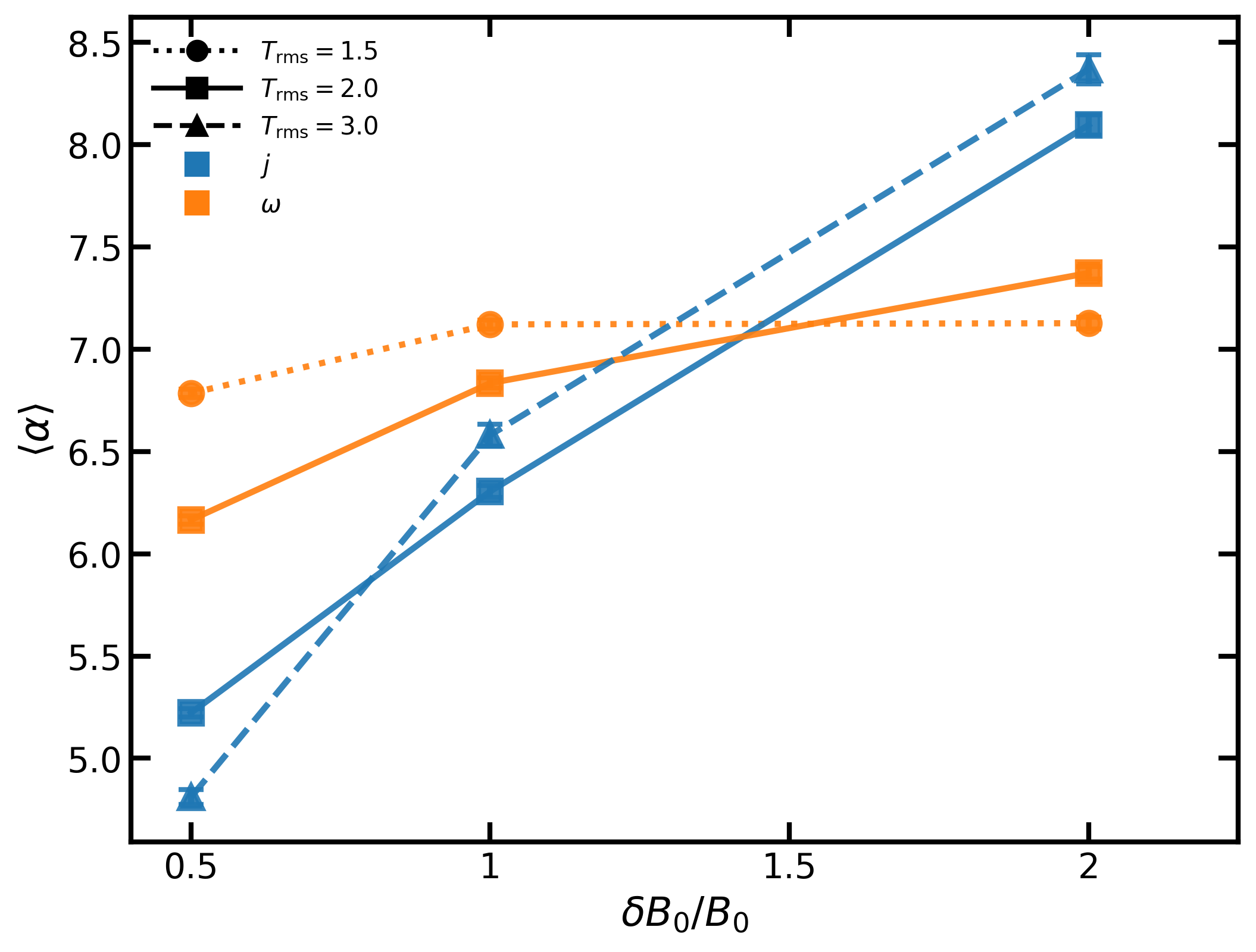}
    \caption{Mean values of the aspect ratio $\alpha$ for current sheets (blue) and vorticity sheets (orange) for different values of $\sigma$ (top) and \db\ (bottom). Linestyles indicate different values of $T_{\text{rms}}$: dashed for $T_{\text{rms}}=3$, solid for $T_{\text{rms}}=2$, and dotted for $T_{\text{rms}} = 1.5$.}
    \label{fig:aspect_ratio_mean}
\end{figure}
With $\sigma$ the mean values of $\alpha$ in current sheets decrease rapidly compared to those in vorticity sheets. The \db \ case sees $\alpha$ increase with \db\ for current sheets much more rapidly than for vorticity sheets. These results are shown in Figure \ref{fig:aspect_ratio_mean}. 
\begin{figure}
    \centering
    \includegraphics[width=\columnwidth]{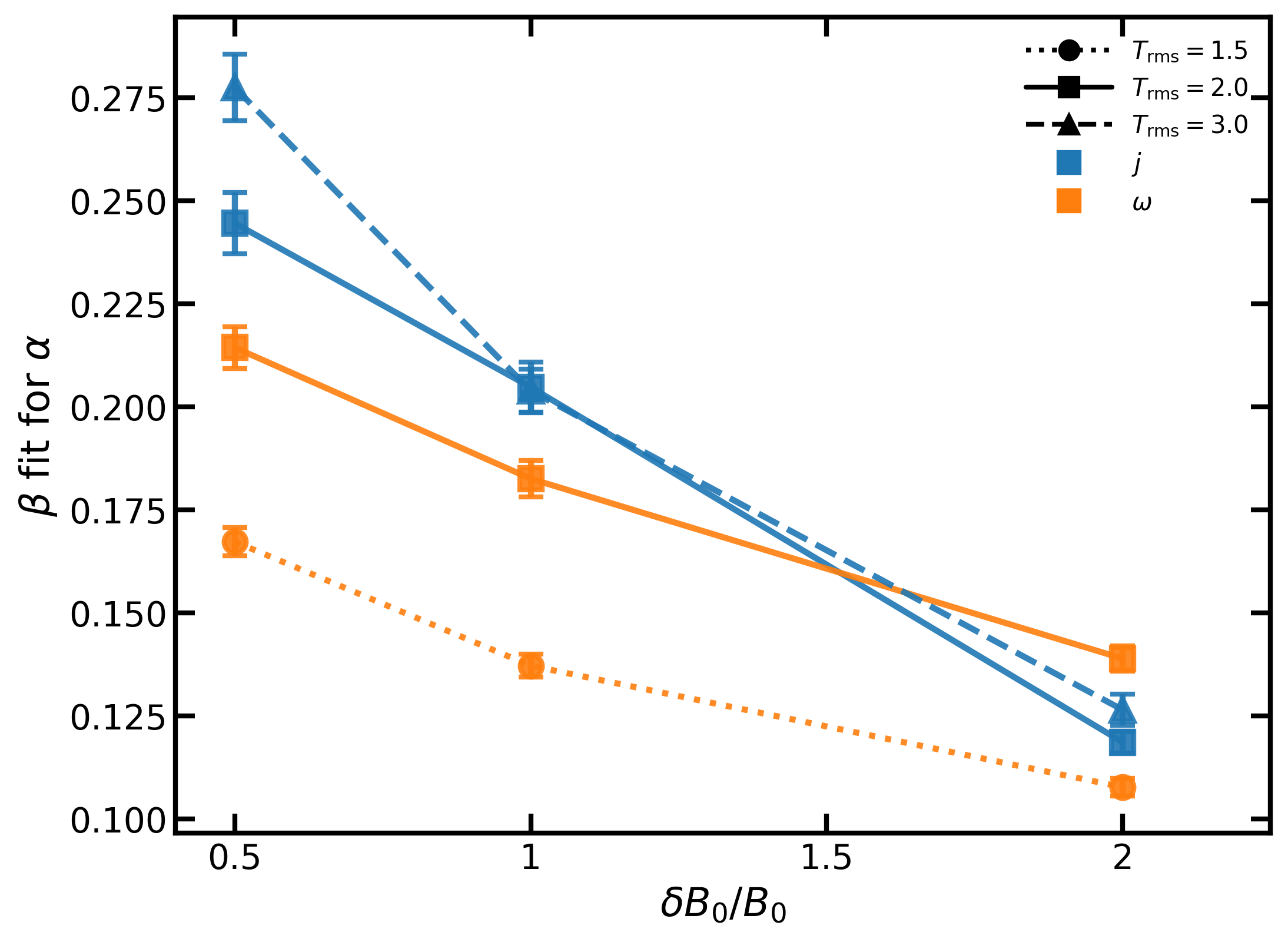}
    \caption{Values of the exponential fit parameters $\beta$ for the measurements $\alpha$. Current sheets are shown in blue and vorticity sheets in orange for different values of $\sigma$ (top) and \db\ (bottom). Linestyles indicate different values of $T_{\text{rms}}$: dashed for $T_{\text{rms}}=3$, solid for $T_{\text{rms}}=2$, and dotted for $T_{\text{rms}} = 1.5$.}
    \label{fig:beta_for_aspect_ratio_vs_db}
\end{figure}
This increase with \db\ is, similarly to the mean values of $l_\perp$, mostly due to the strength of the exponential decay decreasing with \db as seen in Figure \ref{fig:beta_for_aspect_ratio_vs_db}. It is worth noting that mean values for $\alpha$ ranging from 5-8 are consistent with fast reconnecting sheets \citep{Liu2017,Mbarek2022}. 
\begin{figure}[htbp]
    \centering
    
    \includegraphics[width=\columnwidth]{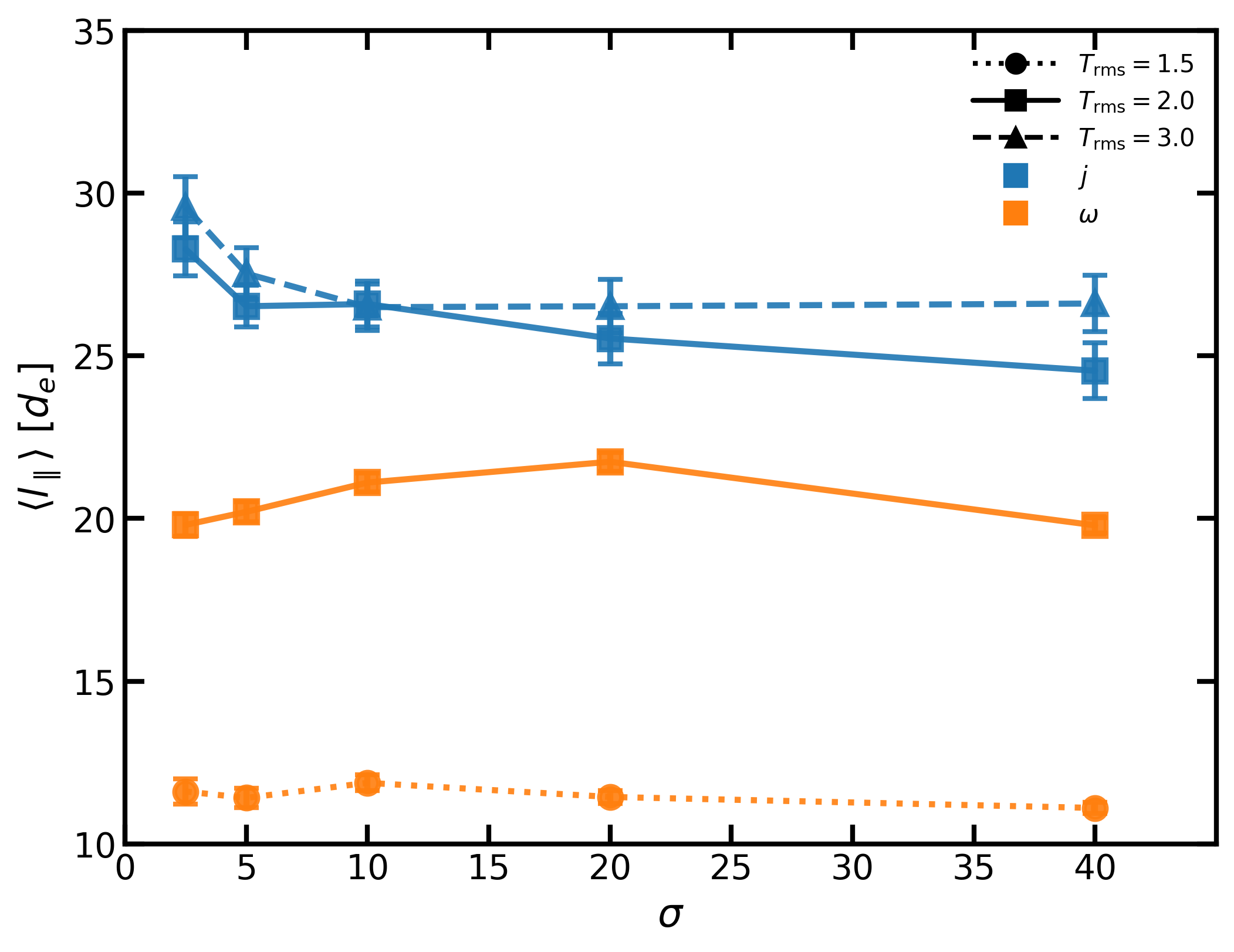}
    
    \includegraphics[width=\columnwidth]{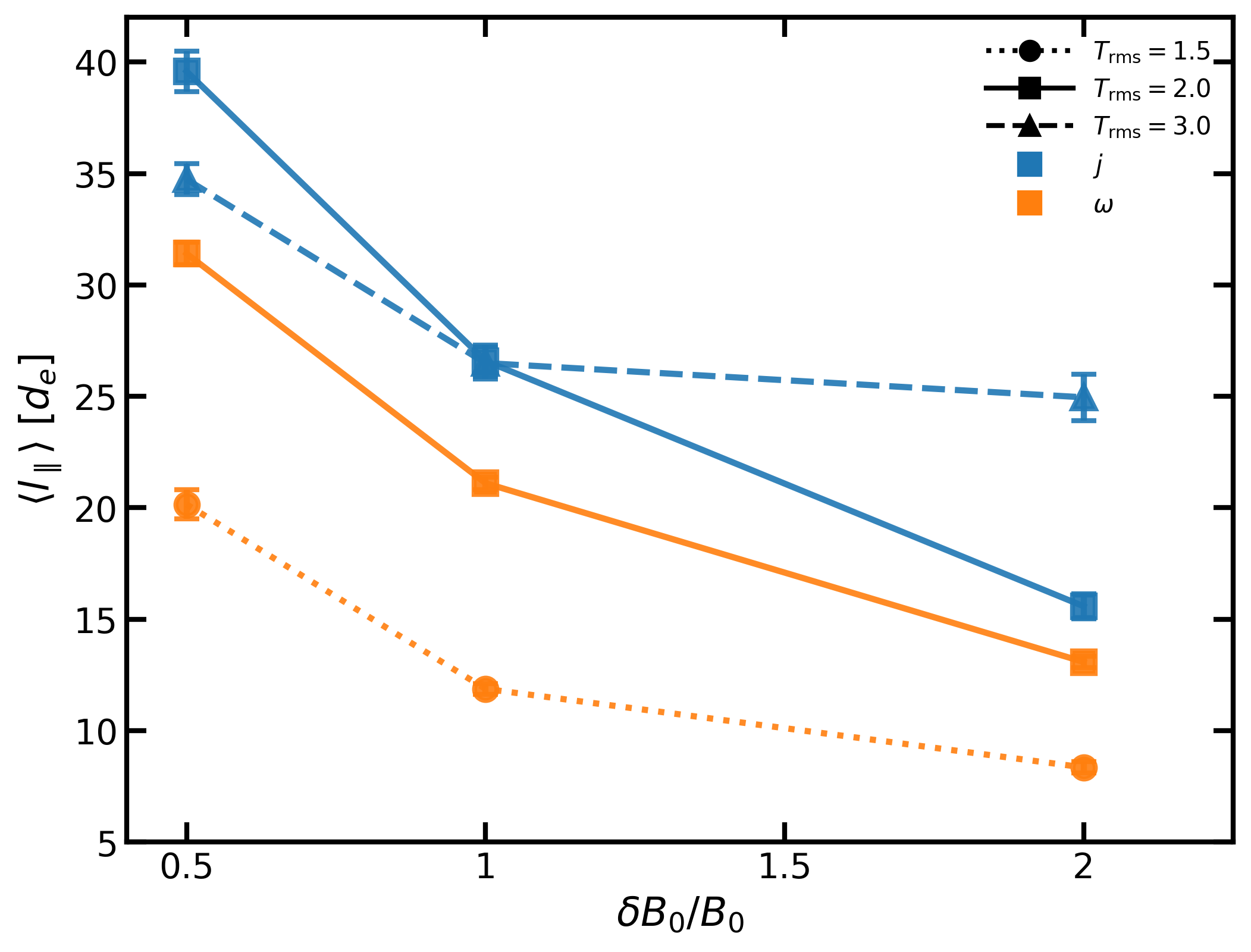}
    \caption{Mean values of $l_\parallel$ for current sheets (blue) and vorticity sheets (orange) for different values of $\sigma$ (top) and \db\ (bottom). Linestyles indicate different values of $T_{\text{rms}}$: dashed for $T_{\text{rms}}=3$, solid for $T_{\text{rms}}=2$, and dotted for $T_{\text{rms}} = 1.5$.}
    \label{fig:z_length_mean}
\end{figure}
The mean of $l_\parallel$ has minimal dependence on $\sigma$, only appearing to drop initially from $\sigma = 2.5$ to $\sigma = 5$ before flattening for higher values of $\sigma$. For vorticity sheets, the mean of $l_\parallel$ remains nearly constant. When compared across \db, both are shown to decrease sharply as \db{} increases. 
\begin{figure}
    \centering
    \includegraphics[width=\columnwidth]{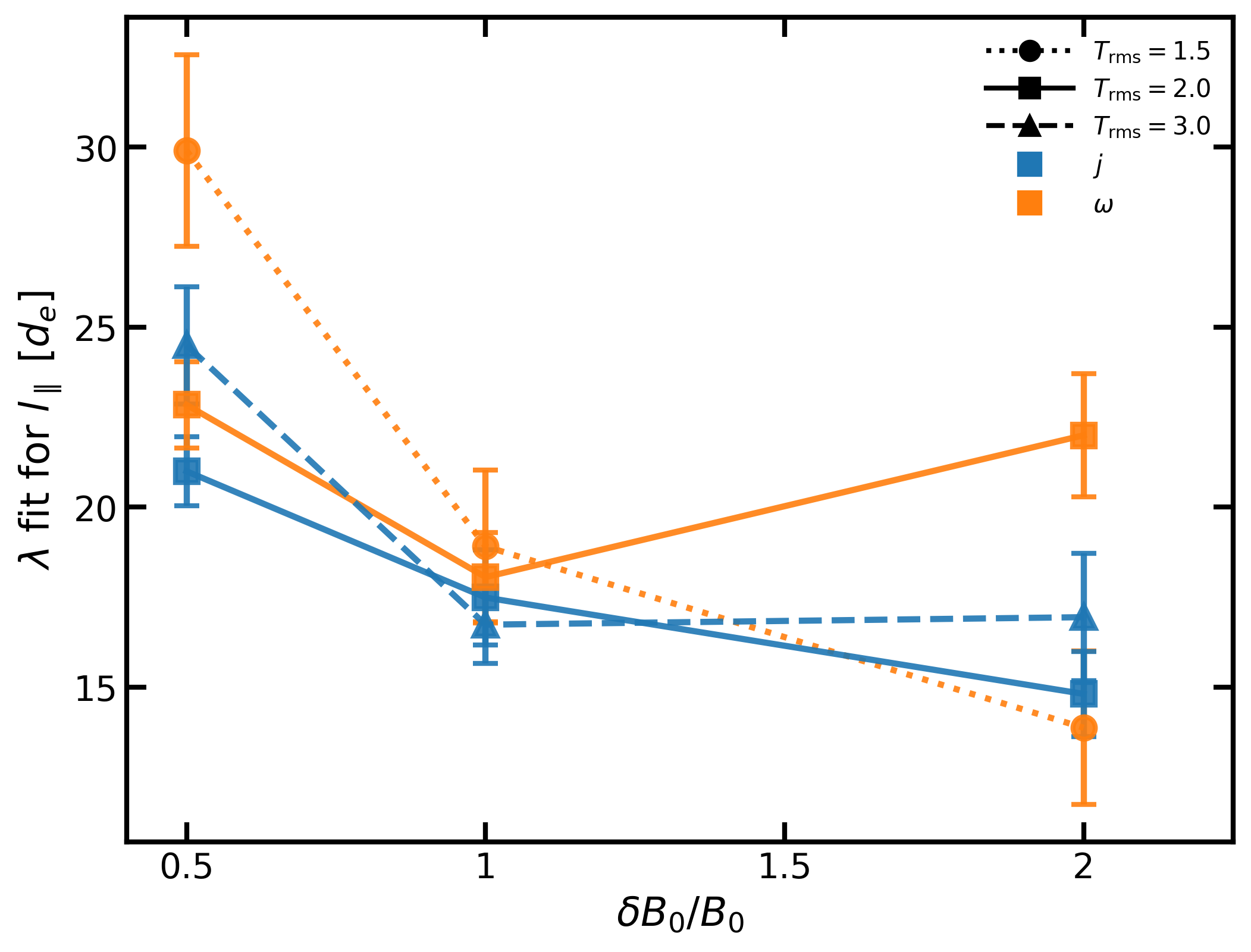}
    \caption{Values of the power-law break parameter $\lambda$ for the measurements $l_\parallel$. Current sheets are shown in blue and vorticity sheets in orange for different values of $\sigma$ (top) and \db\ (bottom). Linestyles indicate different values of $T_{\text{rms}}$: dashed for $T_{\text{rms}}=3$, solid for $T_{\text{rms}}=2$, and dotted for $T_{\text{rms}} = 1.5$.}
    \label{fig:lambda_z_length_vs_db}
\end{figure}
The trend of the mean $l_\parallel$ with \db\ is primarily due to the shifting peak, as can be seen in the PDF at the bottom of Figure \ref{fig:cs_db_primary} or in the fitted value of $\lambda$ for $l_\parallel$ shown in Figure \ref{fig:lambda_z_length_vs_db}.
\begin{figure}[htbp]
    \centering
    
    \includegraphics[width=\columnwidth]{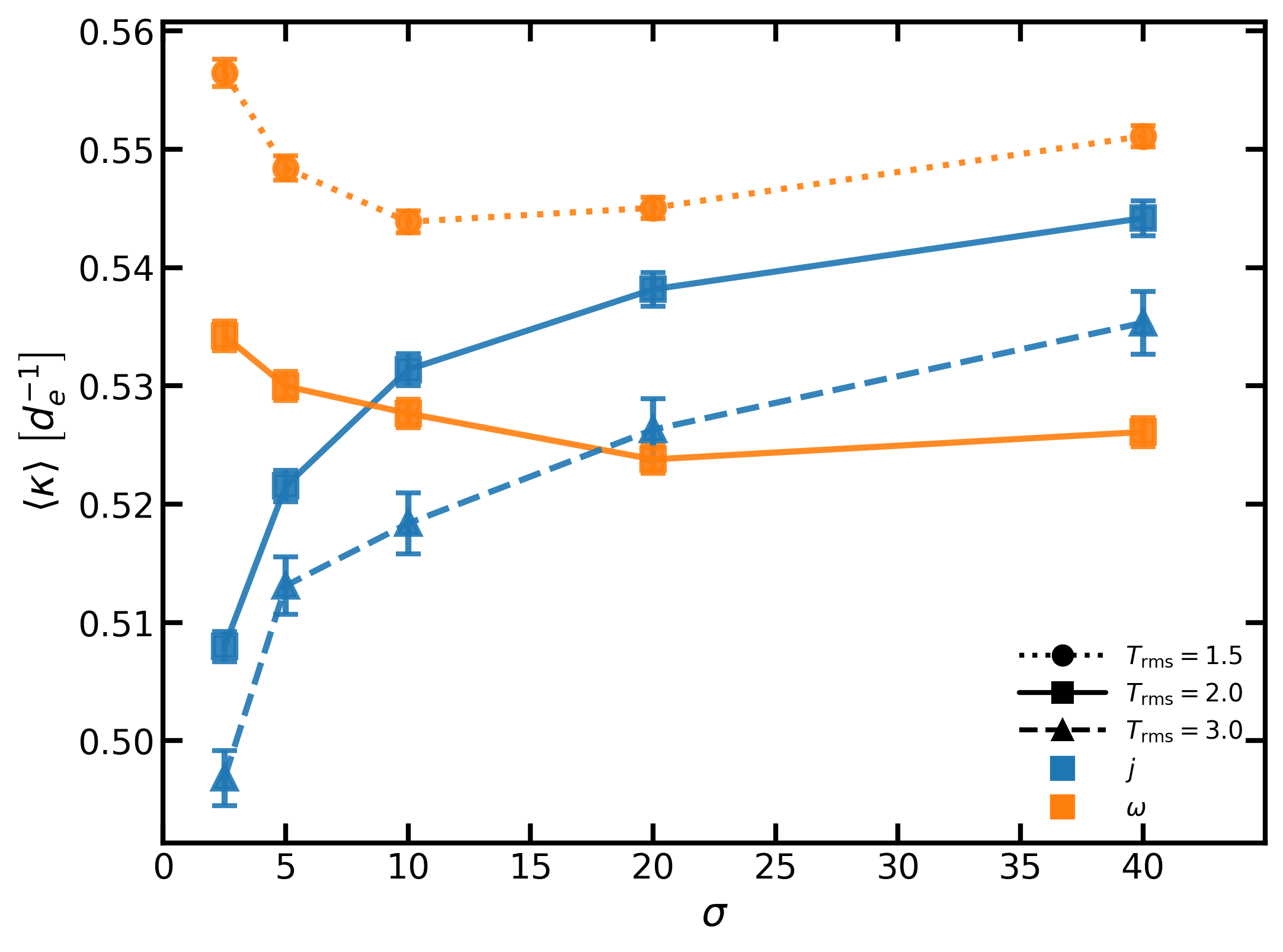}
    
    \includegraphics[width=\columnwidth]{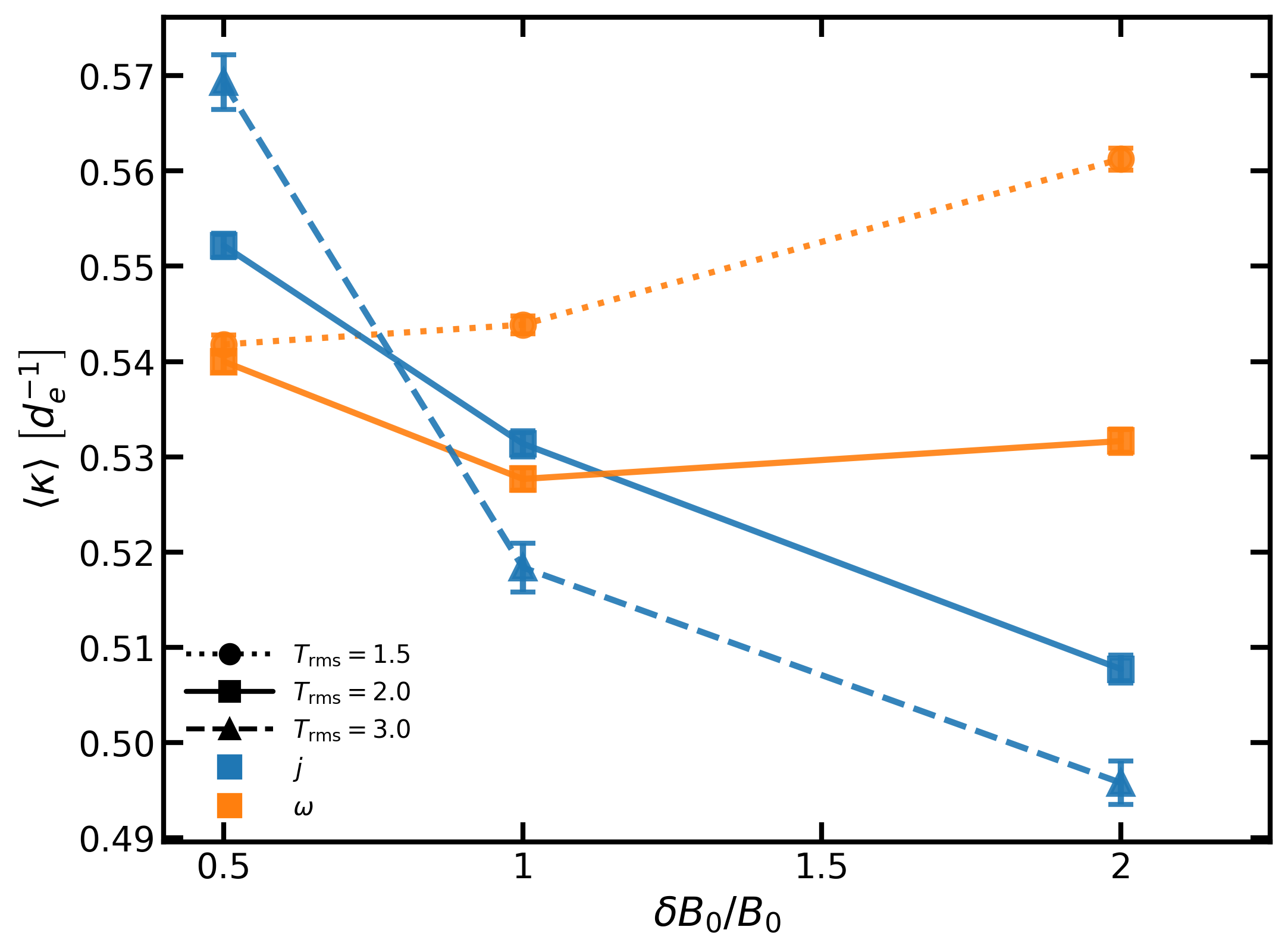}
    \caption{Mean values of $\kappa$ for current sheets (blue) and vorticity sheets (orange) for different values of $\sigma$ (top) and \db\ (bottom). Linestyles indicate different values of $T_{\text{rms}}$: dashed for $T_{\text{rms}}=3$, solid for $T_{\text{rms}}=2$, and dotted for $T_{\text{rms}} = 1.5$.}
    \label{fig:curv_segs_mean}
\end{figure}
The mean values of $\kappa$ remain fairly constant, showing little variation with $\sigma$ or \db as shown in Figure \ref{fig:curv_segs_mean}. The averages hover around $1/\kappa \approx 2 \ d_e$, suggesting that local changes in the current sheet occur on the scale of the current sheet width. 

\begin{figure}[htbp]
    \centering
    
    \includegraphics[width=\columnwidth]{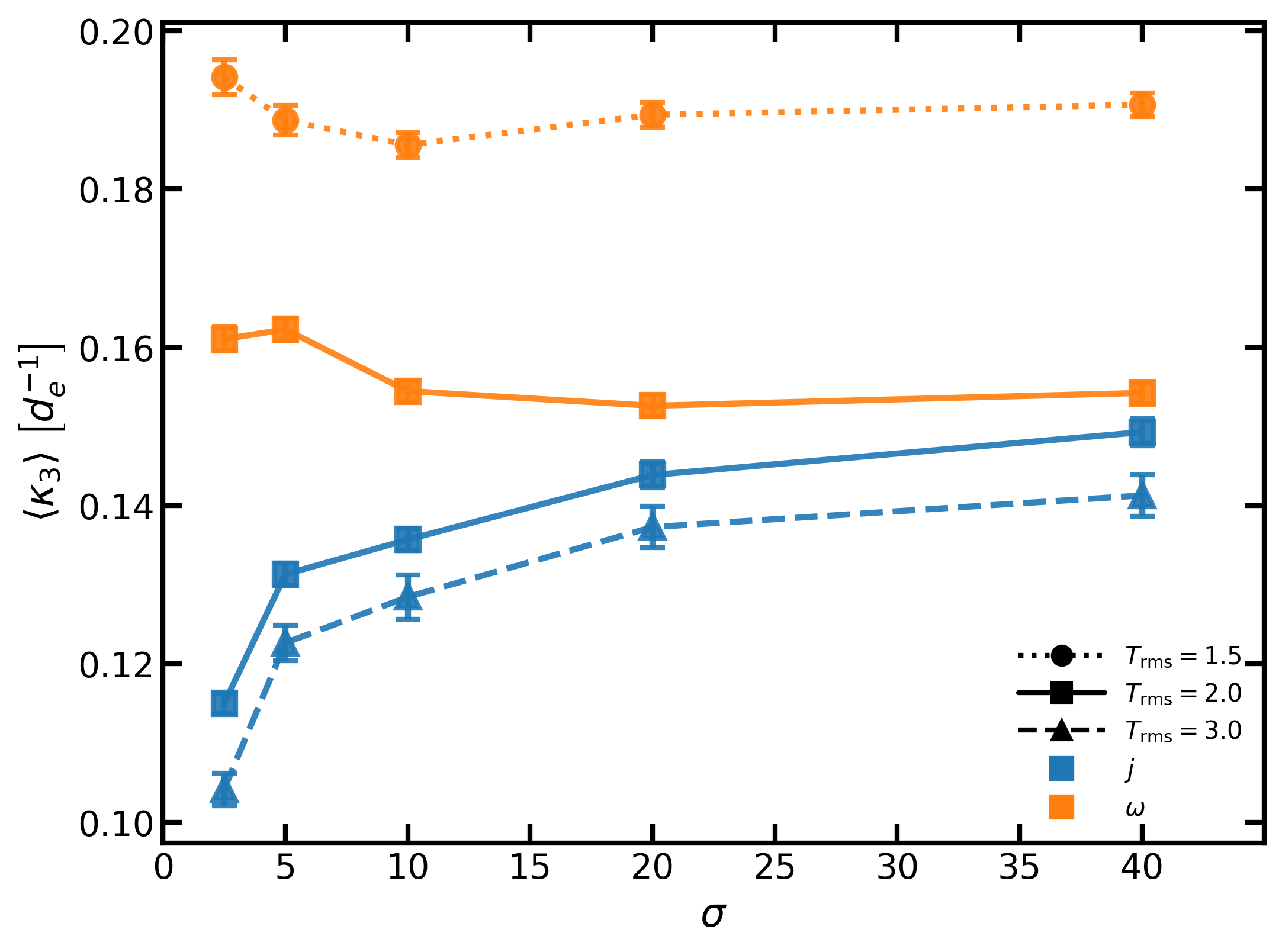}
    
    \includegraphics[width=\columnwidth]{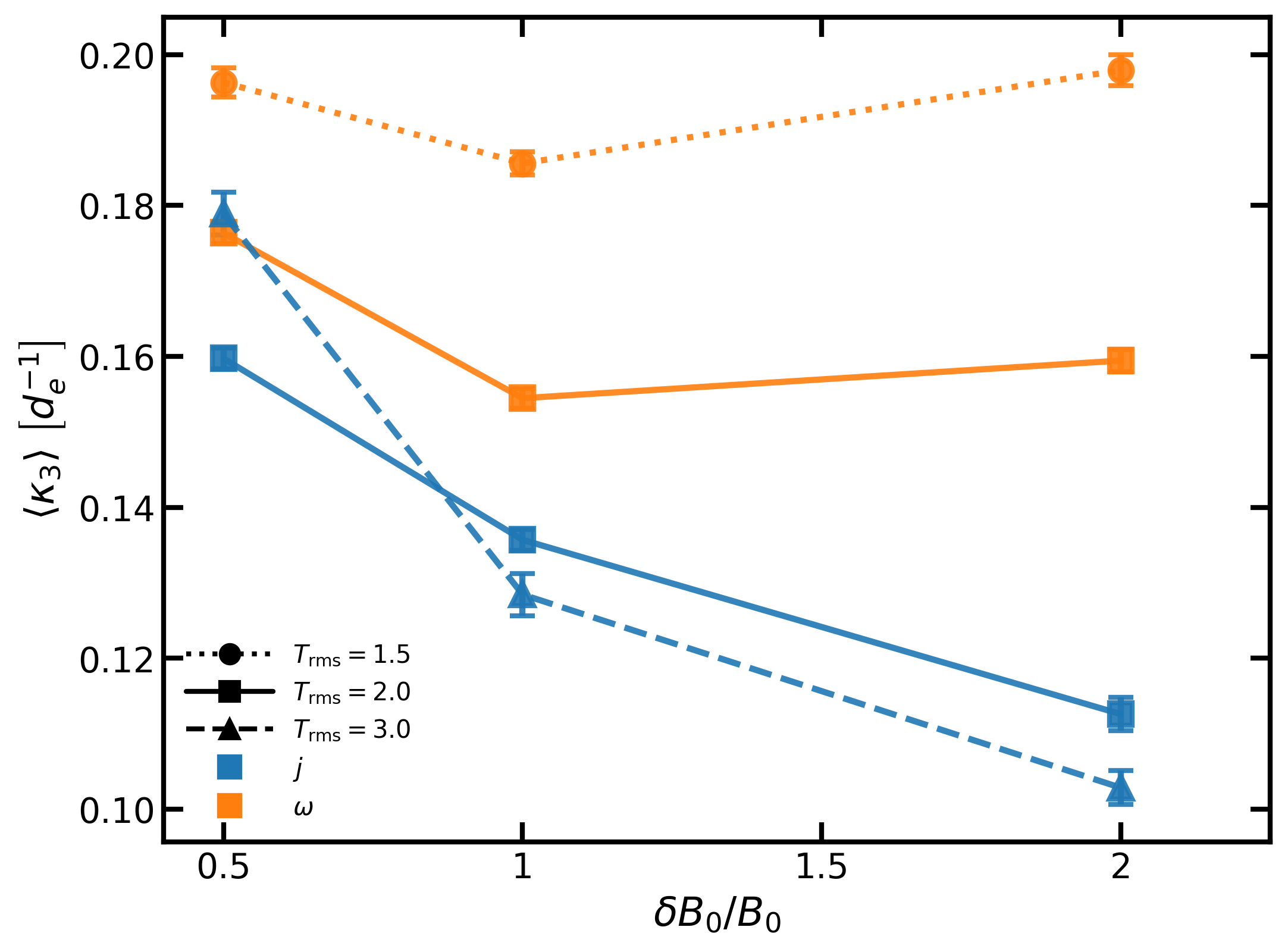}
    \caption{Mean values of $\kappa_3$ for current sheets (blue) and vorticity sheets (orange) for different values of $\sigma$ (top) and \db\ (bottom). Linestyles indicate different values of $T_{\text{rms}}$: dashed for $T_{\text{rms}}=3$, solid for $T_{\text{rms}}=2$, and dotted for $T_{\text{rms}} = 1.5$.}
    \label{fig:curv_3p_mean}
\end{figure}

In Figure \ref{fig:curv_3p_mean}, we show the mean of $\kappa_3$ and its dependence on $\sigma$ and \db, respectively. Vorticity sheets exhibit little to no change with either $\sigma$ or \db, despite the PDF of $\kappa_3$ having a noticeable dependence on \db. For current sheets, $\kappa_3$ shows a stronger dependence on $\sigma$, generally increasing asymptotically with $\sigma$, and also shows a general decrease with \db. 
Next, we examine the dependence of measurements on each other. Due to the large number of plots required for this, we show 2D PDFs only for the fiducial case ($T_{\mathrm{rms}} = 2$) and discuss a select few that exhibit strong trends. All figures are provided in Appendix \ref{ap:extra_pdfs}. In each figure in Appendix \ref{ap:extra_pdfs}, the top left plots show the 2D PDF for current sheets, while the top right plots show the corresponding PDF for vorticity sheets. Each PDF corresponds to the $\sigma = 10$, \db $= 1$ case. In the bottom plots, we show the slope dependence on $\sigma$ (left) and \db\ (right). In Figure \ref{fig:curv_3p_vs_alpha}, we can see a general trend, independent of $\sigma$ and \db, where increasingly large curvature leads to smaller aspect ratio sheets. This trend is seen in both current and vorticity sheets, although current sheets have a slight deviation at $\kappa_3 = 10^{-2}$, where the dependence appears to steepen. Figure \ref{fig:l_perp_vs_w} shows that current sheets' $w$ has a slight dependence on $l_\perp$ where larger values of $l_\perp$ are more likely to coincide with larger values of $w$. Though this trend is weak, it all but disappears for vorticity sheets where $w$ seemingly has no dependence on $l_\perp$. 

When constructing 2D PDFs that include $l_\parallel$, we need to average the other measurements over the structure (previously, they were averaged over a segment), since $l_\parallel$ is only a single measurement per structure. To indicate this, we introduce the notation $\langle X \rangle_s$, where $X$ is a measurement placeholder and the angle brackets with subscript $s$ indicate that all of the measurements within a structure are averaged together. With this, we first look at Figure \ref{fig:stuct_ave_l_perp_vs_l_para} where we show $\langle l_\perp \rangle_s$ vs $l_\parallel$. This allows us to investigate the structures’ critical balance, for which we may expect that $l_\parallel \propto \langle l_\perp \rangle_s^{2/3}$ \citep{goldreich1995}, a relationship generally recovered in Figure \ref{fig:stuct_ave_l_perp_vs_l_para}. In particular, for vorticity sheets, the slope varies between 0.59 and 0.67, with no noticeable dependence on \db\ and a slight decrease with increasing $\sigma$. For current sheets, the slope spans a wider range, from 0.61 to 0.86, and shows a slight increasing trend with $\sigma$ but decreases with \db.

\subsection{Spatial Correlation Between Current and Vorticity Sheets}
\label{cs_vs_relationship}
\begin{figure}[htbp]
    \centering
    \includegraphics[width=\columnwidth]{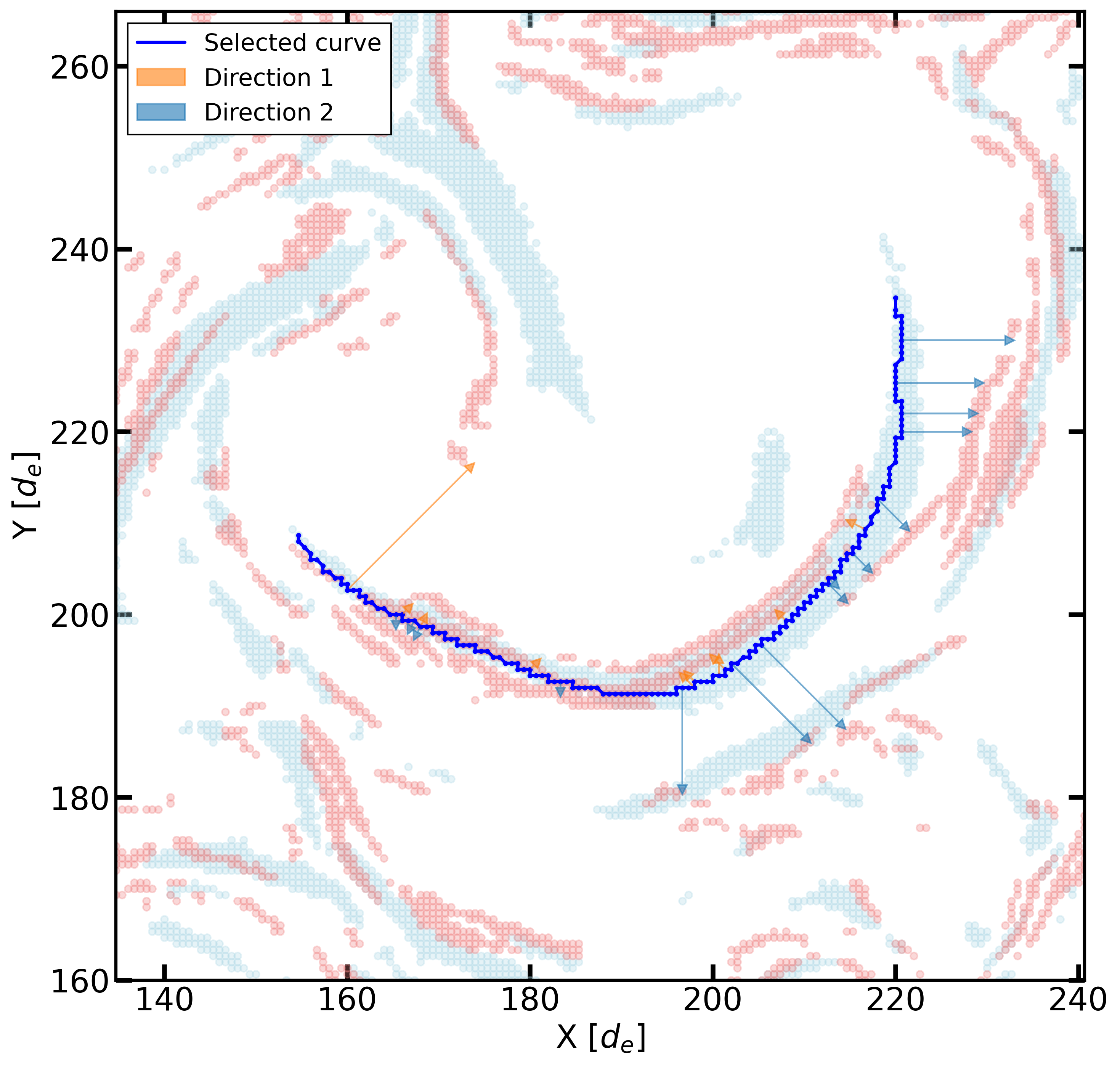}
    \caption{Illustration of the distance calculation method. The current sheet (light blue) is shown with its spline fit (blue line), normal vectors (arrows), and the measured distances to neighboring vorticity sheets (red) on both sides. The distance $d_w$ is measured from the current sheet edge along the normal direction.}
    \label{fig:distance_computation_illustration}
\end{figure}

As seen in Figure \ref{fig:10_largest}, current sheets are often surrounded by vorticity sheets. In order to quantify this, we use the splines of structures to measure the distance to the nearest vorticity sheet on both sides of each current sheet. Specifically, for every segment in a given current sheet, at the same point where we measure the width of the segment, we move along the normal direction until reaching the vorticity sheet. We repeat this procedure for both positive and negative normal directions and then average over the segment to obtain the distance measure $d_w$. An example of this procedure for a single structure in a given slice is shown in Figure \ref{fig:distance_computation_illustration}. 

\begin{figure}[htbp]
    \centering
    \includegraphics[width=\columnwidth]{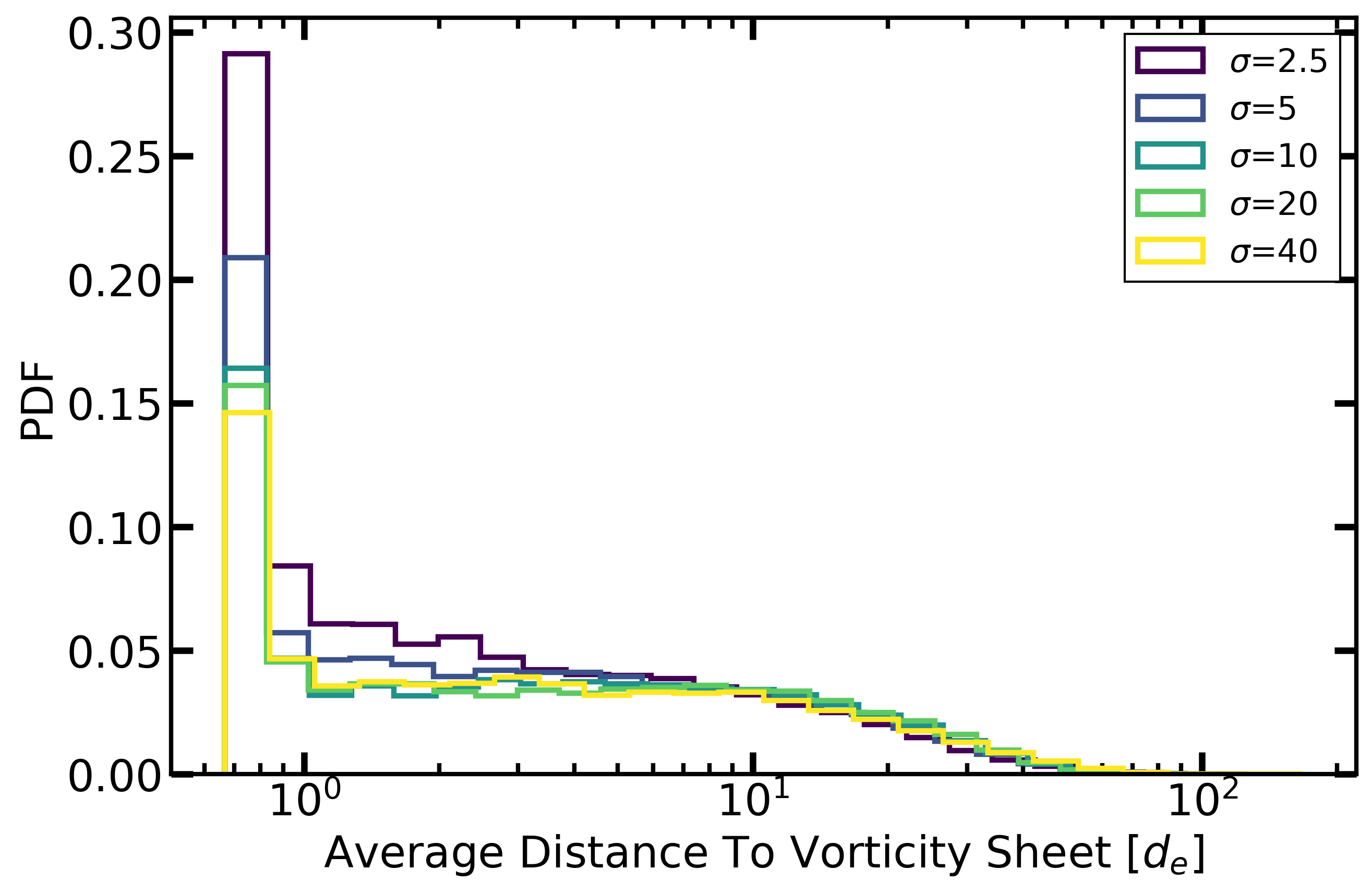}
    % \caption{Caption}
    \includegraphics[width=\columnwidth]{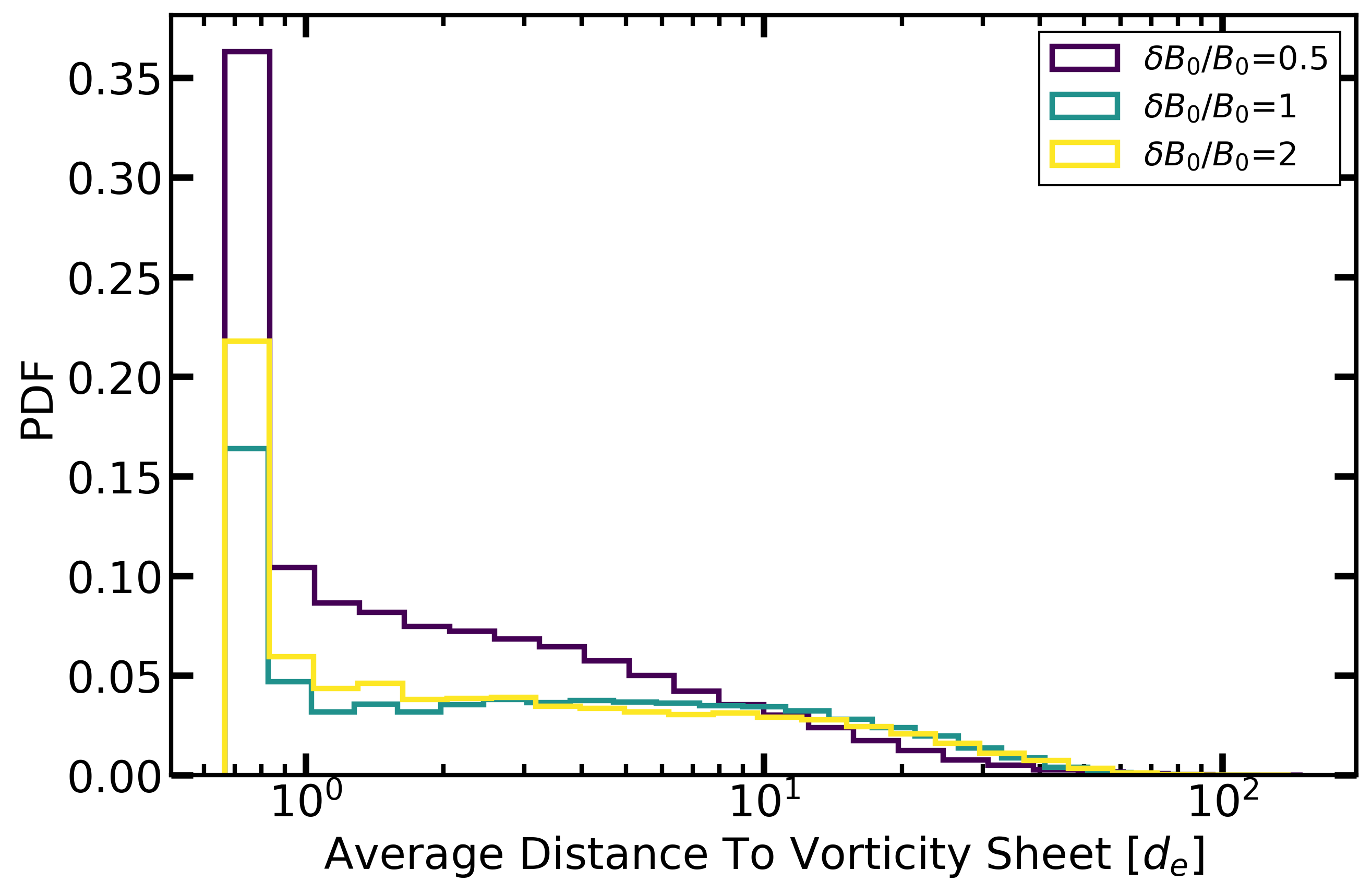}
    \caption{Probability density functions of the distance $d_w$ from current sheet edges to the nearest vorticity sheet. Top: variation with magnetization $\sigma$ for fixed \db $= 1$. Bottom: variation with magnetic fluctuation strength \db for fixed $\sigma = 10$. }
    \label{fig:average_dw}
\end{figure}

We plot the PDF of $d_w$ for different values of $\sigma$ in the top of Figure \ref{fig:average_dw} and for different values of \db\ in the bottom of Figure \ref{fig:average_dw}. First, from Figure \ref{fig:average_dw}, we can see that current sheets are often located within a few $d_e$ of a vorticity sheet. Beyond this range, the distribution appears to become noise where the average distance is just the distance to a nearby unassociated sheet. A small trend can be seen with $\sigma$ in Figure \ref{fig:average_dw}, where as $\sigma$ increases, the probability of reaching a nearby vorticity sheet within a few $d_e$ decreases. However, after $\sigma = 10$, there is no distinguishable difference. In the case of \db, there is a large jump from \db $= 0.5$ to \db $= 1$, but after that there is a small, non-monotonic change. 

\begin{figure}[htbp]
    \centering
    \includegraphics[width=\columnwidth]{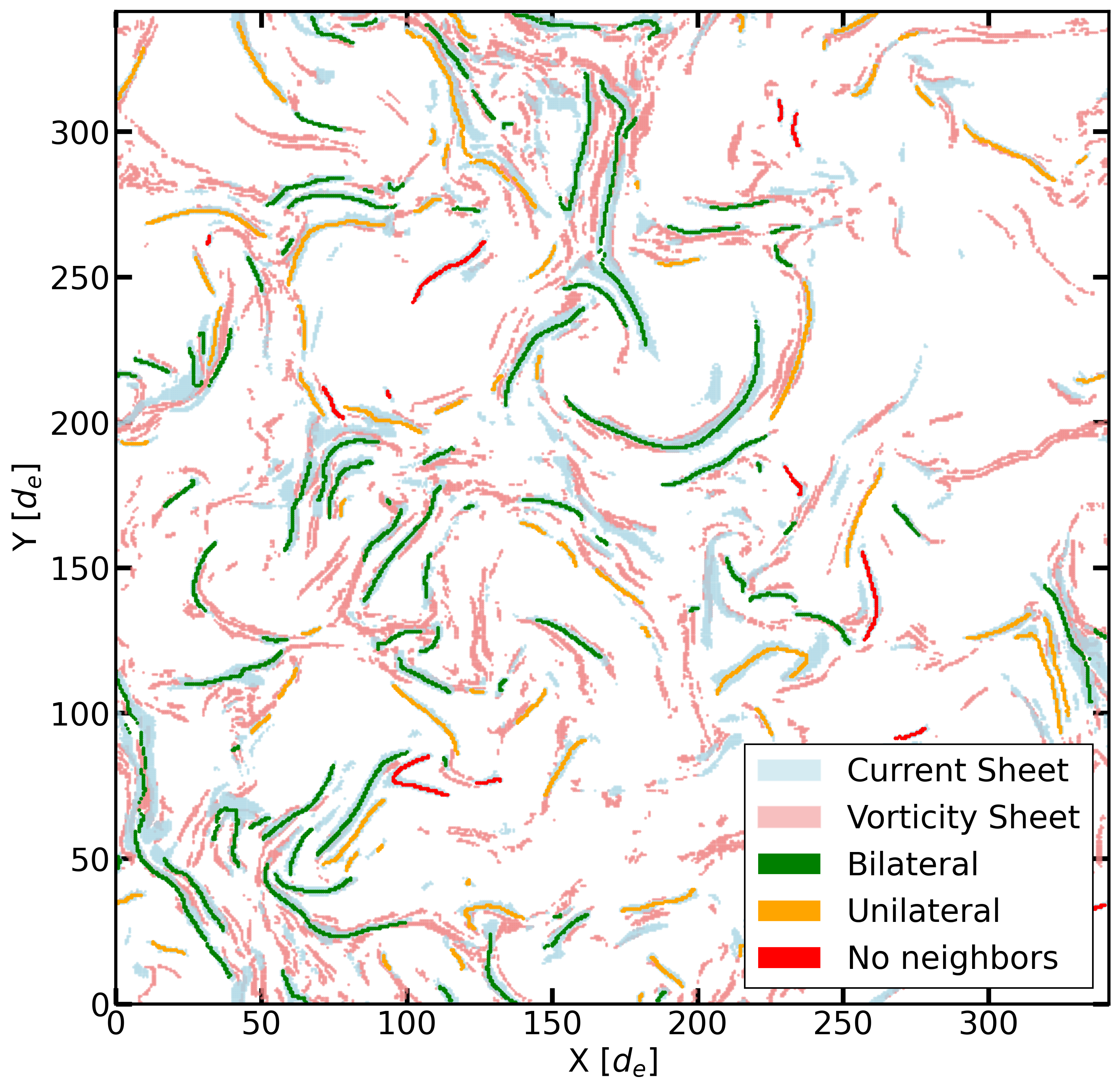}
    \caption{Illustration of the classification scheme for current sheet neighbors. Green: bilateral configuration with vorticity sheets on both sides. Yellow: unilateral configuration with a vorticity sheet on one side only. Red: isolated current sheet with no nearby vorticity sheets.}
    \label{fig:bilateral_distance_example}
\end{figure}

\begin{figure}[htbp]
    \centering
    \includegraphics[width=\columnwidth]{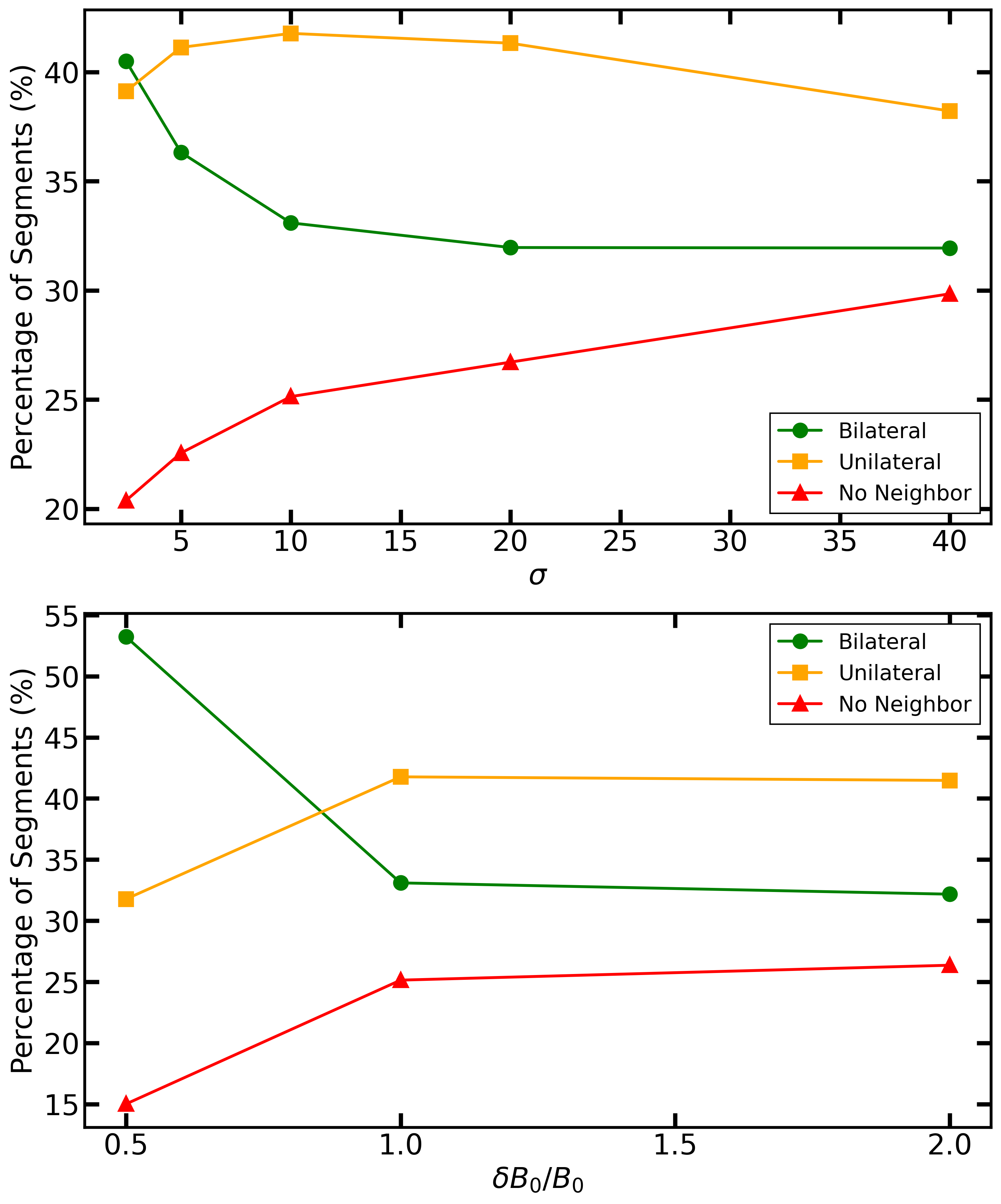}
    
    \caption{Percentage of current sheet segments classified by their proximity to vorticity sheets. Blue: bilateral (neighbors on both sides), orange: unilateral, green: no neighbors. Top: dependence on $\sigma$. Bottom: dependence on \db.}
    \label{fig:bilateral_curve_perc}
\end{figure}

Given that current sheets are often close to vorticity sheets, we set out to quantify whether a given segment has one, two, or no neighbors. To do this, we set a threshold value of $d_w \leq 14 \ d_e$, seven times the current sheet width. Thus, for all measurements on one side of the segment, we average and compare with the threshold value. If the average is below the threshold for both sides, the segment is classified as a bilateral sheet with two neighboring vorticity sheets on both sides. If one side's average is below the threshold, we consider this a unilateral sheet with a neighbor on one side. Otherwise, the sheet is considered to have no neighbors. An illustration of this classification is provided in Figure \ref{fig:bilateral_distance_example}. The value of $d_w$ is chosen to create results that in testing consistently showed qualitative neighbors. Since the value is computed over the average of the segment, changing the value affects to what extent the current sheet can have a direct neighbor.
We computed the percentage of these configurations for each simulation and show the results for $\sigma$ and \db\ at the top and bottom of Figure \ref{fig:bilateral_curve_perc}, respectively. Overall, most sheets are shown to have neighbors, with a smaller percentage, between $\approx 30\%$ and $50\%$, having neighbors on both sides. For $\sigma$, the fraction of unilateral neighbors increase until $\sigma \approx 5$ before declining. The number of bilateral neighbors decreases with increasing $\sigma$ but appears to approach an asymptote for large $\sigma$. For \db, the fraction of bilateral neighbors tends to decrease with \db\ while unilateral neighbors increase. 

\section{Discussion and Conclusions}  \label{sec:discussion}

In this work we developed and applied a framework to identify and measure current and vorticity sheets in relativistic turbulence. Our analysis has focused on quantifying the spatial distributions and geometric properties of these structures as functions of $\sigma$ and \db. These results provide a systematic characterization of the morphology and occurrence of coherent structures, which can serve as a foundation for building transport coefficients relevant to models of particle heating and acceleration in turbulent astrophysical systems. In future work, the same methods can be used to go beyond the coherent structures' spatial statistics and investigate properties in and around the current sheets concerning reconnection and dissipation, where current sheets have been shown to dominate Poynting flux dissipation \citep{comisso2018} and vorticity sheets are expected to play a role in viscous dissipation \citep{Yang2017,Yang2017NC}. Additionally, since current sheets are associated with non-thermal particle injection in turbulence, with many possible mechanisms for injection \citep{French2023}, we may be able to understand each mechanism's association with current sheets and how dominant a given mechanism is. Similarly for acceleration, in future work we may be able to investigate large increases in energy in and around coherent structures to understand the dominant mechanism of acceleration.

Our method for measuring structures relies on the fact that, in our parameter domain, structures are typically elongated along the mean magnetic field. However, this assumption has limitations, and improvements can be made in future work. For current sheets, for example, one can define a path along the ``length'' of the current sheet that follows the mean field line of the current density. Slices could then be created along the field line that preserve the L-M-N directions \citep{Sonnerup1998}. This approach would also improve the measurement of $l_\parallel$, as it accounts for bends and twists by measuring the length along the actual current density field line of the structure.

For this work, we restrict the results to a single time slice once turbulence is fully developed. This allows us to focus on the regime where dissipation is peaking, but it also prevents us from understanding the dynamics of the coherent structures. However, performing such an analysis on a dynamical timescale would require significant developments in analysis methods' computational efficiency or a drastically smaller system. 

We use \awesom to train a SOM to identify regions of current or vorticity that exceed a sufficient $T_{\mathrm{rms}}$. SOMs, when used in this way without combining multiple feature sets into data to look for nonlinear trends, are very similar to simple thresholding. We make this choice to gain an initial understanding of coherent structures based on this simplified definition. In future work, we may refine our definition by incorporating additional features in the SOM model, thereby making fuller use of \awesom.

Previous works on relativistic turbulence have shown a dependence of the non-thermal power-law slope on $\sigma$, with larger $\sigma$ producing a harder power-law spectrum \citep{comisso2018,comisso2019,Joonas2021}. This trend generally appears to approach an asymptote, but still extends to large values of $\sigma$. In this work, aside from the initial increase in $\sigma$ between 2.5 and 5, most statistical measurements show little change, except for the width $w$ and, by definition, the aspect ratio $\alpha$. This suggests that, at least for the case of current and vorticity sheets, increasing past $\sigma \gg 1$ leads to minimal differences in the geometry of structures within turbulence. Accordingly, turbulent scattering theories for relativistic plasmas should reflect only a weak dependence on $\sigma$ for scattering regions in current and vorticity sheets. Similarly, previous studies have also observed a hardening of the particles' non-thermal power law with increasing \db \citep{comisso2018,comisso2019,comisso2020}. In our analysis, structural properties exhibit a strong dependence on \db. Thus, when parameterizing scattering regions based on structures in turbulence, theories are expected to reflect a correspondingly strong dependence.  

Previous work \citep{Davis2024} explored how intermittency modeling may link statistical fluctuations in relativistic turbulence to regions of high-energy dissipation. To investigate this connection, \citet{Davis2024} measured the structure function coefficients ($\zeta_p$) and fitted them using a She–Lévêque-like prescription for $\zeta_p$ \citep{she1994}, which was further simplified as \citep{dubrulle1994}:
\begin{equation}
    \label{eq:sf_coefficient}
    \zeta_p = (p/g)(1-x) + C_0[1-(1-x/C_0)^{p/g}],
\end{equation}
where $C_0$ is the co-dimension of the dissipative structures, $g$ characterizes the fluctuation scaling, and $x$ encodes the cascade time. \citet{Davis2024} adopted $g = 2/3$ and $x = 2/3$ as simplifying assumptions that correspond to a strong \citet{goldreich1995}-like cascade. Once fitted, the value of $C_0$ was inferred from the magnetic fluctuations $(\delta b)$ for the same set of simulations studied here. It was found in \citet{Davis2024} that, under these assumptions, $C_0$ increased with $\sigma$ and minimally changed with \db. Furthermore, when looking for structures with the same $C_0$ by finding those with large Zenitani parameter \citep{zenitani2011}, it was found that the filling fraction of these structures decreased with increasing $C_0$. Given that in Figure \ref{fig:cs_simulation_trends} we do not see these trends through a more direct means of observing current sheets, we propose several possibilities for this discrepancy:
\begin{enumerate}
    \item Current sheets do not fully represent the dissipative structures at the end of the cascade.
    \item Our simplifying assumptions need to be revised, as the values of $g$ and $x$ may not accurately represent the turbulent cascade in these simulations. 
    \item Current sheets are general dissipative structures described in cascade theories but are not generally described by fluctuations in $\delta b$.
\end{enumerate}
Further investigation is needed to clarify this issue, but some supporting evidence for point 3 comes from comparing the $C_0$ values obtained for current sheets in this work with those inferred from current fluctuations $\delta j$ in \citet{Davis2024}, where the trend in $C_0$ derived from $\delta j$ fluctuations is much more consistent with the one seen in Figure \ref{fig:cs_simulation_trends}.

For the measured quantities the structure width ($w$), perpendicular length ($l_\perp$), aspect ratio ($\alpha$), and length along the mean field ($l_\parallel$), we observe that their dependence on $\sigma$ is nearly inverted for current sheets versus vorticity sheets. This behavior might arises from the high compressibility of relativistic turbulence, which causes deviations from the scalings expected in incompressible MHD. This divergence is particularly evident when comparing quantities associated with velocity fluctuations to those associated with magnetic fluctuations, as also noted in \citet{zrake2012}.

Our results show that relativistic turbulence often produces current sheets with nearby vorticity sheet neighbors. As seen in Figure \ref{fig:bilateral_curve_perc}, as much as $\approx 80\%$ of current sheets have at least one neighbor within an average distance of $14 \ d_e$. This can be appreciated more qualitatively in Figure \ref{fig:bilateral_distance_example}. The proximity of vorticity structures to current sheets is expected since large regions of current are also directly tied to strong vorticity generation \citep{Matthaeus1982}. We find, as in other works \citep{Parashar2016}, that vorticity tends to occur at the borders of current sheets, where it is expected to dissipate energy from the fluid flow via work done by the pressure tensor \citep{Yang2017,Yang2017NC,Parashar2016}. In particular, the case of bilateral neighbors, where we may expect shear flows around the current sheet, could be especially dissipative, if results from non-relativistic shear flow reconnection, which show increased heating with stronger shear \citep{Haggerty2025}, also apply in the relativistic regime.
It is also interesting to speculate if the vorticity sheets are connected to the double-current-sheet structures discovered in \citet{Ha2025}.
Unlike regular (single) current sheets, the double sheets did not show signs of active magnetic reconnection and resembled more a local magnetic compression, possibly originating from the non-linear Alfven wave interactions. 
Such a connection is an interesting future avenue to be studied.

\begin{figure}[htbp]
    \centering
    \includegraphics[width=\columnwidth]{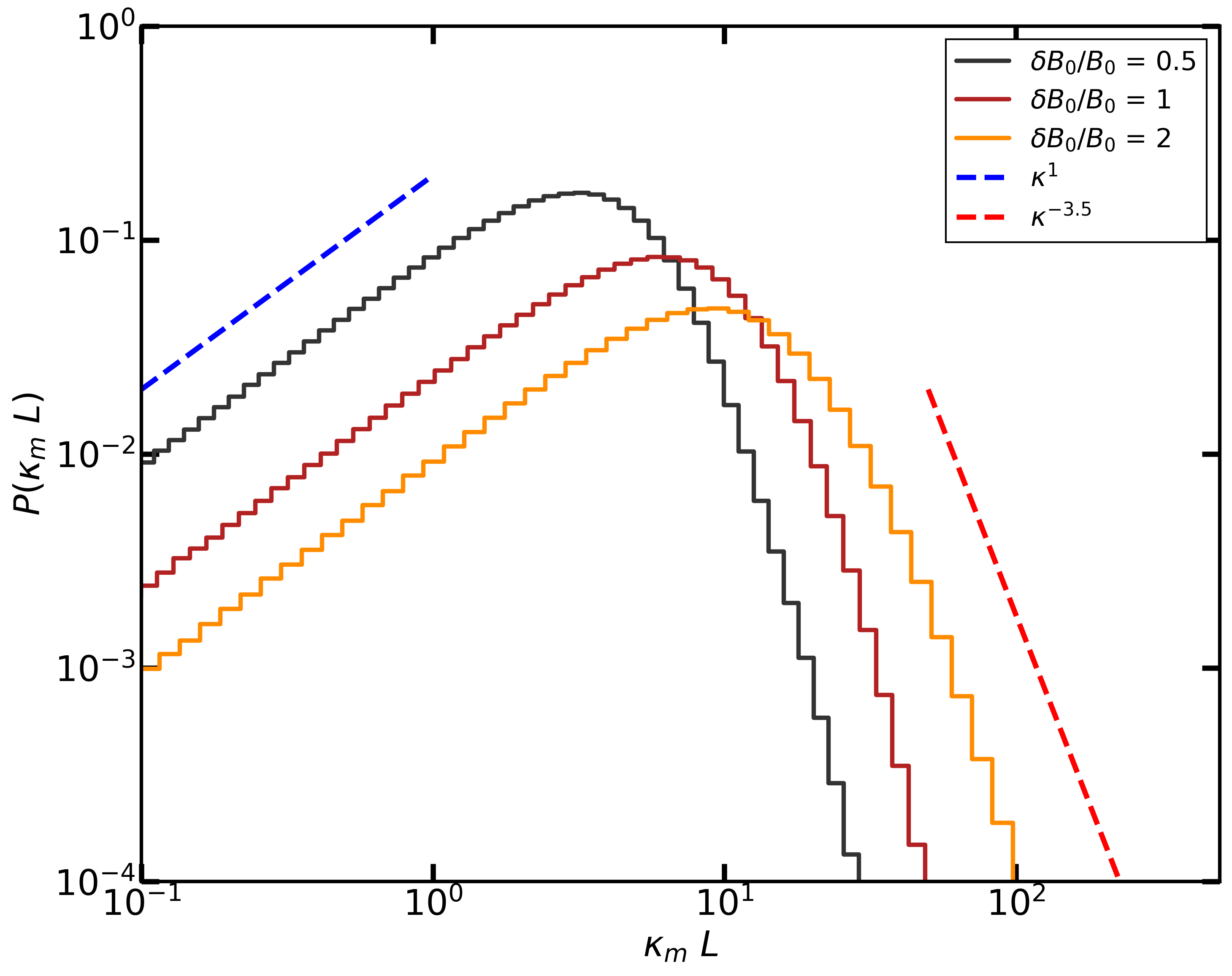}
    \caption{Probability density function of the magnetic field curvature $\kappa_m = |\mathbf{b} \cdot \nabla \mathbf{b}|$ for different values of magnetic fluctuation strength \db\ with fixed $\sigma = 10$. The red reference line shows the power-law slope observed in the high-curvature tail of the structural curvature $\kappa$ and the blue shows a reference fit to the low $\kappa_m$ slope.}
    \label{fig:magnetic_curvature}
\end{figure}

The magnetic curvature in MHD plays an important role in both understanding and detecting structures such as flux ropes \citep{Sun2019} and in particle heating and acceleration \citep{Dahlin2014}. Recent studies have also emphasized the importance of mirror acceleration \citep{Das2025}, which, however, requires strong curvature events to rapidly change the particle pitch angle for sustained acceleration. In this work, we measure the curvature not only to define the structures but also to explore the relationship between the structures' curvature and the magnetic field curvature. Here we employ two measures of curvature: $\kappa$ to measure the local curvature and $\kappa_3$ to measure the large-scale curvature. In the case that $\kappa_3 = \kappa$, we would expect to see perfect circles with radius of curvature $= 1/\kappa_3 = 1/\kappa$. However, given that the structures are subject to perturbations, $\kappa$ sees continuous change in the local curvature. In these small local curvature scenarios, $\kappa$ may be probing the largest values of curvature in the simulation, such as bent exhaust lines or plasmoids formed during reconnection. Similarly, $\kappa_3$ may be a response to larger fluctuations in the magnetic field that form the boundaries of the structures. To illustrate this point, in Figure \ref{fig:magnetic_curvature} we plot the magnetic field curvature $\kappa_m = |\mathbf{b} \cdot \nabla \mathbf{b}|$ for the different values of \db\ where we see the most change. 
Figure \ref{fig:magnetic_curvature} shows that $\kappa_m$ follows a broken power-law distribution, consistent with previous studies \citep{yang2019curv,yuen2020} and with observations in the magnetosheath \citep{bandyopadhyay2020}. The mean curvature, normalized to the simulation size $L$, increases with \db. Additionally, we can see that for large values of $\kappa_m$, the slope matches well with the high-tail slope observed for $\kappa$ (the red reference slope is the same reference slope used in the PDFs of $\kappa$). Though the low-tail slope of $\kappa_m$ does not directly agree with the low-tail slope of $\kappa_3$, the low-side tail of $\kappa_3$ may be limited by system size constraints that are not relevant to $\kappa_m$. Of course, this is only speculation at present, and future work will need to perform an in-depth analysis of the individual structures and the local magnetic field around them.

\begin{figure}[htbp]
    \centering
    \includegraphics[width=\columnwidth]{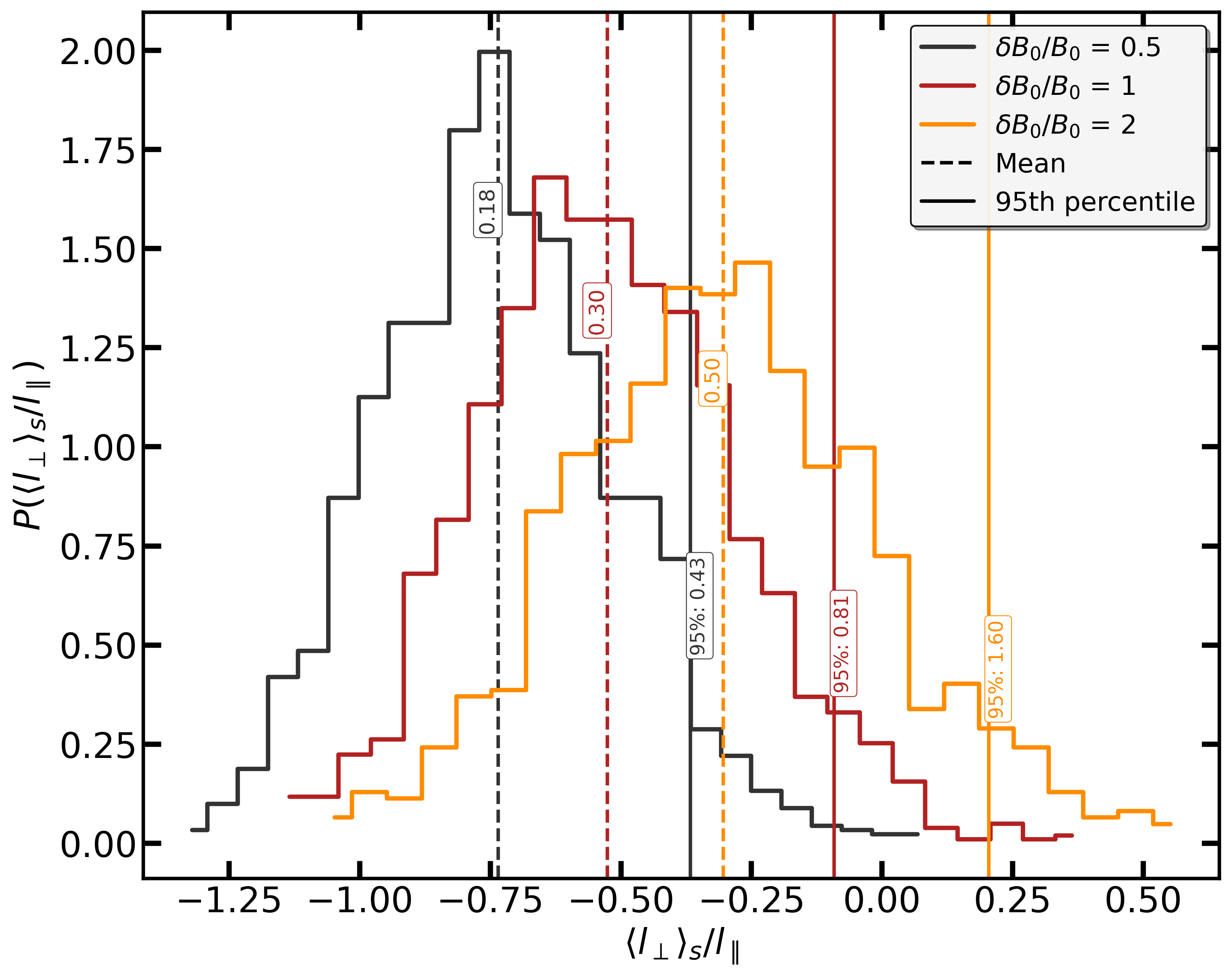}
    \caption{Probability density function for $\langle l_\perp \rangle_s / l_\parallel$ for different values of magnetic fluctuation strength \db\ with fixed $\sigma = 10$. The dotted reference lines show the mean values and the solid reference lines show the 95 percentile.}
    \label{fig:step_hist_z_length_times_struct_average_arc_length_vs_db}
\end{figure}

Our results show that $l_\parallel$ increases as the guide field increases. This is consistent with our understanding of current sheets where a guide field tends to stabilize current sheets by suppressing the flux-rope kink instability \citep{Barkov2016}. For the flux-rope, modes are heavily suppressed that violate the Kruskal-Shafranov condition, $k_z D_f \gtrsim B_r/B_0$ \citep{Bateman1978}, where $k_z$ is the kink instability wave number, $D_f$ is the diameter of the flux rope, $B_r$ is the reconnecting field and $B_0$ is the guide field. If our current sheets exist on the edge of large flux ropes, we can express this condition in terms of our variables as $\langle l_\perp \rangle_s / l_\parallel \gtrsim $ \db. In figure \ref{fig:step_hist_z_length_times_struct_average_arc_length_vs_db} we plot the histogram for $\langle l_\perp \rangle_s / l_\parallel$ for each value of \db, where we see general agreement with the Kruskal-Shafranov condition with the 95th percentile, illustrated by the solid reference lines, both bounded by and scaling with \db. Similarly, the means also tend to scale with \db. This may suggest that flux-rope stability is the primary limitation to the structure size in these simulations.
       
Our results provide comprehensive information on the statistical properties of vorticity and current structures in turbulent plasmas to help inform future heating and acceleration theories of their spatial structure and dependence on plasma parameters $\sigma$ and \db. For this, we fit all measurements to simplified functions that can be easily applied in other works. Beyond their relevance to theories of particle acceleration and heating, our results can also be directly applied in flaring models. For example, by constraining the probability of a large-scale current sheet producing gamma-ray flares in a turbulent Crab model \citep{Lyutikov2019}, or by incorporating the statistics of current sheets into \citet{sobacchi2023}'s turbulent lighthouse theory, one can directly link ultra-fast AGN jet variability to the statistics of current sheets in the plasma and thus to global parameters of the emitting region.

\begin{acknowledgments} 
The work of ZD and CCH was supported in part by NSF/DOE Grant PHY-2205991, NSF-FDSS Grant AGS-1936393, NSF-CAREER Grant AGS-2338131, NASA Grant HTMS-80NSSC24K0173 and HSR-80NSSC23K0099. 
LC acknowledges support from NSF Grant PHY-2308944 and NASA ATP Grant 80NSSC24K1230. JN is supported by ERC grant (ILLUMINATOR, 101114623).
This research was supported in part by grant NSF PHY-2309135 to the Kavli Institute for Theoretical Physics (KITP)
\end{acknowledgments}

\appendix

\section{Skeleton-Based Curve Fitting}
\label{sec:ap_spline_fitting}

To accurately measure coherent structures in this work, we first exploit the natural elongation of structures along the $z$-axis by creating 2-D slices at each $z$ level. Each 2-D slice contains many individual segments that resemble current sheets in 2-D. To measure the segments, we reduce them to 1-D by skeletonizing the segments to find the best line that represents the overall structure. Given the complexity of the task and the large number of measurements required, the ability to skeletonize arbitrary 2-D shapes efficiently while maintaining accuracy is a priority.

This process starts by mapping an individual segment to a simplified grid ($G$). The grid size is defined as $G_x = \max(64,\min(L,S_{p,x}))$ and $G_y = \max(64,\min(L,S_{p,y}))$, where $L$ is the simulation size and $S_p=[S_{p,x},S_{p,y}]=[S_x + 4, S_y + 4]$ is the segment's maximum $x$ and maximum $y$ extent, padded with 4 extra cells to reduce boundary effects. Thus, the grid has a minimum resolution of $64\times64$ and a maximum resolution of the system size $L\times L$.

When mapping to the grid, gaps may appear in the data. To address this, we perform a binary image smoothing operation using a 3-pixel circular region that first erodes the mapped segment before dilating it. This process fills small gaps created during grid mapping without dramatically changing the boundary of the segment. This operation is referred to as binary closing and is implemented using \citep{scikit-image}.

After binary closing, we use the skeletonization algorithm described in \citet{Chen1988} to find the segment's centerline. \citet{Chen1988} accomplish this by iteratively removing boundary pixels until a thin center-line remains. During this process, we check to ensure that the ratio of skeleton points to segment points is less than 0.9. If not, we assume the skeletonization has failed and handle it in subsequent steps.

Irregularities in the shape of the segments often produce many branches. As a first pass through these branches, we remove any branch shorter than 5\% of the total skeleton length. After this, we process the skeleton using the Skan library \citep{skan-library} to find the longest continuous path through the skeleton. If a segment has multiple disconnected paths, we merge them if their respective endpoints are within 5 pixels of each other. This results in a single elongated curve that runs through the segment.

To ensure the final curve is a reasonable representation of the segment, we apply three additional checks. First, we require that at least $95\%$ of the curve points lie within the segment. Second, the maximum extent of the curve must be within 0.3-1.1 times the maximum extent of the segment's point cloud. Third, we check for loops by ensuring the curve does not intersect itself. If at any point the skeleton process fails, we attempt a linear fit of the segment. The linear fit undergoes the same checks, and if it still fails, the segment is excluded from analysis.

When mapping the skeleton back to the simulation coordinates, we apply Gaussian smoothing to reduce noise in the curve with a Gaussian smoothing parameter $\sigma = 2$. The skeleton points are parameterized and upsampled so that the parameterized curve has the same number of points as the original segment. We use SciPy's cubic spline interpolation for the upsampling \citep{2020SciPy}. Finally, the points in the parameterized curve are mapped back to the original data by finding the closest points in the original segment to each curve point. We remove duplicate points and preserve the ordering of the points. This produces a final curve with a number of points proportional to the segment size.

\section{2D PDFs}
\label{ap:extra_pdfs}

\begin{figure}[H]
    \centering
    
    \includegraphics[width=\columnwidth]{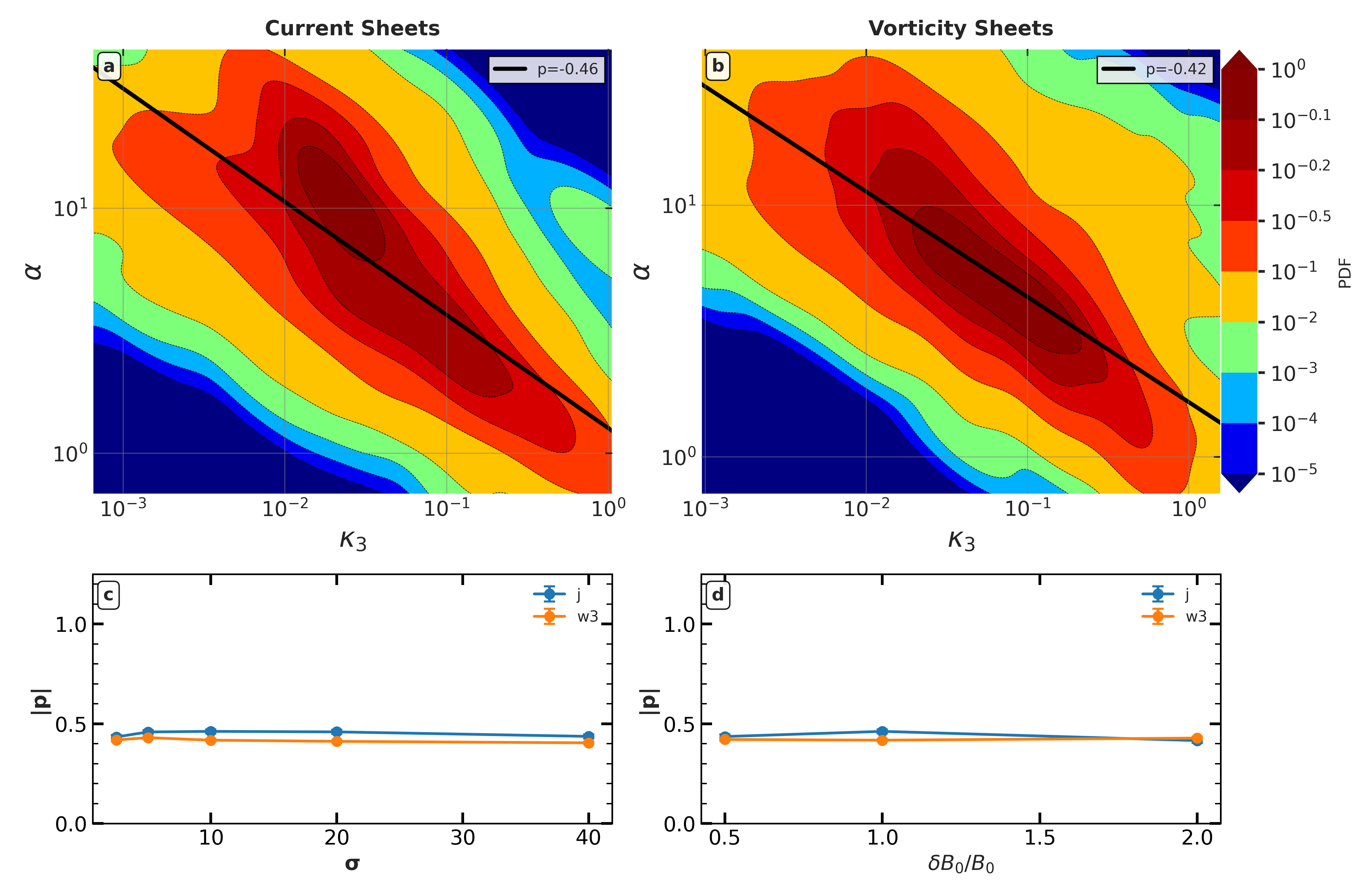}

    \caption{$\kappa_3$ versus $\alpha$ for $\sigma=10$ and \db$=1$. All other simulations are shown with power-law fits (see legend) and the slopes are shown in the insets. Blue for $\sigma$ slopes and green for \db.The top plot shows the results for current sheets and the bottom for vorticity sheets.}
    \label{fig:curv_3p_vs_alpha}
\end{figure}
\begin{figure}[H]
    \centering
    
    \includegraphics[width=\columnwidth]{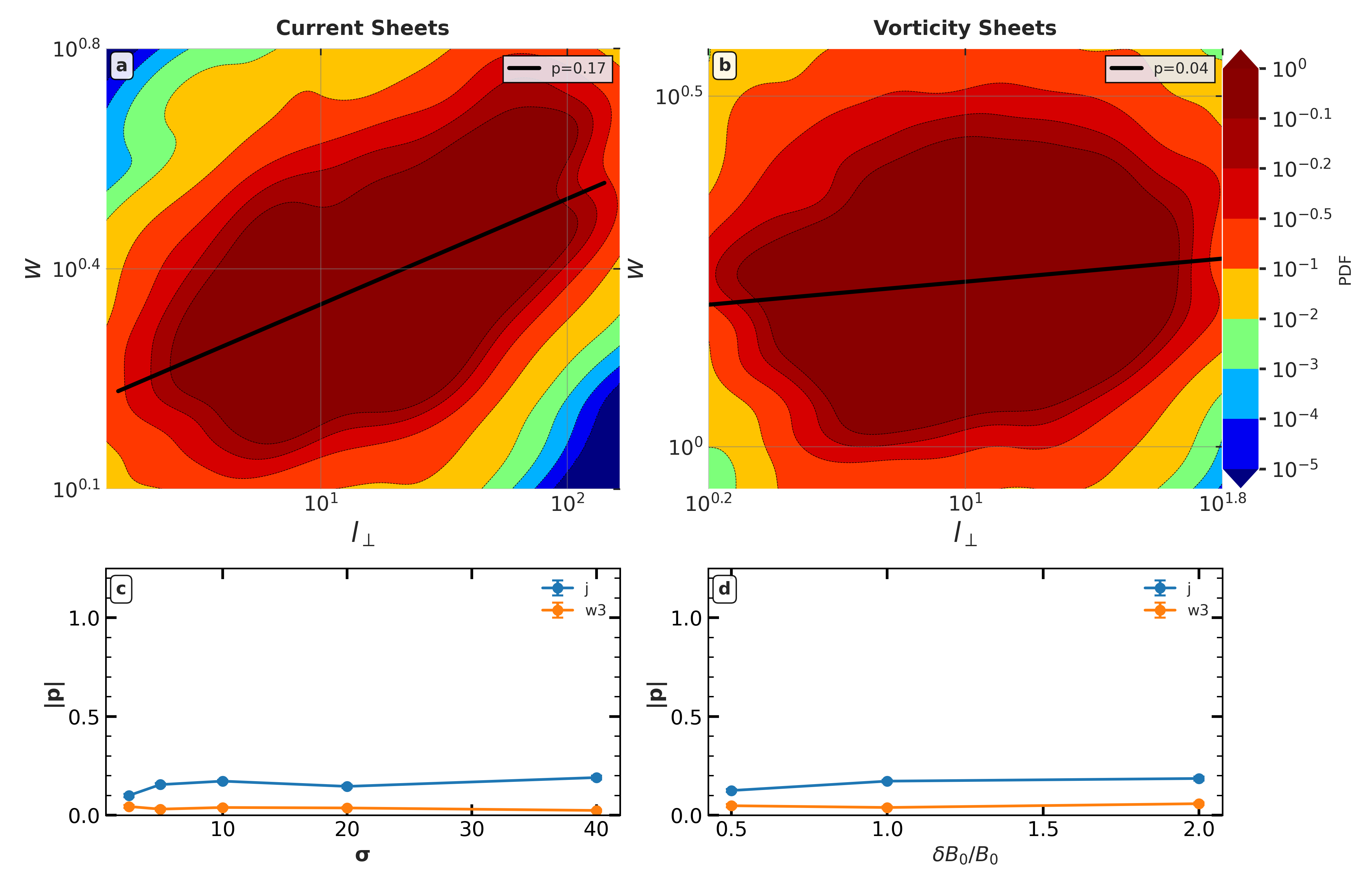}

    \caption{$l_\perp$ versus $w$ for $\sigma=10$ and \db$=1$. All other simulations are shown with power-law fits (see legend) and the slopes are shown in the insets. Blue for $\sigma$ slopes and green for \db. The top plot shows the results for current sheets and the bottom for vorticity sheets. }
    \label{fig:l_perp_vs_w}
\end{figure}
\begin{figure}[H]
    \centering
    
    \includegraphics[width=\columnwidth]{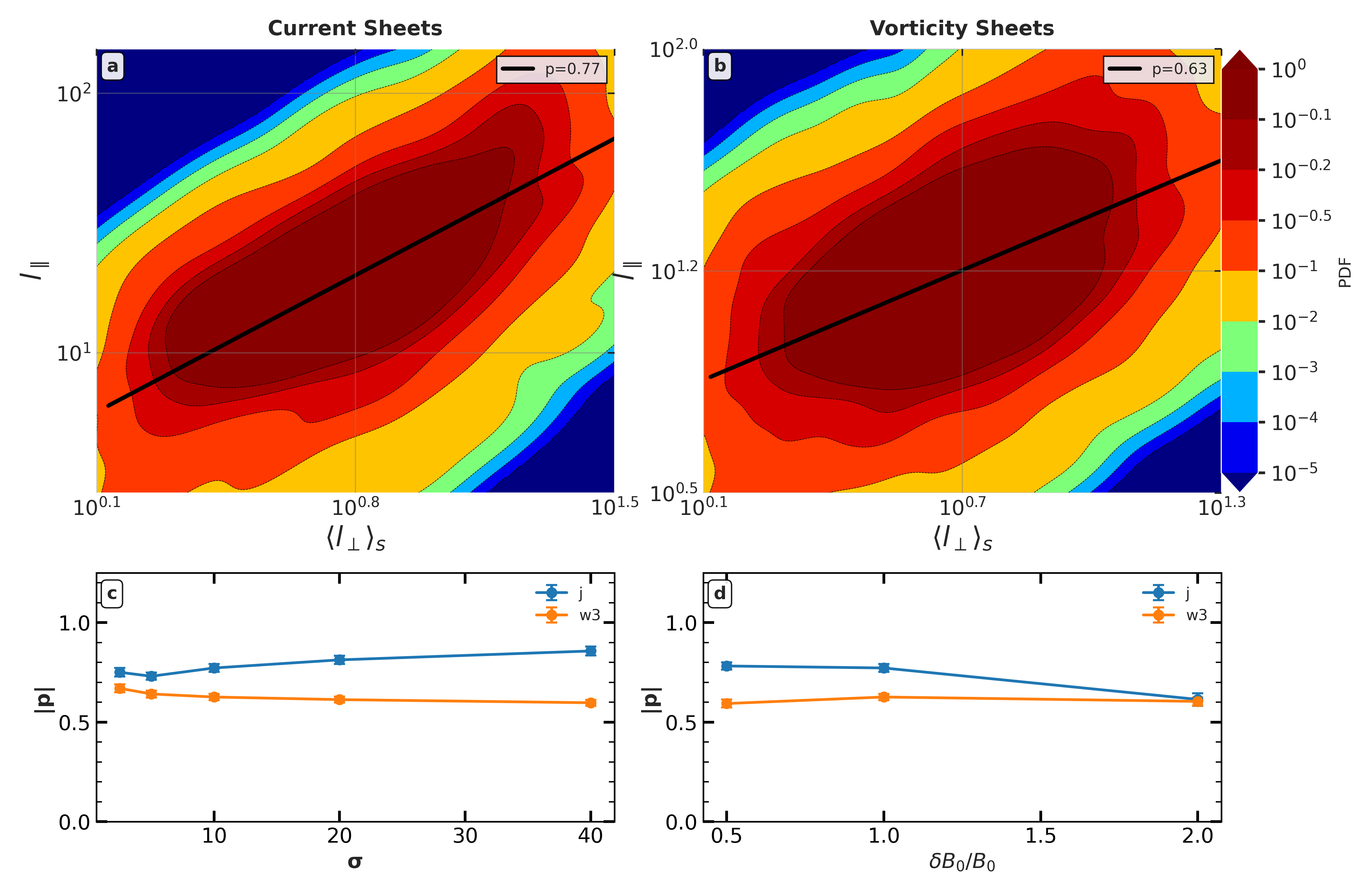}

    \caption{Structure averaged $l_\perp$ versus $l_\parallel$ for $\sigma=10$ and \db$=1$. All other simulations are shown with power-law fits (see legend) and the slopes are shown in the insets. Blue for $\sigma$ slopes and green for \db. The top plot shows the results for current sheets and the bottom for vorticity sheets.}
    \label{fig:stuct_ave_l_perp_vs_l_para}
\end{figure}
\begin{figure}[H]
    \centering
    
    \includegraphics[width=\columnwidth]{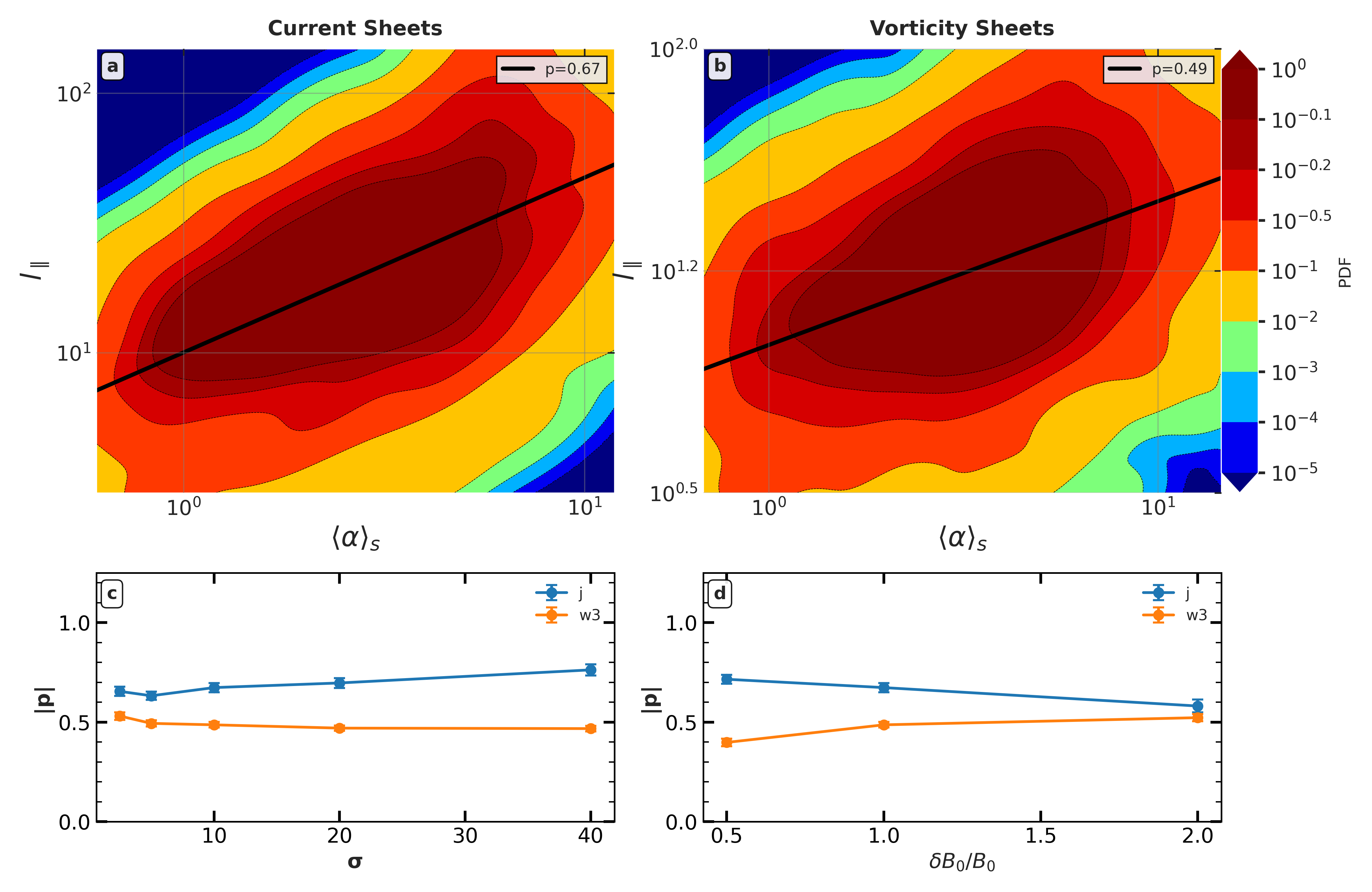}

    \caption{Structure averaged $\alpha$ vs $\parallel$ for $\sigma=10$ and \db$=1$. All other simulations are shown with power-law fits (see legend) and the slopes are shown in the insets. Blue for $\sigma$ slopes and green for \db. The top plot shows the results for current sheets and the bottom for vorticity sheets.}
    \label{fig:stuct_ave_alpha_vs_l_para}
\end{figure}
\section{Additional 3D Renderings}
\subsection{Current Sheets}
\begin{figure}[H]
\centering 
\includegraphics[width=0.49\textwidth]{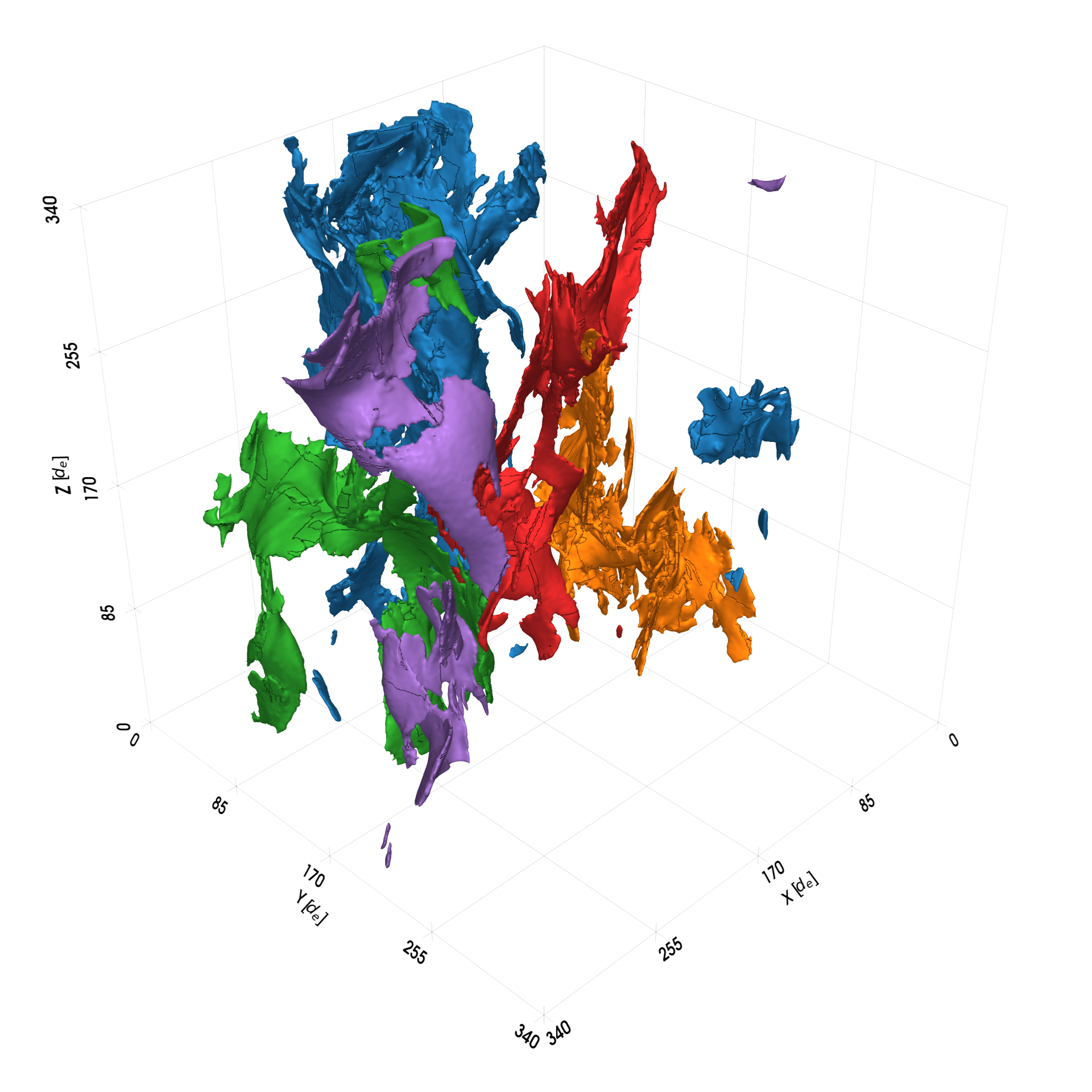}\hfill% 
\includegraphics[width=0.49\textwidth]{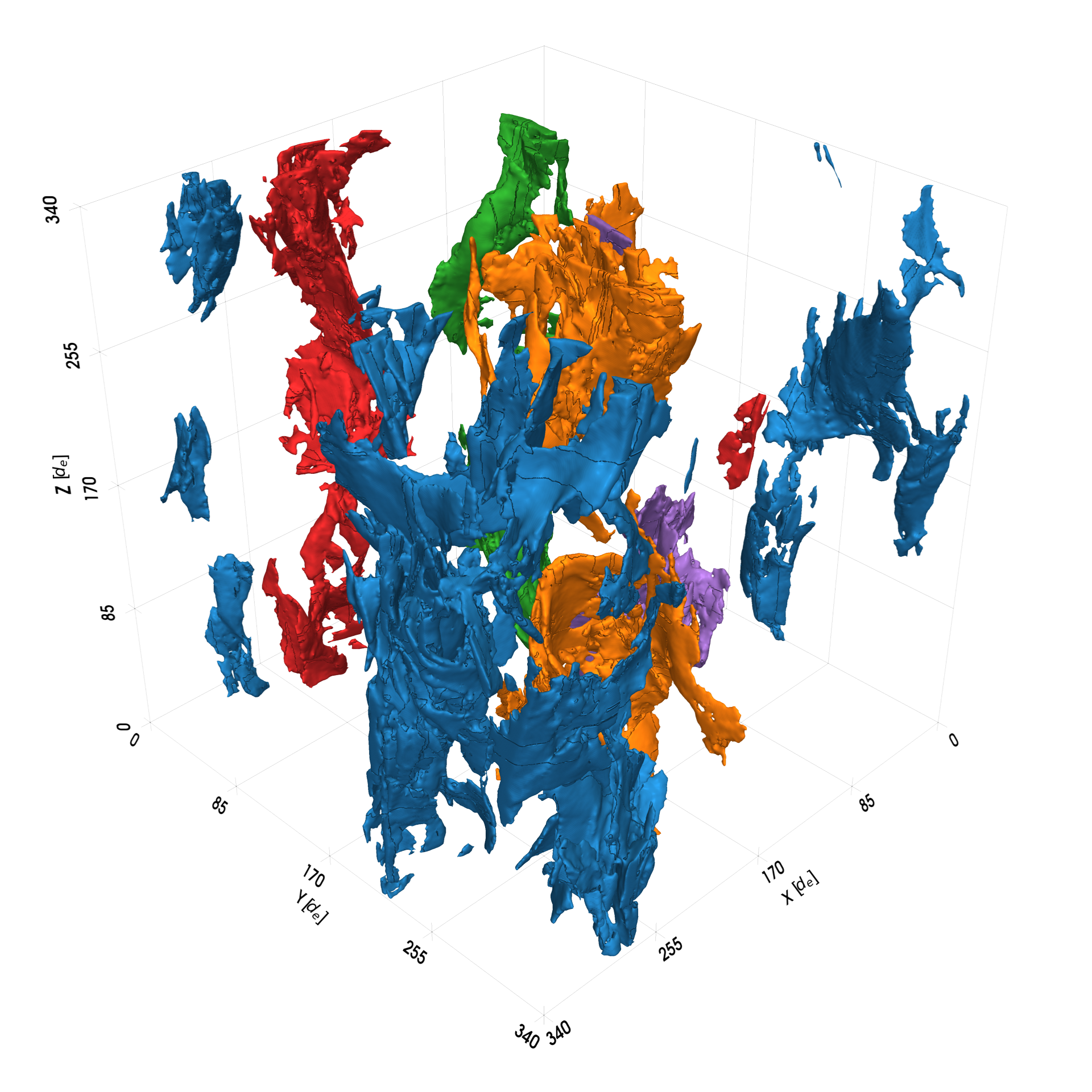}\hfill% 
\caption{Three-dimensional visualization of the largest current sheets for different $\sigma$. Left: the 5 largest current sheets for $\sigma = 2.5$. Right: the 5 largest current sheets for $\sigma = 40$.} 
\end{figure}
\begin{figure}[H]
\centering 
\includegraphics[width=0.49\textwidth]{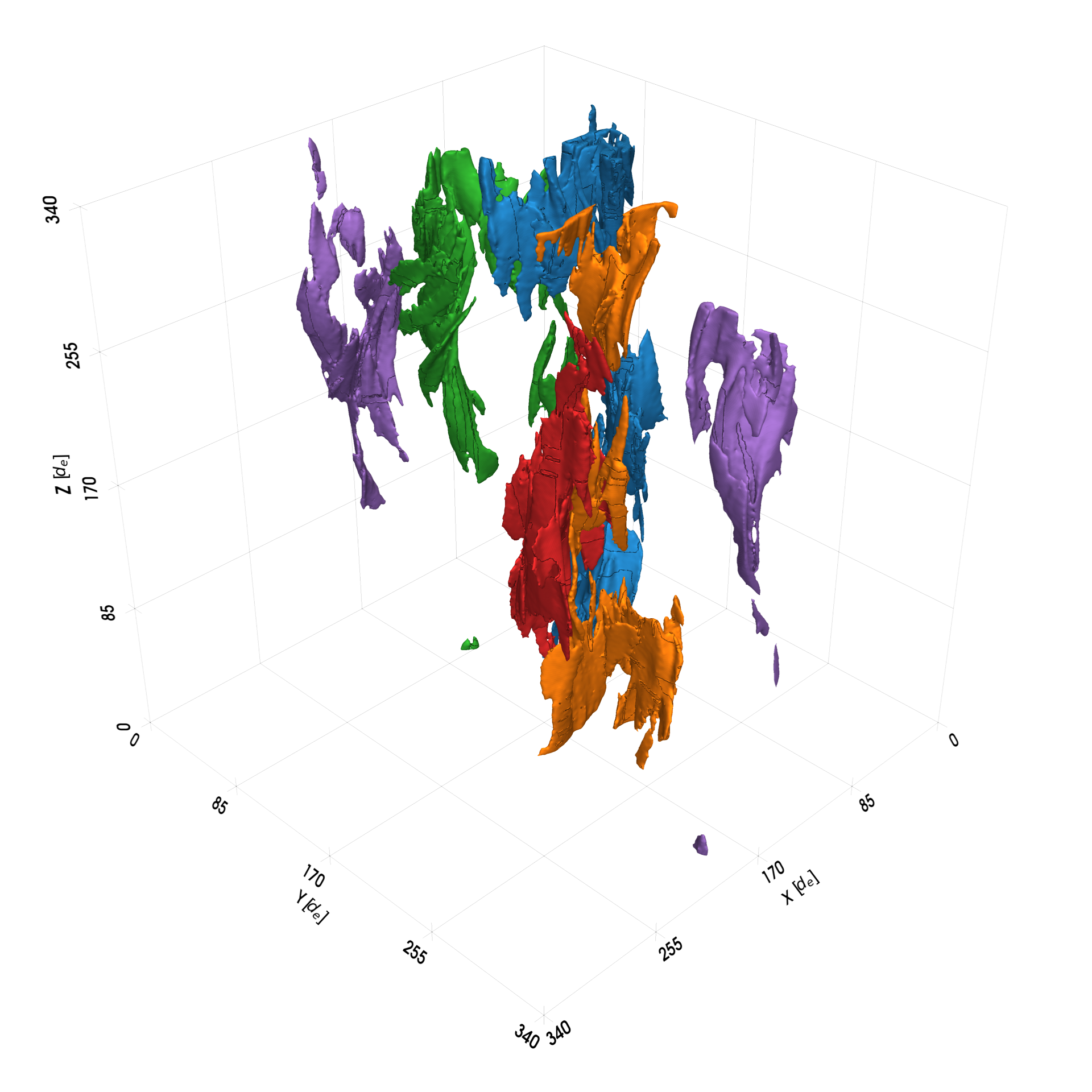}\hfill% 
\includegraphics[width=0.49\textwidth]{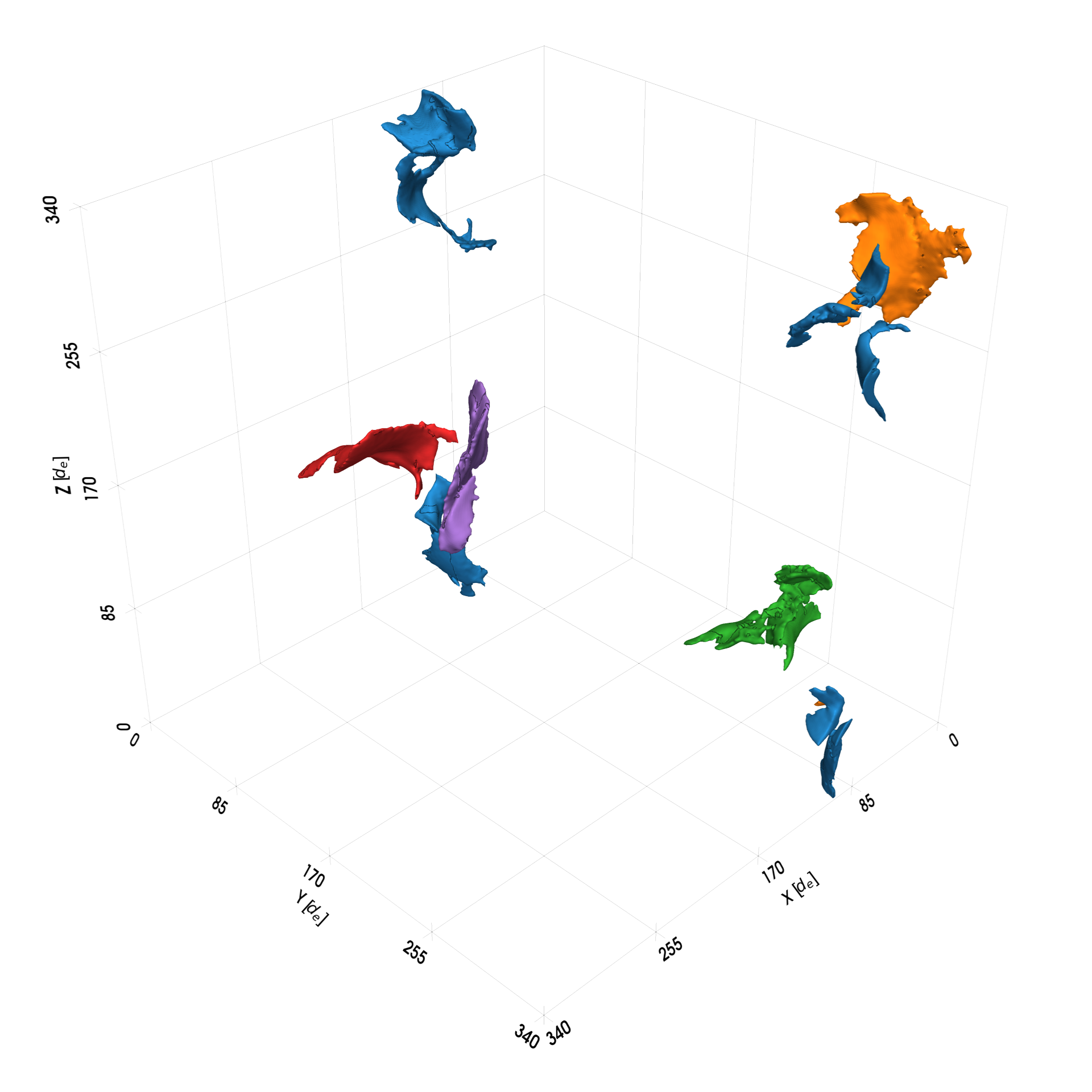}\hfill% 
\caption{Three-dimensional visualization of the largest current sheets for different values of \db. Left: the 5 largest current sheets for \db$= 0.5$. Right: the 5 largest current sheets for \db$=2$.} 
\end{figure}
\subsection{Vorticity Sheets}
\begin{figure}[H]
\centering 
\includegraphics[width=0.49\textwidth]{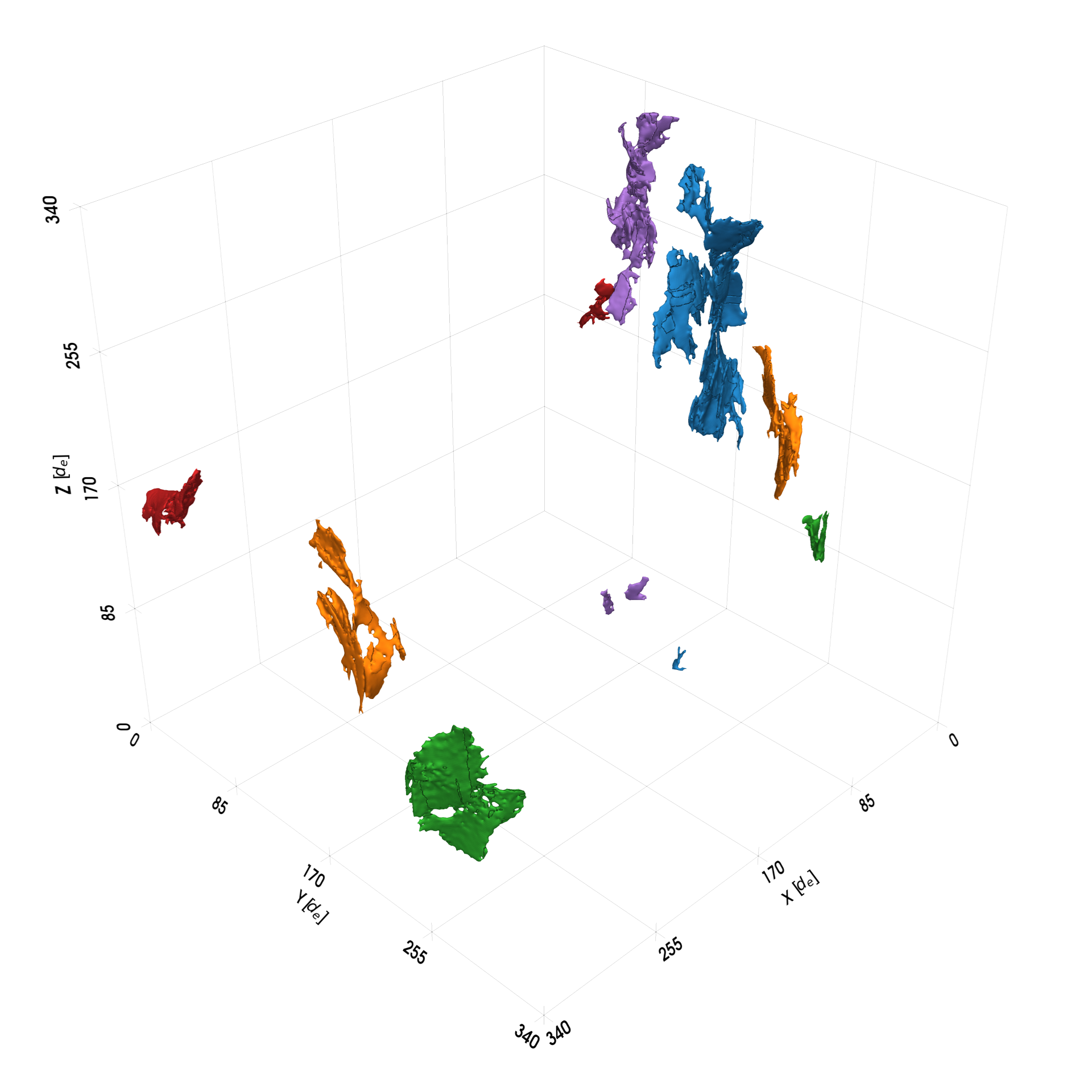}\hfill% 
\includegraphics[width=0.49\textwidth]{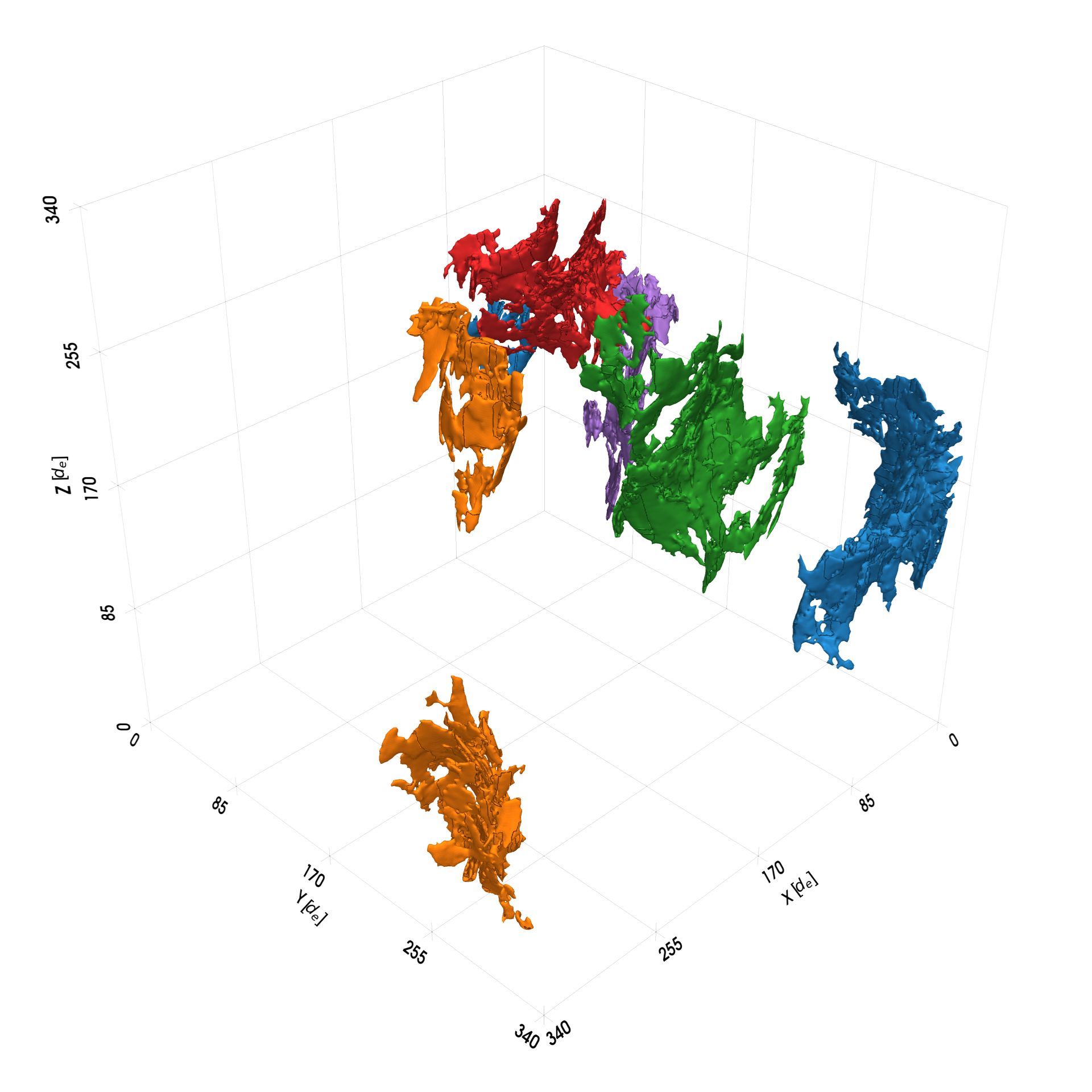}\hfill% 
\caption{Three-dimensional visualization of the largest vorticity sheets for different $\sigma$. Left: the 5 largest vorticity sheets for $\sigma = 2.5$. Right: the 5 largest vorticity sheets for $\sigma = 40$.} 
\end{figure}
\begin{figure}[H]
\centering 
\includegraphics[width=0.49\textwidth]{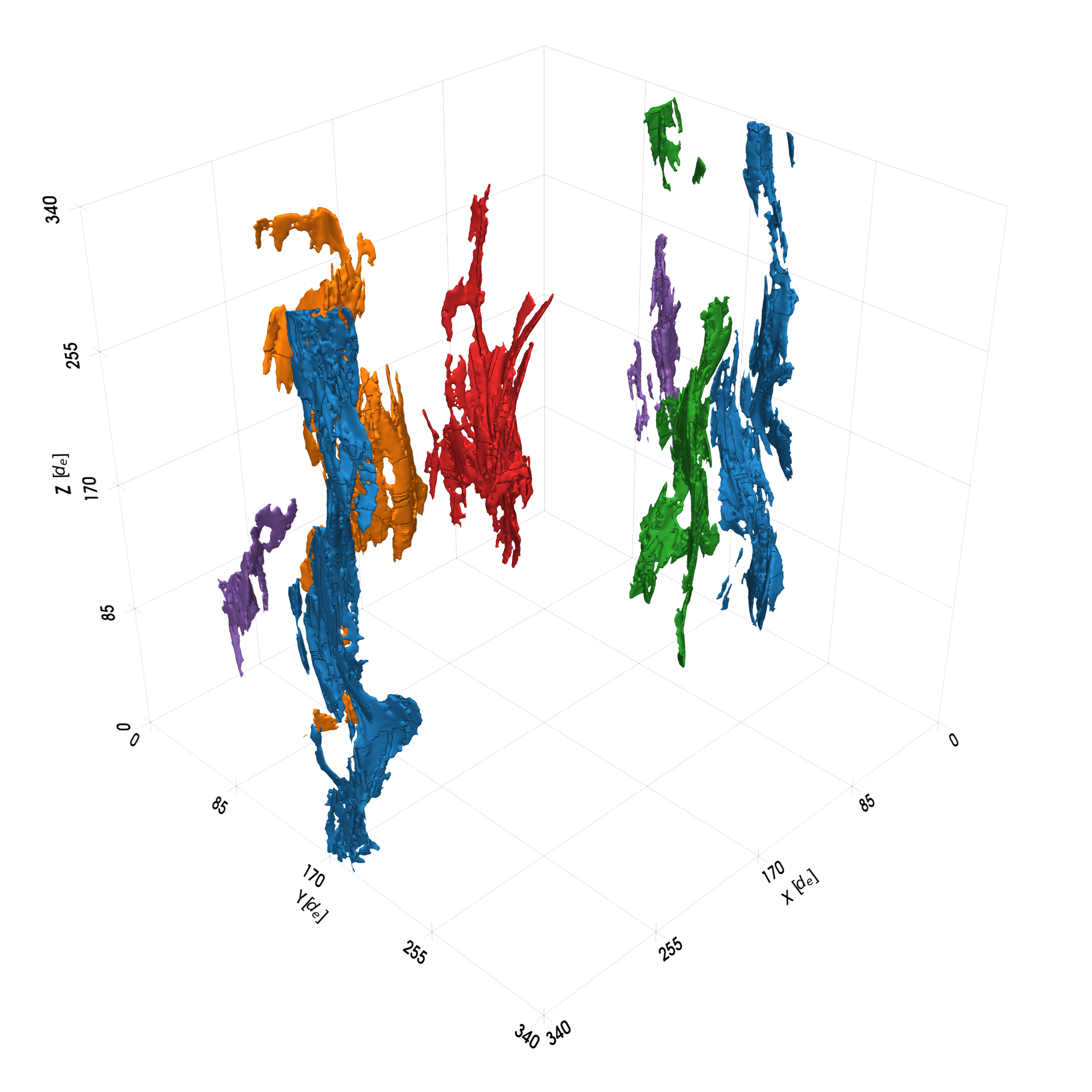}\hfill% 
\includegraphics[width=0.49\textwidth]{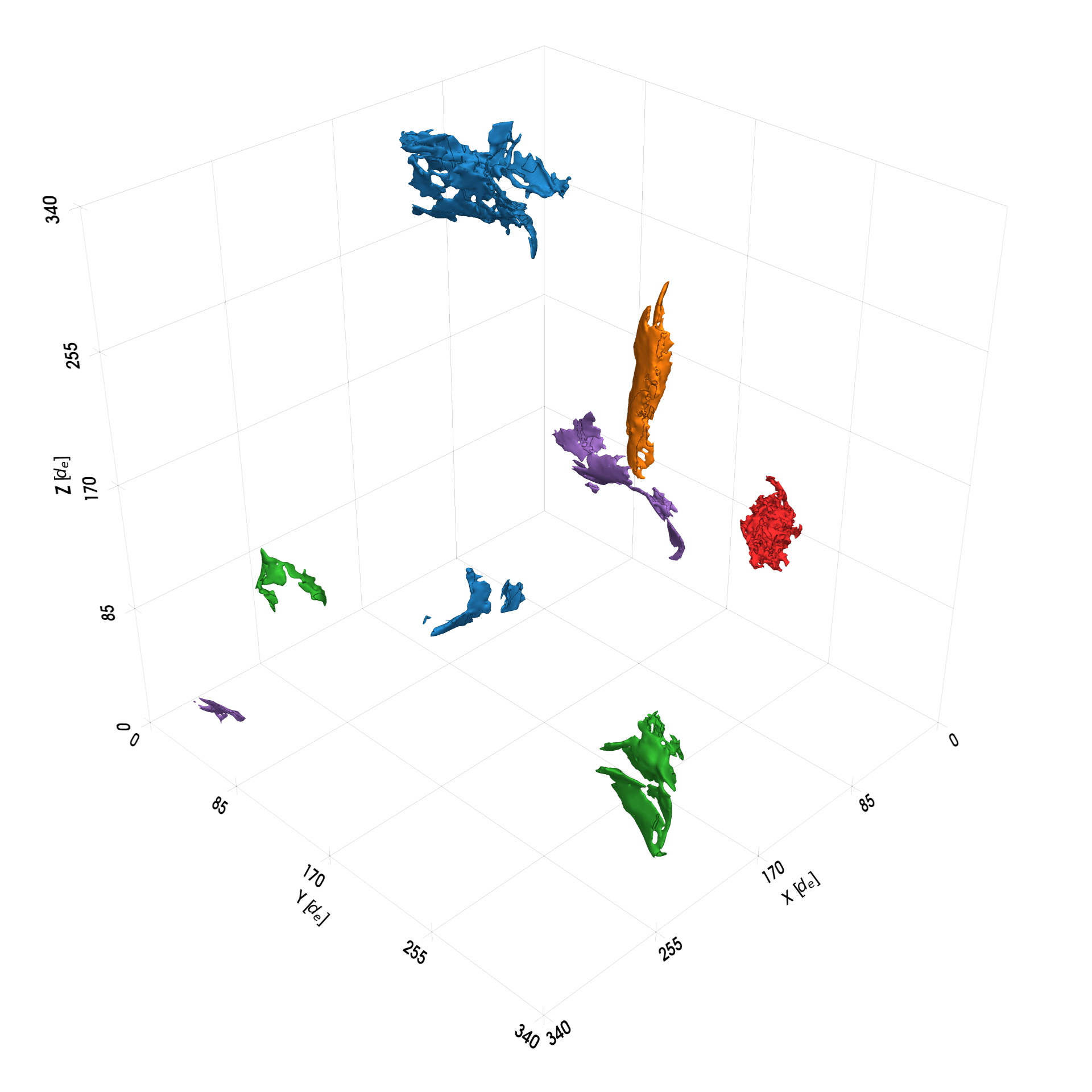}\hfill% 
\caption{Three-dimensional visualization of the largest vorticity sheets for different values of \db. Left: the 5 largest vorticity sheets for \db$= 0.5$. Right: the 5 largest vorticity sheets for \db$=2$.} 
\end{figure}
\section{Mean Summary}
\label{ap:mean_tables}

\begin{longtable}{|l|c|c|c|c|c|}
\caption{Statistical Summary of Measurements} \label{tab:statistics_summary} \\
\hline
Measurement & Feature & $T_{\text{rms}}$ & $\sigma$ & $\delta B_0/B_0$ & Mean $\pm$ Std Error \\
\hline
\endfirsthead
\multicolumn{6}{c}{\tablename\ \thetable{} -- continued from previous page} \\
\hline
Measurement & Feature & $T_{\text{rms}}$ & $\sigma$ & $\delta B_0/B_0$ & Mean $\pm$ Std Error \\
\hline
\endhead
\hline
\multicolumn{6}{r}{Continued on next page} \\
\hline
\endfoot
\hline
\endlastfoot
$w$ & j & 2 & 2.5 & 1 & 2.434 ± 0.004 \\
$w$ & j & 2 & 5 & 1 & 2.509 ± 0.004 \\
$w$ & j & 2 & 10 & 0.5 & 2.618 ± 0.004 \\
$w$ & j & 2 & 10 & 1 & 2.621 ± 0.005 \\
$w$ & j & 2 & 10 & 2 & 2.888 ± 0.007 \\
$w$ & j & 2 & 20 & 1 & 2.769 ± 0.005 \\
$w$ & j & 2 & 40 & 1 & 2.971 ± 0.007 \\
$w$ & j & 3 & 2.5 & 1 & 2.123 ± 0.006 \\
$w$ & j & 3 & 5 & 1 & 2.199 ± 0.006 \\
$w$ & j & 3 & 10 & 0.5 & 2.261 ± 0.006 \\
$w$ & j & 3 & 10 & 1 & 2.272 ± 0.007 \\
$w$ & j & 3 & 10 & 2 & 2.398 ± 0.008 \\
$w$ & j & 3 & 20 & 1 & 2.370 ± 0.008 \\
$w$ & j & 3 & 40 & 1 & 2.515 ± 0.009 \\
$w$ & $\omega$ & 1.5 & 2.5 & 1 & 2.502 ± 0.004 \\
$w$ & $\omega$ & 1.5 & 5 & 1 & 2.399 ± 0.003 \\
$w$ & $\omega$ & 1.5 & 10 & 0.5 & 2.297 ± 0.003 \\
$w$ & $\omega$ & 1.5 & 10 & 1 & 2.281 ± 0.003 \\
$w$ & $\omega$ & 1.5 & 10 & 2 & 2.565 ± 0.004 \\
$w$ & $\omega$ & 1.5 & 20 & 1 & 2.192 ± 0.002 \\
$w$ & $\omega$ & 1.5 & 40 & 1 & 2.165 ± 0.002 \\
$w$ & $\omega$ & 2 & 2.5 & 1 & 1.965 ± 0.003 \\
$w$ & $\omega$ & 2 & 5 & 1 & 1.843 ± 0.002 \\
$w$ & $\omega$ & 2 & 10 & 0.5 & 1.832 ± 0.002 \\
$w$ & $\omega$ & 2 & 10 & 1 & 1.741 ± 0.002 \\
$w$ & $\omega$ & 2 & 10 & 2 & 2.054 ± 0.003 \\
$w$ & $\omega$ & 2 & 20 & 1 & 1.696 ± 0.002 \\
$w$ & $\omega$ & 2 & 40 & 1 & 1.710 ± 0.002 \\
$l_{\perp}$ & j & 2 & 2.5 & 1 & 18.61 ± 0.10 \\
$l_{\perp}$ & j & 2 & 5 & 1 & 16.95 ± 0.09 \\
$l_{\perp}$ & j & 2 & 10 & 0.5 & 13.68 ± 0.06 \\
$l_{\perp}$ & j & 2 & 10 & 1 & 16.72 ± 0.09 \\
$l_{\perp}$ & j & 2 & 10 & 2 & 23.9 ± 0.2 \\
$l_{\perp}$ & j & 2 & 20 & 1 & 16.61 ± 0.09 \\
$l_{\perp}$ & j & 2 & 40 & 1 & 16.92 ± 0.10 \\
$l_{\perp}$ & j & 3 & 2.5 & 1 & 16.1 ± 0.1 \\
$l_{\perp}$ & j & 3 & 5 & 1 & 14.7 ± 0.1 \\
$l_{\perp}$ & j & 3 & 10 & 0.5 & 10.47 ± 0.08 \\
$l_{\perp}$ & j & 3 & 10 & 1 & 14.6 ± 0.1 \\
$l_{\perp}$ & j & 3 & 10 & 2 & 19.3 ± 0.2 \\
$l_{\perp}$ & j & 3 & 20 & 1 & 14.2 ± 0.1 \\
$l_{\perp}$ & j & 3 & 40 & 1 & 13.7 ± 0.1 \\
$l_{\perp}$ & $\omega$ & 1.5 & 2.5 & 1 & 19.77 ± 0.10 \\
$l_{\perp}$ & $\omega$ & 1.5 & 5 & 1 & 18.73 ± 0.08 \\
$l_{\perp}$ & $\omega$ & 1.5 & 10 & 0.5 & 16.61 ± 0.07 \\
$l_{\perp}$ & $\omega$ & 1.5 & 10 & 1 & 17.10 ± 0.07 \\
$l_{\perp}$ & $\omega$ & 1.5 & 10 & 2 & 19.9 ± 0.1 \\
$l_{\perp}$ & $\omega$ & 1.5 & 20 & 1 & 16.13 ± 0.06 \\
$l_{\perp}$ & $\omega$ & 1.5 & 40 & 1 & 15.26 ± 0.06 \\
$l_{\perp}$ & $\omega$ & 2 & 2.5 & 1 & 13.45 ± 0.06 \\
$l_{\perp}$ & $\omega$ & 2 & 5 & 1 & 12.04 ± 0.05 \\
$l_{\perp}$ & $\omega$ & 2 & 10 & 0.5 & 10.77 ± 0.04 \\
$l_{\perp}$ & $\omega$ & 2 & 10 & 1 & 11.29 ± 0.04 \\
$l_{\perp}$ & $\omega$ & 2 & 10 & 2 & 14.52 ± 0.06 \\
$l_{\perp}$ & $\omega$ & 2 & 20 & 1 & 11.14 ± 0.04 \\
$l_{\perp}$ & $\omega$ & 2 & 40 & 1 & 11.38 ± 0.04 \\
$l_{\parallel}$ & j & 2 & 2.5 & 1 & 28.3 ± 0.8 \\
$l_{\parallel}$ & j & 2 & 5 & 1 & 26.5 ± 0.6 \\
$l_{\parallel}$ & j & 2 & 10 & 0.5 & 39.6 ± 0.9 \\
$l_{\parallel}$ & j & 2 & 10 & 1 & 26.6 ± 0.7 \\
$l_{\parallel}$ & j & 2 & 10 & 2 & 15.6 ± 0.6 \\
$l_{\parallel}$ & j & 2 & 20 & 1 & 25.5 ± 0.8 \\
$l_{\parallel}$ & j & 2 & 40 & 1 & 24.5 ± 0.9 \\
$l_{\parallel}$ & j & 3 & 2.5 & 1 & 29.6 ± 0.9 \\
$l_{\parallel}$ & j & 3 & 5 & 1 & 27.5 ± 0.8 \\
$l_{\parallel}$ & j & 3 & 10 & 0.5 & 34.7 ± 0.7 \\
$l_{\parallel}$ & j & 3 & 10 & 1 & 26.5 ± 0.7 \\
$l_{\parallel}$ & j & 3 & 10 & 2 & 25 ± 1 \\
$l_{\parallel}$ & j & 3 & 20 & 1 & 26.5 ± 0.8 \\
$l_{\parallel}$ & j & 3 & 40 & 1 & 26.6 ± 0.9 \\
$l_{\parallel}$ & $\omega$ & 1.5 & 2.5 & 1 & 11.6 ± 0.4 \\
$l_{\parallel}$ & $\omega$ & 1.5 & 5 & 1 & 11.4 ± 0.3 \\
$l_{\parallel}$ & $\omega$ & 1.5 & 10 & 0.5 & 20.2 ± 0.7 \\
$l_{\parallel}$ & $\omega$ & 1.5 & 10 & 1 & 11.9 ± 0.2 \\
$l_{\parallel}$ & $\omega$ & 1.5 & 10 & 2 & 8.4 ± 0.3 \\
$l_{\parallel}$ & $\omega$ & 1.5 & 20 & 1 & 11.4 ± 0.2 \\
$l_{\parallel}$ & $\omega$ & 1.5 & 40 & 1 & 11.1 ± 0.2 \\
$l_{\parallel}$ & $\omega$ & 2 & 2.5 & 1 & 19.8 ± 0.4 \\
$l_{\parallel}$ & $\omega$ & 2 & 5 & 1 & 20.2 ± 0.3 \\
$l_{\parallel}$ & $\omega$ & 2 & 10 & 0.5 & 31.4 ± 0.5 \\
$l_{\parallel}$ & $\omega$ & 2 & 10 & 1 & 21.1 ± 0.3 \\
$l_{\parallel}$ & $\omega$ & 2 & 10 & 2 & 13.1 ± 0.2 \\
$l_{\parallel}$ & $\omega$ & 2 & 20 & 1 & 21.7 ± 0.3 \\
$l_{\parallel}$ & $\omega$ & 2 & 40 & 1 & 19.8 ± 0.2 \\
$\kappa$ & j & 2 & 2.5 & 1 & 0.508 ± 0.001 \\
$\kappa$ & j & 2 & 5 & 1 & 0.522 ± 0.001 \\
$\kappa$ & j & 2 & 10 & 0.5 & 0.552 ± 0.001 \\
$\kappa$ & j & 2 & 10 & 1 & 0.531 ± 0.001 \\
$\kappa$ & j & 2 & 10 & 2 & 0.508 ± 0.001 \\
$\kappa$ & j & 2 & 20 & 1 & 0.538 ± 0.001 \\
$\kappa$ & j & 2 & 40 & 1 & 0.544 ± 0.001 \\
$\kappa$ & j & 3 & 2.5 & 1 & 0.497 ± 0.002 \\
$\kappa$ & j & 3 & 5 & 1 & 0.513 ± 0.002 \\
$\kappa$ & j & 3 & 10 & 0.5 & 0.569 ± 0.003 \\
$\kappa$ & j & 3 & 10 & 1 & 0.518 ± 0.003 \\
$\kappa$ & j & 3 & 10 & 2 & 0.496 ± 0.002 \\
$\kappa$ & j & 3 & 20 & 1 & 0.526 ± 0.003 \\
$\kappa$ & j & 3 & 40 & 1 & 0.535 ± 0.003 \\
$\kappa$ & $\omega$ & 1.5 & 2.5 & 1 & 0.556 ± 0.001 \\
$\kappa$ & $\omega$ & 1.5 & 5 & 1 & 0.548 ± 0.001 \\
$\kappa$ & $\omega$ & 1.5 & 10 & 0.5 & 0.5418 ± 0.0010 \\
$\kappa$ & $\omega$ & 1.5 & 10 & 1 & 0.5439 ± 0.0009 \\
$\kappa$ & $\omega$ & 1.5 & 10 & 2 & 0.561 ± 0.001 \\
$\kappa$ & $\omega$ & 1.5 & 20 & 1 & 0.5450 ± 0.0009 \\
$\kappa$ & $\omega$ & 1.5 & 40 & 1 & 0.5511 ± 0.0009 \\
$\kappa$ & $\omega$ & 2 & 2.5 & 1 & 0.534 ± 0.001 \\
$\kappa$ & $\omega$ & 2 & 5 & 1 & 0.530 ± 0.001 \\
$\kappa$ & $\omega$ & 2 & 10 & 0.5 & 0.540 ± 0.001 \\
$\kappa$ & $\omega$ & 2 & 10 & 1 & 0.528 ± 0.001 \\
$\kappa$ & $\omega$ & 2 & 10 & 2 & 0.532 ± 0.001 \\
$\kappa$ & $\omega$ & 2 & 20 & 1 & 0.524 ± 0.001 \\
$\kappa$ & $\omega$ & 2 & 40 & 1 & 0.526 ± 0.001 \\
$\kappa_3$ & j & 2 & 2.5 & 1 & 0.115 ± 0.001 \\
$\kappa_3$ & j & 2 & 5 & 1 & 0.131 ± 0.001 \\
$\kappa_3$ & j & 2 & 10 & 0.5 & 0.160 ± 0.001 \\
$\kappa_3$ & j & 2 & 10 & 1 & 0.136 ± 0.001 \\
$\kappa_3$ & j & 2 & 10 & 2 & 0.113 ± 0.002 \\
$\kappa_3$ & j & 2 & 20 & 1 & 0.144 ± 0.002 \\
$\kappa_3$ & j & 2 & 40 & 1 & 0.149 ± 0.002 \\
$\kappa_3$ & j & 3 & 2.5 & 1 & 0.104 ± 0.002 \\
$\kappa_3$ & j & 3 & 5 & 1 & 0.123 ± 0.002 \\
$\kappa_3$ & j & 3 & 10 & 0.5 & 0.179 ± 0.003 \\
$\kappa_3$ & j & 3 & 10 & 1 & 0.128 ± 0.003 \\
$\kappa_3$ & j & 3 & 10 & 2 & 0.103 ± 0.002 \\
$\kappa_3$ & j & 3 & 20 & 1 & 0.137 ± 0.003 \\
$\kappa_3$ & j & 3 & 40 & 1 & 0.141 ± 0.003 \\
$\kappa_3$ & $\omega$ & 1.5 & 2.5 & 1 & 0.194 ± 0.002 \\
$\kappa_3$ & $\omega$ & 1.5 & 5 & 1 & 0.189 ± 0.002 \\
$\kappa_3$ & $\omega$ & 1.5 & 10 & 0.5 & 0.196 ± 0.002 \\
$\kappa_3$ & $\omega$ & 1.5 & 10 & 1 & 0.186 ± 0.002 \\
$\kappa_3$ & $\omega$ & 1.5 & 10 & 2 & 0.198 ± 0.002 \\
$\kappa_3$ & $\omega$ & 1.5 & 20 & 1 & 0.189 ± 0.002 \\
$\kappa_3$ & $\omega$ & 1.5 & 40 & 1 & 0.191 ± 0.002 \\
$\kappa_3$ & $\omega$ & 2 & 2.5 & 1 & 0.161 ± 0.002 \\
$\kappa_3$ & $\omega$ & 2 & 5 & 1 & 0.162 ± 0.001 \\
$\kappa_3$ & $\omega$ & 2 & 10 & 0.5 & 0.176 ± 0.001 \\
$\kappa_3$ & $\omega$ & 2 & 10 & 1 & 0.154 ± 0.001 \\
$\kappa_3$ & $\omega$ & 2 & 10 & 2 & 0.159 ± 0.002 \\
$\kappa_3$ & $\omega$ & 2 & 20 & 1 & 0.153 ± 0.001 \\
$\kappa_3$ & $\omega$ & 2 & 40 & 1 & 0.154 ± 0.001 \\
$\alpha$ & j & 2 & 2.5 & 1 & 7.50 ± 0.03 \\
$\alpha$ & j & 2 & 5 & 1 & 6.68 ± 0.03 \\
$\alpha$ & j & 2 & 10 & 0.5 & 5.22 ± 0.02 \\
$\alpha$ & j & 2 & 10 & 1 & 6.31 ± 0.03 \\
$\alpha$ & j & 2 & 10 & 2 & 8.10 ± 0.05 \\
$\alpha$ & j & 2 & 20 & 1 & 5.92 ± 0.03 \\
$\alpha$ & j & 2 & 40 & 1 & 5.59 ± 0.03 \\
$\alpha$ & j & 3 & 2.5 & 1 & 7.82 ± 0.06 \\
$\alpha$ & j & 3 & 5 & 1 & 6.89 ± 0.05 \\
$\alpha$ & j & 3 & 10 & 0.5 & 4.81 ± 0.04 \\
$\alpha$ & j & 3 & 10 & 1 & 6.58 ± 0.05 \\
$\alpha$ & j & 3 & 10 & 2 & 8.37 ± 0.07 \\
$\alpha$ & j & 3 & 20 & 1 & 6.14 ± 0.05 \\
$\alpha$ & j & 3 & 40 & 1 & 5.56 ± 0.05 \\
$\alpha$ & $\omega$ & 1.5 & 2.5 & 1 & 7.19 ± 0.03 \\
$\alpha$ & $\omega$ & 1.5 & 5 & 1 & 7.26 ± 0.02 \\
$\alpha$ & $\omega$ & 1.5 & 10 & 0.5 & 6.79 ± 0.02 \\
$\alpha$ & $\omega$ & 1.5 & 10 & 1 & 7.12 ± 0.02 \\
$\alpha$ & $\omega$ & 1.5 & 10 & 2 & 7.13 ± 0.03 \\
$\alpha$ & $\omega$ & 1.5 & 20 & 1 & 7.09 ± 0.02 \\
$\alpha$ & $\omega$ & 1.5 & 40 & 1 & 6.79 ± 0.02 \\
$\alpha$ & $\omega$ & 2 & 2.5 & 1 & 7.08 ± 0.03 \\
$\alpha$ & $\omega$ & 2 & 5 & 1 & 6.89 ± 0.03 \\
$\alpha$ & $\omega$ & 2 & 10 & 0.5 & 6.17 ± 0.02 \\
$\alpha$ & $\omega$ & 2 & 10 & 1 & 6.83 ± 0.02 \\
$\alpha$ & $\omega$ & 2 & 10 & 2 & 7.37 ± 0.03 \\
$\alpha$ & $\omega$ & 2 & 20 & 1 & 6.92 ± 0.02 \\
$\alpha$ & $\omega$ & 2 & 40 & 1 & 6.96 ± 0.03 \\
\end{longtable}

\clearpage
\newpage
\onecolumngrid
\section{Fit Summary}
\label{ap:trends_tables}

\begin{longtable}{|l|l|c|c|c|c|c|}
\caption{Power Law with Exponential Cutoff Fit Results} \label{tab:fit_power_law_exp_cutoff} \\
\hline
Measurement & Feature & $T_{\text{rms}}$ & $\sigma$ & $\delta B_0/B_0$ & $\beta$ & $p_1$ \\
\hline
\endfirsthead
\multicolumn{7}{c}{{\tablename\ \thetable{} -- continued from previous page}} \\
\hline
Measurement & Feature & $T_{\text{rms}}$ & $\sigma$ & $\delta B_0/B_0$ & $\beta$ & $p_1$ \\
\hline
\endhead
\hline
\multicolumn{7}{r}{{Continued on next page}} \\
\hline
\endfoot
\hline
\endlastfoot
$l_\perp$ & j & 2 & 2.5 & 1 & 0.043 ± 0.002 & 0.28 ± 0.07 \\
$l_\perp$ & j & 2 & 5 & 1 & 0.051 ± 0.002 & 0.26 ± 0.06 \\
$l_\perp$ & j & 2 & 10 & 0.5 & 0.063 ± 0.002 & 0.30 ± 0.07 \\
$l_\perp$ & j & 2 & 10 & 1 & 0.061 ± 0.005 & 0.0 ± 0.2 \\
$l_\perp$ & j & 2 & 10 & 2 & 0.034 ± 0.004 & 0.2 ± 0.1 \\
$l_\perp$ & j & 2 & 20 & 1 & 0.049 ± 0.002 & 0.31 ± 0.08 \\
$l_\perp$ & j & 2 & 40 & 1 & 0.047 ± 0.002 & 0.33 ± 0.07 \\
$l_\perp$ & j & 3 & 2.5 & 1 & 0.067 ± 0.003 & 0.01 ± 0.07 \\
$l_\perp$ & j & 3 & 5 & 1 & 0.080 ± 0.006 & -0.2 ± 0.1 \\
$l_\perp$ & j & 3 & 10 & 0.5 & 0.14 ± 0.02 & -0.7 ± 0.3 \\
$l_\perp$ & j & 3 & 10 & 1 & 0.081 ± 0.005 & -0.2 ± 0.1 \\
$l_\perp$ & j & 3 & 10 & 2 & 0.044 ± 0.003 & 0.2 ± 0.1 \\
$l_\perp$ & j & 3 & 20 & 1 & 0.087 ± 0.006 & -0.2 ± 0.1 \\
$l_\perp$ & j & 3 & 40 & 1 & 0.098 ± 0.006 & -0.3 ± 0.1 \\
$l_\perp$ & $\omega$ & 1.5 & 2.5 & 1 & 0.024 ± 0.002 & 0.71 ± 0.07 \\
$l_\perp$ & $\omega$ & 1.5 & 5 & 1 & 0.028 ± 0.003 & 0.6 ± 0.1 \\
$l_\perp$ & $\omega$ & 1.5 & 10 & 0.5 & 0.032 ± 0.003 & 0.6 ± 0.1 \\
$l_\perp$ & $\omega$ & 1.5 & 10 & 1 & 0.031 ± 0.002 & 0.63 ± 0.07 \\
$l_\perp$ & $\omega$ & 1.5 & 10 & 2 & 0.023 ± 0.002 & 0.72 ± 0.07 \\
$l_\perp$ & $\omega$ & 1.5 & 20 & 1 & 0.035 ± 0.003 & 0.5 ± 0.1 \\
$l_\perp$ & $\omega$ & 1.5 & 40 & 1 & 0.036 ± 0.003 & 0.6 ± 0.1 \\
$l_\perp$ & $\omega$ & 2 & 2.5 & 1 & 0.058 ± 0.002 & 0.40 ± 0.06 \\
$l_\perp$ & $\omega$ & 2 & 5 & 1 & 0.071 ± 0.003 & 0.37 ± 0.09 \\
$l_\perp$ & $\omega$ & 2 & 10 & 0.5 & 0.078 ± 0.004 & 0.4 ± 0.1 \\
$l_\perp$ & $\omega$ & 2 & 10 & 1 & 0.091 ± 0.004 & 0.1 ± 0.1 \\
$l_\perp$ & $\omega$ & 2 & 10 & 2 & 0.057 ± 0.002 & 0.30 ± 0.07 \\
$l_\perp$ & $\omega$ & 2 & 20 & 1 & 0.078 ± 0.004 & 0.3 ± 0.1 \\
$l_\perp$ & $\omega$ & 2 & 40 & 1 & 0.066 ± 0.003 & 0.48 ± 0.09 \\
$\alpha$ & j & 2 & 2.5 & 1 & 0.159 ± 0.002 & -0.11 ± 0.03 \\
$\alpha$ & j & 2 & 5 & 1 & 0.208 ± 0.006 & -0.32 ± 0.08 \\
$\alpha$ & j & 2 & 10 & 0.5 & 0.245 ± 0.007 & -0.13 ± 0.08 \\
$\alpha$ & j & 2 & 10 & 1 & 0.205 ± 0.006 & -0.20 ± 0.08 \\
$\alpha$ & j & 2 & 10 & 2 & 0.119 ± 0.003 & 0.14 ± 0.03 \\
$\alpha$ & j & 2 & 20 & 1 & 0.186 ± 0.005 & 0.03 ± 0.06 \\
$\alpha$ & j & 2 & 40 & 1 & 0.178 ± 0.005 & 0.18 ± 0.06 \\
$\alpha$ & j & 3 & 2.5 & 1 & 0.190 ± 0.005 & -0.41 ± 0.06 \\
$\alpha$ & j & 3 & 5 & 1 & 0.179 ± 0.004 & -0.13 ± 0.04 \\
$\alpha$ & j & 3 & 10 & 0.5 & 0.277 ± 0.008 & -0.18 ± 0.06 \\
$\alpha$ & j & 3 & 10 & 1 & 0.204 ± 0.005 & -0.26 ± 0.05 \\
$\alpha$ & j & 3 & 10 & 2 & 0.126 ± 0.004 & 0.04 ± 0.05 \\
$\alpha$ & j & 3 & 20 & 1 & 0.216 ± 0.005 & -0.21 ± 0.05 \\
$\alpha$ & j & 3 & 40 & 1 & 0.28 ± 0.01 & -0.4 ± 0.1 \\
$\alpha$ & $\omega$ & 1.5 & 2.5 & 1 & 0.122 ± 0.002 & 0.25 ± 0.03 \\
$\alpha$ & $\omega$ & 1.5 & 5 & 1 & 0.132 ± 0.003 & 0.15 ± 0.03 \\
$\alpha$ & $\omega$ & 1.5 & 10 & 0.5 & 0.167 ± 0.003 & -0.07 ± 0.04 \\
$\alpha$ & $\omega$ & 1.5 & 10 & 1 & 0.137 ± 0.003 & 0.13 ± 0.04 \\
$\alpha$ & $\omega$ & 1.5 & 10 & 2 & 0.108 ± 0.002 & 0.39 ± 0.03 \\
$\alpha$ & $\omega$ & 1.5 & 20 & 1 & 0.144 ± 0.003 & 0.08 ± 0.04 \\
$\alpha$ & $\omega$ & 1.5 & 40 & 1 & 0.145 ± 0.003 & 0.13 ± 0.04 \\
$\alpha$ & $\omega$ & 2 & 2.5 & 1 & 0.165 ± 0.002 & -0.08 ± 0.03 \\
$\alpha$ & $\omega$ & 2 & 5 & 1 & 0.166 ± 0.004 & -0.06 ± 0.05 \\
$\alpha$ & $\omega$ & 2 & 10 & 0.5 & 0.214 ± 0.005 & -0.30 ± 0.06 \\
$\alpha$ & $\omega$ & 2 & 10 & 1 & 0.183 ± 0.004 & -0.23 ± 0.06 \\
$\alpha$ & $\omega$ & 2 & 10 & 2 & 0.139 ± 0.003 & 0.07 ± 0.04 \\
$\alpha$ & $\omega$ & 2 & 20 & 1 & 0.178 ± 0.004 & -0.19 ± 0.06 \\
$\alpha$ & $\omega$ & 2 & 40 & 1 & 0.161 ± 0.005 & -0.07 ± 0.06 \\
\end{longtable}

\begin{longtable}{|l|l|c|c|c|c|c|c|}
    \caption{Broken Power Law Fit Results} \label{tab:fit_broken_power_law} \\
    \hline
    Measurement & Feature & $T_{\text{rms}}$ & $\sigma$ & $\delta B_0/B_0$ & $p_1$ & $p_2$ & $\lambda$ \\
    \hline
    \endfirsthead
    \multicolumn{8}{c}{{\tablename\ \thetable{} -- continued from previous page}} \\
    \hline
    Measurement & Feature & $T_{\text{rms}}$ & $\sigma$ & $\delta B_0/B_0$ & $p_1$ & $p_2$ & $\lambda$ \\
    \hline
    \endhead
    \hline
    \multicolumn{8}{r}{{Continued on next page}} \\
    \hline
    \endfoot
    \hline
    \endlastfoot
    $\kappa_3$ & j & 2 & 2.5 & 1 & -0.21 ± 0.05 & 1.45 ± 0.06 & 0.015 ± 0.002 \\
$\kappa_3$ & j & 2 & 5 & 1 & -0.16 ± 0.04 & 1.43 ± 0.06 & 0.018 ± 0.002 \\
$\kappa_3$ & j & 2 & 10 & 0.5 & -0.25 ± 0.05 & 1.36 ± 0.09 & 0.023 ± 0.003 \\
$\kappa_3$ & j & 2 & 10 & 1 & -0.22 ± 0.05 & 1.37 ± 0.06 & 0.016 ± 0.002 \\
$\kappa_3$ & j & 2 & 10 & 2 & -0.10 ± 0.07 & 1.51 ± 0.09 & 0.015 ± 0.003 \\
$\kappa_3$ & j & 2 & 20 & 1 & -0.21 ± 0.05 & 1.36 ± 0.06 & 0.017 ± 0.002 \\
$\kappa_3$ & j & 2 & 40 & 1 & -0.27 ± 0.07 & 1.34 ± 0.08 & 0.016 ± 0.003 \\
$\kappa_3$ & j & 3 & 2.5 & 1 & -0.22 ± 0.06 & 1.48 ± 0.07 & 0.014 ± 0.002 \\
$\kappa_3$ & j & 3 & 5 & 1 & -0.49 ± 0.07 & 1.42 ± 0.08 & 0.012 ± 0.002 \\
$\kappa_3$ & j & 3 & 10 & 0.5 & -0.32 ± 0.08 & 1.4 ± 0.1 & 0.028 ± 0.006 \\
$\kappa_3$ & j & 3 & 10 & 1 & -0.20 ± 0.07 & 1.48 ± 0.08 & 0.017 ± 0.003 \\
$\kappa_3$ & j & 3 & 10 & 2 & -0.11 ± 0.06 & 1.55 ± 0.07 & 0.014 ± 0.002 \\
$\kappa_3$ & j & 3 & 20 & 1 & -0.29 ± 0.07 & 1.37 ± 0.10 & 0.015 ± 0.003 \\
$\kappa_3$ & j & 3 & 40 & 1 & -0.19 ± 0.07 & 1.36 ± 0.10 & 0.019 ± 0.004 \\
$\kappa_3$ & $\omega$ & 1.5 & 2.5 & 1 & -0.17 ± 0.05 & 1.24 ± 0.06 & 0.017 ± 0.003 \\
$\kappa_3$ & $\omega$ & 1.5 & 5 & 1 & -0.15 ± 0.05 & 1.29 ± 0.07 & 0.021 ± 0.003 \\
$\kappa_3$ & $\omega$ & 1.5 & 10 & 0.5 & -0.16 ± 0.04 & 1.6 ± 0.1 & 0.041 ± 0.006 \\
$\kappa_3$ & $\omega$ & 1.5 & 10 & 1 & -0.20 ± 0.04 & 1.32 ± 0.07 & 0.023 ± 0.003 \\
$\kappa_3$ & $\omega$ & 1.5 & 10 & 2 & -0.06 ± 0.06 & 1.26 ± 0.08 & 0.020 ± 0.004 \\
$\kappa_3$ & $\omega$ & 1.5 & 20 & 1 & -0.12 ± 0.05 & 1.41 ± 0.09 & 0.029 ± 0.004 \\
$\kappa_3$ & $\omega$ & 1.5 & 40 & 1 & -0.19 ± 0.05 & 1.34 ± 0.08 & 0.024 ± 0.004 \\
$\kappa_3$ & $\omega$ & 2 & 2.5 & 1 & -0.21 ± 0.06 & 1.30 ± 0.08 & 0.018 ± 0.003 \\
$\kappa_3$ & $\omega$ & 2 & 5 & 1 & -0.31 ± 0.05 & 1.30 ± 0.07 & 0.018 ± 0.003 \\
$\kappa_3$ & $\omega$ & 2 & 10 & 0.5 & -0.32 ± 0.05 & 1.6 ± 0.1 & 0.038 ± 0.006 \\
$\kappa_3$ & $\omega$ & 2 & 10 & 1 & -0.30 ± 0.07 & 1.6 ± 0.1 & 0.025 ± 0.004 \\
$\kappa_3$ & $\omega$ & 2 & 10 & 2 & -0.27 ± 0.05 & 1.27 ± 0.06 & 0.014 ± 0.002 \\
$\kappa_3$ & $\omega$ & 2 & 20 & 1 & -0.28 ± 0.06 & 1.55 ± 0.09 & 0.024 ± 0.004 \\
$\kappa_3$ & $\omega$ & 2 & 40 & 1 & -0.30 ± 0.06 & 1.34 ± 0.08 & 0.017 ± 0.003 \\
$\kappa$ & j & 2 & 2.5 & 1 & -2.8 ± 0.1 & 4.1 ± 0.1 & 0.43 ± 0.01 \\
$\kappa$ & j & 2 & 5 & 1 & -2.8 ± 0.1 & 4.1 ± 0.1 & 0.443 ± 0.009 \\
$\kappa$ & j & 2 & 10 & 0.5 & -2.8 ± 0.1 & 4.2 ± 0.2 & 0.48 ± 0.01 \\
$\kappa$ & j & 2 & 10 & 1 & -2.9 ± 0.1 & 4.1 ± 0.1 & 0.45 ± 0.01 \\
$\kappa$ & j & 2 & 10 & 2 & -2.8 ± 0.1 & 4.4 ± 0.1 & 0.45 ± 0.01 \\
$\kappa$ & j & 2 & 20 & 1 & -2.9 ± 0.2 & 4.1 ± 0.2 & 0.46 ± 0.01 \\
$\kappa$ & j & 2 & 40 & 1 & -3.0 ± 0.1 & 4.2 ± 0.1 & 0.46 ± 0.01 \\
$\kappa$ & j & 3 & 2.5 & 1 & -2.2 ± 0.1 & 4.3 ± 0.1 & 0.45 ± 0.01 \\
$\kappa$ & j & 3 & 5 & 1 & -2.5 ± 0.1 & 4.2 ± 0.1 & 0.45 ± 0.01 \\
$\kappa$ & j & 3 & 10 & 0.5 & -2.4 ± 0.2 & 3.8 ± 0.2 & 0.47 ± 0.02 \\
$\kappa$ & j & 3 & 10 & 1 & -2.6 ± 0.1 & 4.1 ± 0.1 & 0.45 ± 0.01 \\
$\kappa$ & j & 3 & 10 & 2 & -2.4 ± 0.1 & 4.4 ± 0.2 & 0.45 ± 0.01 \\
$\kappa$ & j & 3 & 20 & 1 & -2.2 ± 0.1 & 4.3 ± 0.2 & 0.48 ± 0.01 \\
$\kappa$ & j & 3 & 40 & 1 & -2.5 ± 0.1 & 4.2 ± 0.1 & 0.47 ± 0.01 \\
$\kappa$ & $\omega$ & 1.5 & 2.5 & 1 & -3.1 ± 0.1 & 4.0 ± 0.2 & 0.45 ± 0.01 \\
$\kappa$ & $\omega$ & 1.5 & 5 & 1 & -3.0 ± 0.1 & 4.2 ± 0.2 & 0.46 ± 0.01 \\
$\kappa$ & $\omega$ & 1.5 & 10 & 0.5 & -2.9 ± 0.1 & 4.4 ± 0.2 & 0.47 ± 0.01 \\
$\kappa$ & $\omega$ & 1.5 & 10 & 1 & -3.0 ± 0.1 & 4.2 ± 0.1 & 0.45 ± 0.01 \\
$\kappa$ & $\omega$ & 1.5 & 10 & 2 & -3.1 ± 0.2 & 3.9 ± 0.2 & 0.45 ± 0.01 \\
$\kappa$ & $\omega$ & 1.5 & 20 & 1 & -2.9 ± 0.1 & 4.0 ± 0.1 & 0.451 ± 0.009 \\
$\kappa$ & $\omega$ & 1.5 & 40 & 1 & -2.9 ± 0.1 & 4.1 ± 0.2 & 0.46 ± 0.01 \\
$\kappa$ & $\omega$ & 2 & 2.5 & 1 & -2.9 ± 0.2 & 3.8 ± 0.1 & 0.43 ± 0.01 \\
$\kappa$ & $\omega$ & 2 & 5 & 1 & -2.8 ± 0.2 & 3.8 ± 0.2 & 0.42 ± 0.02 \\
$\kappa$ & $\omega$ & 2 & 10 & 0.5 & -2.7 ± 0.1 & 4.1 ± 0.2 & 0.46 ± 0.01 \\
$\kappa$ & $\omega$ & 2 & 10 & 1 & -2.4 ± 0.1 & 4.1 ± 0.2 & 0.45 ± 0.01 \\
$\kappa$ & $\omega$ & 2 & 10 & 2 & -2.8 ± 0.2 & 3.9 ± 0.1 & 0.43 ± 0.01 \\
$\kappa$ & $\omega$ & 2 & 20 & 1 & -2.6 ± 0.2 & 4.2 ± 0.2 & 0.44 ± 0.01 \\
$\kappa$ & $\omega$ & 2 & 40 & 1 & -2.7 ± 0.2 & 3.8 ± 0.2 & 0.42 ± 0.02 \\
$l_\parallel$ & j & 2 & 2.5 & 1 & -0.9 ± 0.2 & 2.41 ± 0.09 & 17 ± 1 \\
$l_\parallel$ & j & 2 & 5 & 1 & -1.0 ± 0.3 & 2.7 ± 0.2 & 18 ± 2 \\
$l_\parallel$ & j & 2 & 10 & 0.5 & -1.9 ± 0.2 & 2.2 ± 0.1 & 21.0 ± 1.0 \\
$l_\parallel$ & j & 2 & 10 & 1 & -0.9 ± 0.3 & 2.7 ± 0.1 & 17 ± 1 \\
$l_\parallel$ & j & 2 & 10 & 2 & 0.1 ± 0.2 & 2.8 ± 0.1 & 15 ± 1 \\
$l_\parallel$ & j & 2 & 20 & 1 & -1.5 ± 0.4 & 2.27 ± 0.10 & 13.4 ± 0.9 \\
$l_\parallel$ & j & 2 & 40 & 1 & -0.4 ± 0.2 & 2.37 ± 0.09 & 15 ± 1 \\
$l_\parallel$ & j & 3 & 2.5 & 1 & -2.6 ± 0.4 & 2.6 ± 0.2 & 17 ± 1 \\
$l_\parallel$ & j & 3 & 5 & 1 & -2.1 ± 0.3 & 2.6 ± 0.1 & 16.6 ± 0.8 \\
$l_\parallel$ & j & 3 & 10 & 0.5 & -3.1 ± 0.5 & 3.3 ± 0.3 & 24 ± 2 \\
$l_\parallel$ & j & 3 & 10 & 1 & -2.3 ± 0.4 & 2.8 ± 0.2 & 17 ± 1 \\
$l_\parallel$ & j & 3 & 10 & 2 & -0.3 ± 0.3 & 2.4 ± 0.1 & 17 ± 2 \\
$l_\parallel$ & j & 3 & 20 & 1 & -1.2 ± 0.3 & 2.7 ± 0.1 & 18 ± 1 \\
$l_\parallel$ & j & 3 & 40 & 1 & -2.3 ± 0.5 & 2.5 ± 0.2 & 15 ± 1 \\
$l_\parallel$ & $\omega$ & 1.5 & 2.5 & 1 & 1.0 ± 0.3 & 3.4 ± 0.2 & 18 ± 2 \\
$l_\parallel$ & $\omega$ & 1.5 & 5 & 1 & 0.6 ± 0.4 & 4.8 ± 0.4 & 19 ± 2 \\
$l_\parallel$ & $\omega$ & 1.5 & 10 & 0.5 & 0.4 ± 0.2 & 4.3 ± 0.4 & 30 ± 3 \\
$l_\parallel$ & $\omega$ & 1.5 & 10 & 1 & 0.7 ± 0.4 & 4.6 ± 0.4 & 19 ± 2 \\
$l_\parallel$ & $\omega$ & 1.5 & 10 & 2 & 1.3 ± 0.6 & 3.9 ± 0.3 & 14 ± 2 \\
$l_\parallel$ & $\omega$ & 1.5 & 20 & 1 & 0.8 ± 0.3 & 4.8 ± 0.3 & 19 ± 2 \\
$l_\parallel$ & $\omega$ & 1.5 & 40 & 1 & 1.1 ± 0.3 & 5.5 ± 0.4 & 22 ± 2 \\
$l_\parallel$ & $\omega$ & 2 & 2.5 & 1 & -0.5 ± 0.3 & 3.7 ± 0.2 & 21 ± 2 \\
$l_\parallel$ & $\omega$ & 2 & 5 & 1 & -0.7 ± 0.3 & 3.6 ± 0.2 & 20 ± 1 \\
$l_\parallel$ & $\omega$ & 2 & 10 & 0.5 & -1.9 ± 0.2 & 3.2 ± 0.2 & 23 ± 1 \\
$l_\parallel$ & $\omega$ & 2 & 10 & 1 & -0.9 ± 0.3 & 3.2 ± 0.1 & 18 ± 1 \\
$l_\parallel$ & $\omega$ & 2 & 10 & 2 & 0.9 ± 0.2 & 4.1 ± 0.2 & 22 ± 2 \\
$l_\parallel$ & $\omega$ & 2 & 20 & 1 & -0.8 ± 0.3 & 3.2 ± 0.1 & 18 ± 1 \\
$l_\parallel$ & $\omega$ & 2 & 40 & 1 & -0.4 ± 0.3 & 3.6 ± 0.2 & 20 ± 1 \\
\end{longtable}

\clearpage
\newpage
\onecolumngrid

\begin{deluxetable}{l@{\hspace{2pt}}l@{\hspace{2pt}}c@{\hspace{2pt}}c@{\hspace{2pt}}c@{\hspace{2pt}}c}
\tablewidth{\columnwidth}
\tabletypesize{\scriptsize}
\tablecolumns{6}
\tablecaption{Exponential Fit Results
\label{tab:fit_exponential}}
\tablehead{
\colhead{Measurement} & \colhead{Feature} & \colhead{$T_{\text{rms}}$} & \colhead{$\sigma$} & \colhead{$\delta B_0/B_0$} & \colhead{$\beta$}
}
\startdata
$w$ & j & 2 & 2.5 & 1 & 1.45 ± 0.04 \\
$w$ & j & 2 & 5 & 1 & 1.41 ± 0.04 \\
$w$ & j & 2 & 10 & 0.5 & 1.42 ± 0.04 \\
$w$ & j & 2 & 10 & 1 & 1.25 ± 0.03 \\
$w$ & j & 2 & 10 & 2 & 0.91 ± 0.01 \\
$w$ & j & 2 & 20 & 1 & 1.11 ± 0.03 \\
$w$ & j & 2 & 40 & 1 & 0.92 ± 0.02 \\
$w$ & j & 3 & 2.5 & 1 & 1.80 ± 0.06 \\
$w$ & j & 3 & 5 & 1 & 1.9 ± 0.1 \\
$w$ & j & 3 & 10 & 0.5 & 1.61 ± 0.04 \\
$w$ & j & 3 & 10 & 1 & 1.69 ± 0.06 \\
$w$ & j & 3 & 10 & 2 & 1.22 ± 0.03 \\
$w$ & j & 3 & 20 & 1 & 1.57 ± 0.05 \\
$w$ & j & 3 & 40 & 1 & 1.29 ± 0.05 \\
$w$ & $\omega$ & 1.5 & 2.5 & 1 & 1.26 ± 0.04 \\
$w$ & $\omega$ & 1.5 & 5 & 1 & 1.38 ± 0.03 \\
$w$ & $\omega$ & 1.5 & 10 & 0.5 & 1.21 ± 0.03 \\
$w$ & $\omega$ & 1.5 & 10 & 1 & 1.50 ± 0.04 \\
$w$ & $\omega$ & 1.5 & 10 & 2 & 1.03 ± 0.01 \\
$w$ & $\omega$ & 1.5 & 20 & 1 & 1.62 ± 0.04 \\
$w$ & $\omega$ & 1.5 & 40 & 1 & 1.67 ± 0.03 \\
$w$ & $\omega$ & 2 & 2.5 & 1 & 2.06 ± 0.05 \\
$w$ & $\omega$ & 2 & 5 & 1 & 2.05 ± 0.03 \\
$w$ & $\omega$ & 2 & 10 & 0.5 & 2.03 ± 0.05 \\
$w$ & $\omega$ & 2 & 10 & 1 & 2.27 ± 0.04 \\
$w$ & $\omega$ & 2 & 10 & 2 & 1.41 ± 0.07 \\
$w$ & $\omega$ & 2 & 20 & 1 & 2.3 ± 0.1 \\
$w$ & $\omega$ & 2 & 40 & 1 & 2.2 ± 0.1 \\
\enddata
\end{deluxetable}

\section{Additional Histograms for $T_{\text{rms}}=1.5$ and $T_{\text{rms}}=3$}
\vspace{-7px}
\label{ap:rms_hists}
\subsection{Current Sheets $T_{\text{rms}}=3$}
\begin{figure}[H]
    \centering
    \includegraphics[width=\textwidth]{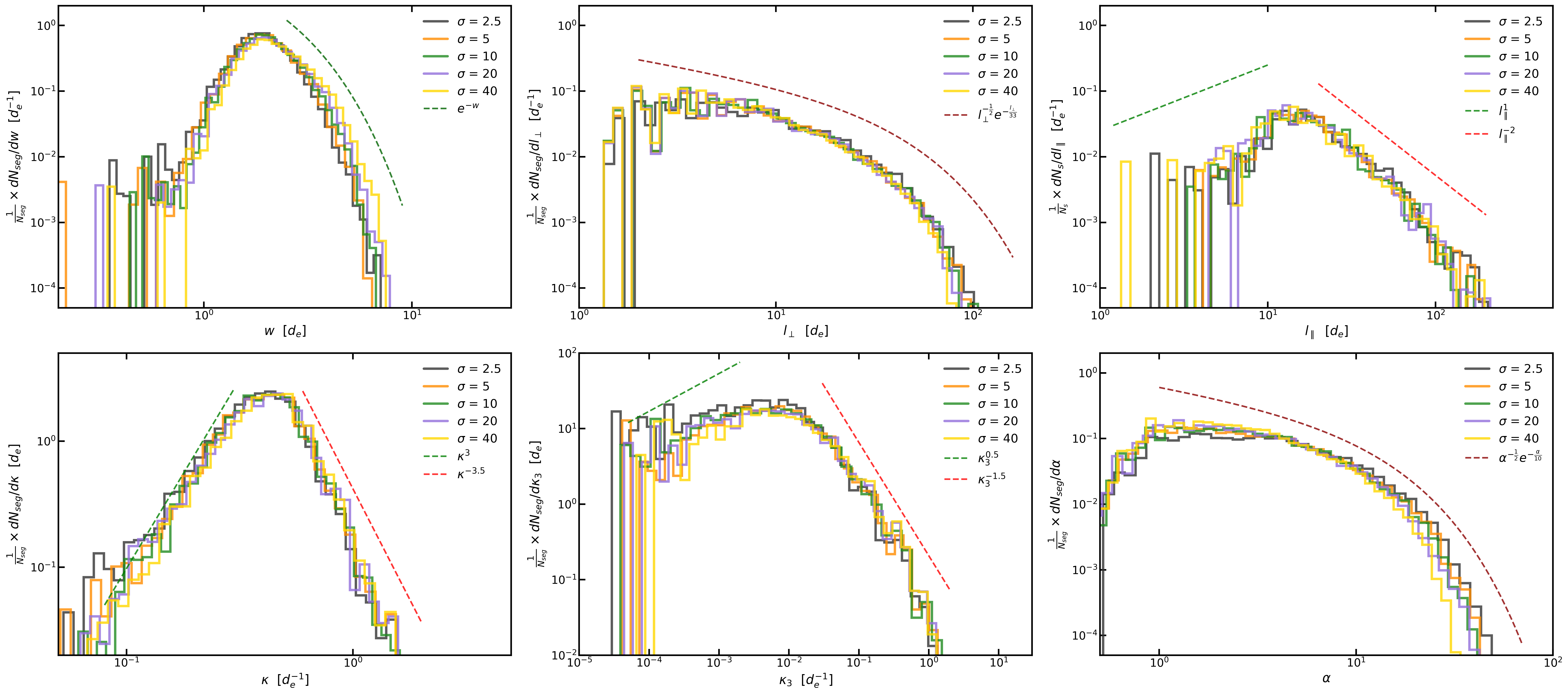}
    \caption{PDFs of current sheet measurements with $T_{\text{rms}}=3$ for different values of $\sigma$ (see legend), with illustrative fits shown as dashed lines (see legend). From Left to right, the top panels show the structure width ($w$), perpendicular length ($l_\perp$), and length along the mean field ($l_\parallel$). From left to right, the bottom panels show  aspect ratio ($\alpha$), the local curvature ($\kappa$), and the three-point curvature ($\kappa_3$).}
    \label{fig:cs_rms3_sigma}
\end{figure}

\begin{figure}[H]
    \centering
    \includegraphics[width=\textwidth]{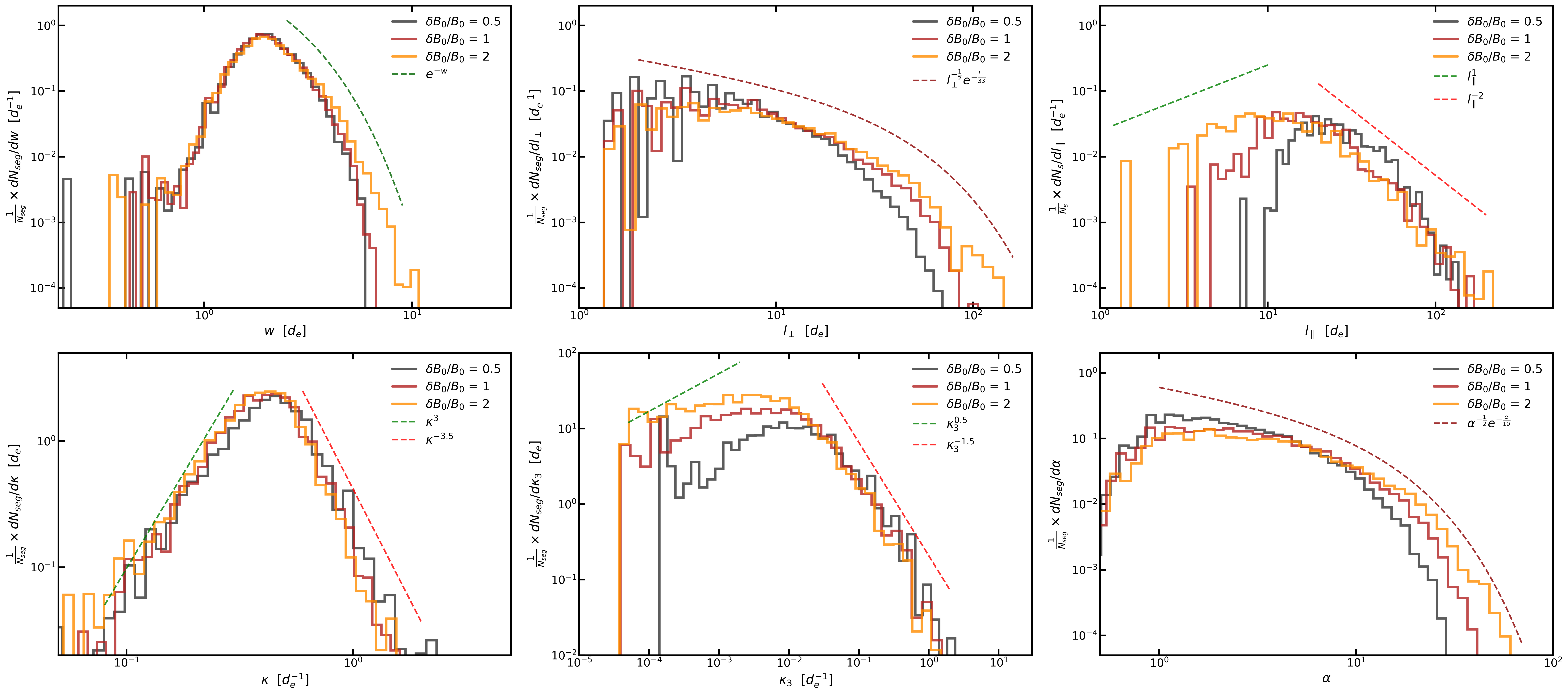}
    \caption{PDFs of current sheet measurements with $T_{\text{rms}}=3$ for different values of \db\ (see legend), with illustrative fits shown as dashed lines (see legend). From Left to right, the top panels show the structure width ($w$), perpendicular length ($l_\perp$), and length along the mean field ($l_\parallel$). From left to right, the bottom panels show  aspect ratio ($\alpha$), the local curvature ($\kappa$), and the three-point curvature ($\kappa_3$).}
    \label{fig:cs_rms3_db}
\end{figure}

\subsection{Vorticity Sheets $T_{\text{rms}}=1.5$}
\begin{figure}[H]
    \centering
    \includegraphics[width=\textwidth]{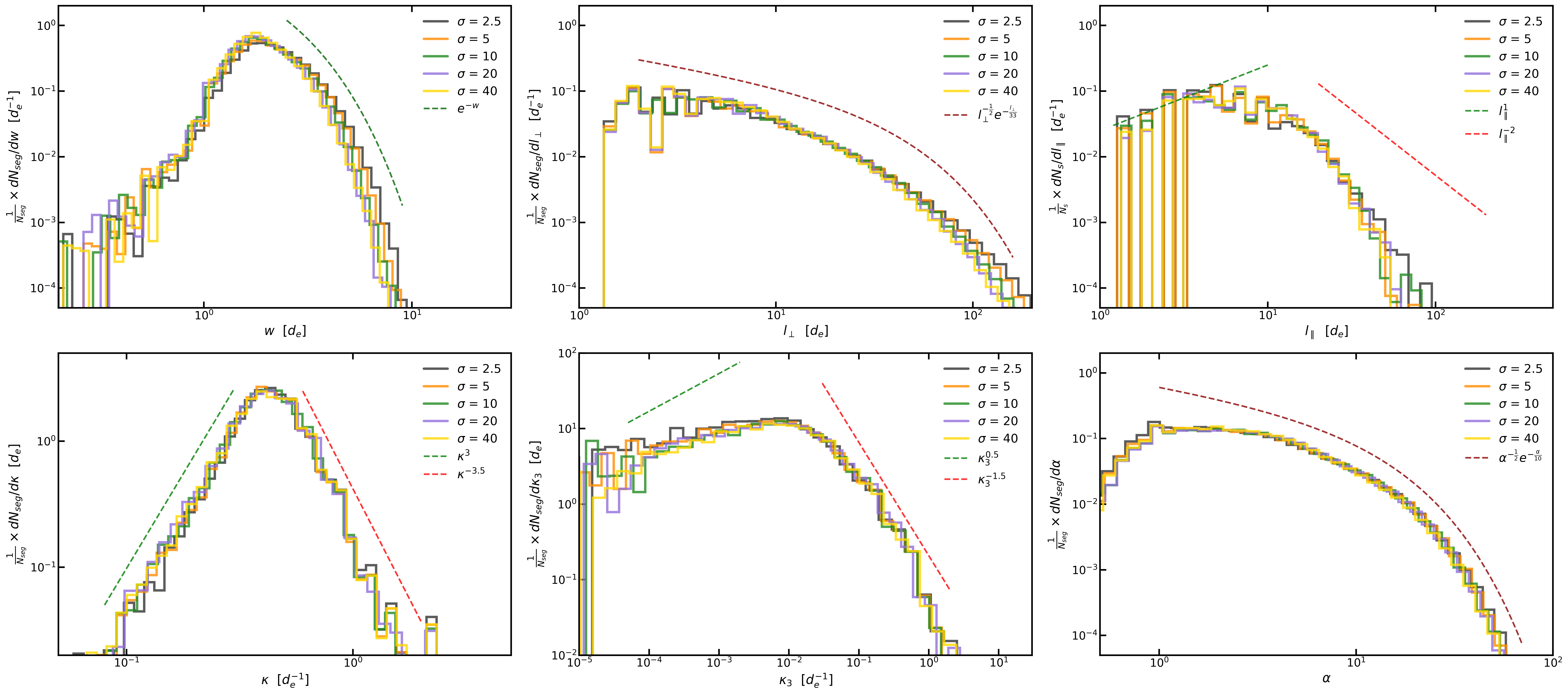}
    \caption{PDFs of vorticity sheet measurements with $T_{\text{rms}}=1.5$ for different values of $\sigma$ (see legend), with illustrative fits shown as dashed lines (see legend). From Left to right, the top panels show the structure width ($w$), perpendicular length ($l_\perp$), and length along the mean field ($l_\parallel$). From left to right, the bottom panels show  aspect ratio ($\alpha$), the local curvature ($\kappa$), and the three-point curvature ($\kappa_3$).}
    \label{fig:vs_rms1.5_sigma}
\end{figure}

\begin{figure}[H]
    \centering
    \includegraphics[width=\textwidth]{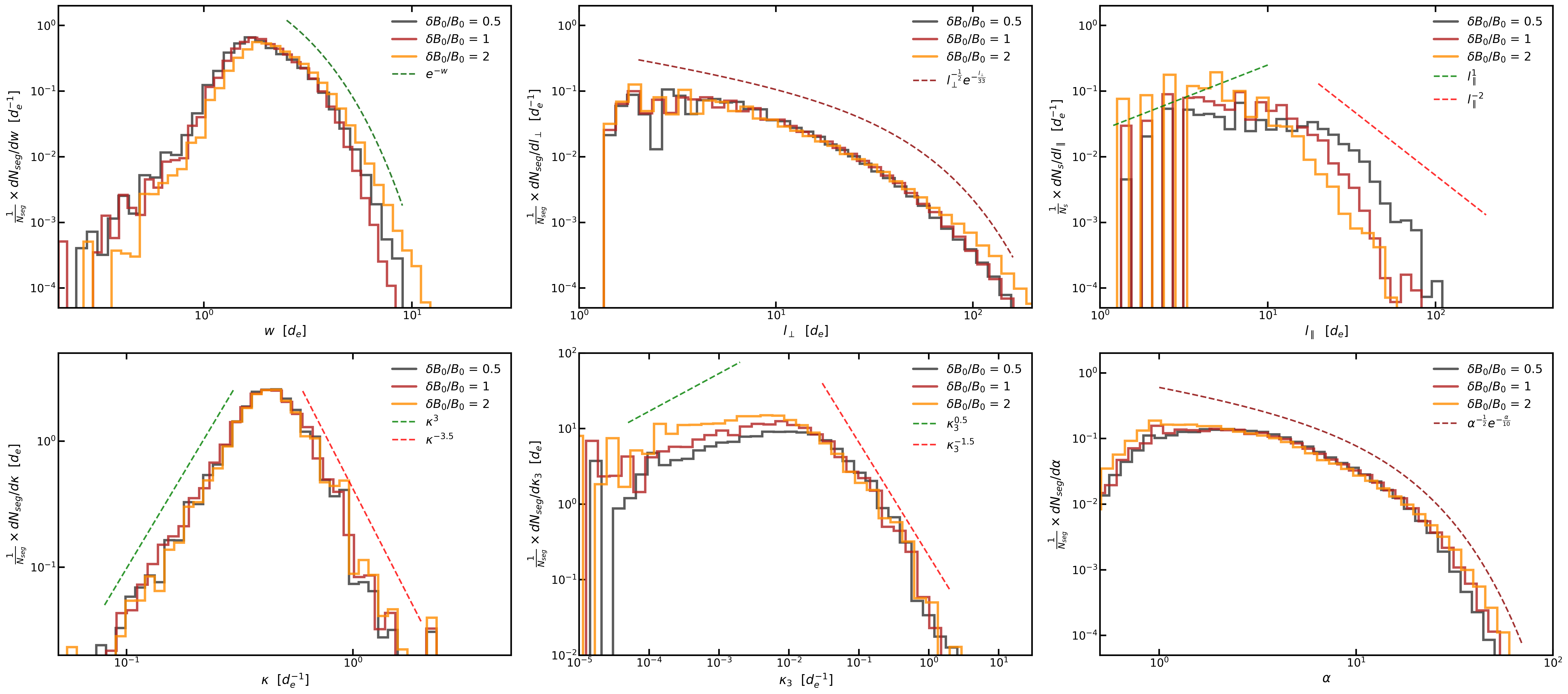}
    \caption{PDFs of vorticity sheet measurements with $T_{\text{rms}}=1.5$ for different values of \db\ (see legend), with illustrative fits shown as dashed lines (see legend). From Left to right, the top panels show the structure width ($w$), perpendicular length ($l_\perp$), and length along the mean field ($l_\parallel$). From left to right, the bottom panels show  aspect ratio ($\alpha$), the local curvature ($\kappa$), and the three-point curvature ($\kappa_3$).}
    \label{fig:vs_rms1.5_db}
\end{figure}

\bibliography{main}{}

\begin{thebibliography}{}
\expandafter\ifx\csname natexlab\endcsname\relax\def\natexlab#1{#1}\fi
\providecommand{\url}[1]{\href{#1}{#1}}
\providecommand{\dodoi}[1]{doi:~\href{http://doi.org/#1}{\nolinkurl{#1}}}
\providecommand{\doeprint}[1]{\href{http://ascl.net/#1}{\nolinkurl{http://ascl.net/#1}}}
\providecommand{\doarXiv}[1]{\href{https://arxiv.org/abs/#1}{\nolinkurl{https://arxiv.org/abs/#1}}}

% type= article
\bibitem[{R. {Bandyopadhyay} {et~al.}(2020){Bandyopadhyay}, {Yang}, {Matthaeus}, {Chasapis}, {Parashar}, {Russell}, {Strangeway}, {Torbert}, {Giles}, {Gershman}, {Pollock}, {Moore}, \& {Burch}}]{bandyopadhyay2020}
{Bandyopadhyay}, R., {Yang}, Y., {Matthaeus}, W.~H., {et~al.} 2020, \bibinfo{title}{{In Situ Measurement of Curvature of Magnetic Field in Turbulent Space Plasmas: A Statistical Study},} \apjl, 893, L25, \dodoi{10.3847/2041-8213/ab846e}

% type= article
\bibitem[{M.~V. {Barkov} \& S.~S. {Komissarov}(2016){Barkov} \& {Komissarov}}]{Barkov2016}
{Barkov}, M.~V., \& {Komissarov}, S.~S. 2016, \bibinfo{title}{{Relativistic tearing and drift-kink instabilities in two-fluid simulations},} \mnras, 458, 1939, \dodoi{10.1093/mnras/stw384}

% type= book
\bibitem[{G. {Bateman}(1978){Bateman}}]{Bateman1978}
{Bateman}, G. 1978, {MHD instabilities}

% type= article
\bibitem[{A.~R. {Bell}(1978{\natexlab{a}}){Bell}}]{Bell1978I}
{Bell}, A.~R. 1978{\natexlab{a}}, \bibinfo{title}{{The acceleration of cosmic rays in shock fronts - I.},} \mnras, 182, 147, \dodoi{10.1093/mnras/182.2.147}

% type= article
\bibitem[{A.~R. {Bell}(1978{\natexlab{b}}){Bell}}]{Bell1978II}
{Bell}, A.~R. 1978{\natexlab{b}}, \bibinfo{title}{{The acceleration of cosmic rays in shock fronts - II.},} \mnras, 182, 443, \dodoi{10.1093/mnras/182.3.443}

% type= article
\bibitem[{A.~R. {Bell}(2004){Bell}}]{Bell2004}
{Bell}, A.~R. 2004, \bibinfo{title}{{Turbulent amplification of magnetic field and diffusive shock acceleration of cosmic rays},} \mnras, 353, 550, \dodoi{10.1111/j.1365-2966.2004.08097.x}

% type= article
\bibitem[{I.~B. {Bernstein} \& F. {Engelmann}(1966){Bernstein} \& {Engelmann}}]{bernstein1966}
{Bernstein}, I.~B., \& {Engelmann}, F. 1966, \bibinfo{title}{{Quasi-Linear Theory of Plasma Waves},} Physics of Fluids, 9, 937, \dodoi{10.1063/1.1761795}

% type= article
\bibitem[{R. Blandford \& D. Eichler(1987)Blandford \& Eichler}]{Blandford1987}
Blandford, R., \& Eichler, D. 1987, \bibinfo{title}{Particle acceleration at astrophysical shocks: A theory of cosmic ray origin,} Physics Reports, 154, 1, \dodoi{https://doi.org/10.1016/0370-1573(87)90134-7}

% type= article
\bibitem[{R.~D. {Blandford} \& R.~L. {Znajek}(1977){Blandford} \& {Znajek}}]{Blandford1977}
{Blandford}, R.~D., \& {Znajek}, R.~L. 1977, \bibinfo{title}{{Electromagnetic extraction of energy from Kerr black holes.},} \mnras, 179, 433, \dodoi{10.1093/mnras/179.3.433}

% type= article
\bibitem[{S. {Boldyrev}(2005){Boldyrev}}]{Boldyrev2005}
{Boldyrev}, S. 2005, \bibinfo{title}{{On the Spectrum of Magnetohydrodynamic Turbulence},} \apjl, 626, L37, \dodoi{10.1086/431649}

% type= article
\bibitem[{N. {Borse} {et~al.}(2021){Borse}, {Acharya}, {Vaidya}, {Mukherjee}, {Bodo}, {Rossi}, \& {Mignone}}]{Borse2021}
{Borse}, N., {Acharya}, S., {Vaidya}, B., {et~al.} 2021, \bibinfo{title}{{Numerical study of the Kelvin-Helmholtz instability and its effect on synthetic emission from magnetized jets},} \aap, 649, A150, \dodoi{10.1051/0004-6361/202140440}

% type= article
\bibitem[{M. {Bussov} \& J. {N{\"a}ttil{\"a}}(2021){Bussov} \& {N{\"a}ttil{\"a}}}]{bussov2021}
{Bussov}, M., \& {N{\"a}ttil{\"a}}, J. 2021, \bibinfo{title}{{Segmentation of turbulent computational fluid dynamics simulations with unsupervised ensemble learning},} Signal Processing: Image Communication, 99, 116450, \dodoi{10.1016/j.image.2021.116450}

% type= article
\bibitem[{C. {Bustard} \& E.~G. {Zweibel}(2021){Bustard} \& {Zweibel}}]{Bustard2021}
{Bustard}, C., \& {Zweibel}, E.~G. 2021, \bibinfo{title}{{Cosmic-Ray Transport, Energy Loss, and Influence in the Multiphase Interstellar Medium},} \apj, 913, 106, \dodoi{10.3847/1538-4357/abf64c}

% type= article
\bibitem[{A.~M. {Bykov} \& P. {Meszaros}(1996){Bykov} \& {Meszaros}}]{Bykov1996}
{Bykov}, A.~M., \& {Meszaros}, P. 1996, \bibinfo{title}{{Electron Acceleration and Efficiency in Nonthermal Gamma-Ray Sources},} \apjl, 461, L37, \dodoi{10.1086/309999}

% type= article
\bibitem[{D. {Caprioli} {et~al.}(2020){Caprioli}, {Haggerty}, \& {Blasi}}]{Caprioli2020}
{Caprioli}, D., {Haggerty}, C.~C., \& {Blasi}, P. 2020, \bibinfo{title}{{Kinetic Simulations of Cosmic-Ray-modified Shocks. II. Particle Spectra},} \apj, 905, 2, \dodoi{10.3847/1538-4357/abbe05}

% type= article
\bibitem[{D. {Caprioli} \& A. {Spitkovsky}(2014){Caprioli} \& {Spitkovsky}}]{Caprioli2014}
{Caprioli}, D., \& {Spitkovsky}, A. 2014, \bibinfo{title}{{Simulations of Ion Acceleration at Non-relativistic Shocks. III. Particle Diffusion},} \apj, 794, 47, \dodoi{10.1088/0004-637X/794/1/47}

% type= article
\bibitem[{Y.-S. {Chen} \& W.-H. {Hsu}(1988){Chen} \& {Hsu}}]{Chen1988}
{Chen}, Y.-S., \& {Hsu}, W.-H. 1988, \bibinfo{title}{{A modified fast parallel algorithm for thinning digital patterns},} Pattern Recognition Letters, 7, 99, \dodoi{10.1016/0167-8655(88)90124-9}

% type= article
\bibitem[{L. {Comisso} {et~al.}(2024){Comisso}, {Farrar}, \& {Muzio}}]{Comisso2024}
{Comisso}, L., {Farrar}, G.~R., \& {Muzio}, M.~S. 2024, \bibinfo{title}{{Ultra-High-Energy Cosmic Rays Accelerated by Magnetically Dominated Turbulence},} \apjl, 977, L18, \dodoi{10.3847/2041-8213/ad955f}

% type= article
\bibitem[{L. {Comisso} \& L. {Sironi}(2018){Comisso} \& {Sironi}}]{comisso2018}
{Comisso}, L., \& {Sironi}, L. 2018, \bibinfo{title}{{Particle Acceleration in Relativistic Plasma Turbulence},} \prl, 121, 255101, \dodoi{10.1103/PhysRevLett.121.255101}

% type= article
\bibitem[{L. {Comisso} \& L. {Sironi}(2019){Comisso} \& {Sironi}}]{comisso2019}
{Comisso}, L., \& {Sironi}, L. 2019, \bibinfo{title}{{The Interplay of Magnetically Dominated Turbulence and Magnetic Reconnection in Producing Nonthermal Particles},} \apj, 886, 122, \dodoi{10.3847/1538-4357/ab4c33}

% type= article
\bibitem[{L. {Comisso} \& L. {Sironi}(2021){Comisso} \& {Sironi}}]{Comisso2021}
{Comisso}, L., \& {Sironi}, L. 2021, \bibinfo{title}{{Pitch-Angle Anisotropy Controls Particle Acceleration and Cooling in Radiative Relativistic Plasma Turbulence},} \prl, 127, 255102, \dodoi{10.1103/PhysRevLett.127.255102}

% type= article
\bibitem[{L. {Comisso} \& L. {Sironi}(2022){Comisso} \& {Sironi}}]{Comisso2022}
{Comisso}, L., \& {Sironi}, L. 2022, \bibinfo{title}{{Ion and Electron Acceleration in Fully Kinetic Plasma Turbulence},} \apjl, 936, L27, \dodoi{10.3847/2041-8213/ac8422}

% type= article
\bibitem[{L. {Comisso} {et~al.}(2020){Comisso}, {Sobacchi}, \& {Sironi}}]{comisso2020}
{Comisso}, L., {Sobacchi}, E., \& {Sironi}, L. 2020, \bibinfo{title}{{Hard Synchrotron Spectra from Magnetically Dominated Plasma Turbulence},} \apjl, 895, L40, \dodoi{10.3847/2041-8213/ab93dc}

% type= article
\bibitem[{S.~R. {Cranmer} {et~al.}(2007){Cranmer}, {van Ballegooijen}, \& {Edgar}}]{Cranmer2007}
{Cranmer}, S.~R., {van Ballegooijen}, A.~A., \& {Edgar}, R.~J. 2007, \bibinfo{title}{{Self-consistent Coronal Heating and Solar Wind Acceleration from Anisotropic Magnetohydrodynamic Turbulence},} \apjs, 171, 520, \dodoi{10.1086/518001}

% type= article
\bibitem[{J.~T. {Dahlin} {et~al.}(2014){Dahlin}, {Drake}, \& {Swisdak}}]{Dahlin2014}
{Dahlin}, J.~T., {Drake}, J.~F., \& {Swisdak}, M. 2014, \bibinfo{title}{{The mechanisms of electron heating and acceleration during magnetic reconnection},} Physics of Plasmas, 21, 092304, \dodoi{10.1063/1.4894484}

% type= article
\bibitem[{S. {Das} {et~al.}(2025){Das}, {Xu}, \& {N{\"a}ttil{\"a}}}]{Das2025}
{Das}, S., {Xu}, S., \& {N{\"a}ttil{\"a}}, J. 2025, \bibinfo{title}{{Studying mirror acceleration via kinetic simulations of relativistic plasma turbulence},} arXiv e-prints, arXiv:2506.04212, \dodoi{10.48550/arXiv.2506.04212}

% type= article
\bibitem[{Z. {Davis} {et~al.}(2024){Davis}, {Comisso}, \& {Giannios}}]{Davis2024}
{Davis}, Z., {Comisso}, L., \& {Giannios}, D. 2024, \bibinfo{title}{{Intermittency and Dissipative Structures Arising from Relativistic Magnetized Turbulence},} \apj, 964, 14, \dodoi{10.3847/1538-4357/ad284c}

% type= article
\bibitem[{Z. {Davis} {et~al.}(2022){Davis}, {Rueda-Becerril}, \& {Giannios}}]{Davis2022}
{Davis}, Z., {Rueda-Becerril}, J.~M., \& {Giannios}, D. 2022, \bibinfo{title}{{Balancing turbulent heating with radiative cooling in blazars},} \mnras, 513, 5766, \dodoi{10.1093/mnras/stac1282}

% type= article
\bibitem[{C. {Demidem} {et~al.}(2020){Demidem}, {Lemoine}, \& {Casse}}]{demidem2020}
{Demidem}, C., {Lemoine}, M., \& {Casse}, F. 2020, \bibinfo{title}{{Particle acceleration in relativistic turbulence: A theoretical appraisal},} \prd, 102, 023003, \dodoi{10.1103/PhysRevD.102.023003}

% type= article
\bibitem[{C. {Dong} {et~al.}(2022){Dong}, {Wang}, {Huang}, {Comisso}, {Sandstrom}, \& {Bhattacharjee}}]{dong2022}
{Dong}, C., {Wang}, L., {Huang}, Y.-M., {et~al.} 2022, \bibinfo{title}{{Reconnection-driven energy cascade in magnetohydrodynamic turbulence},} Science Advances, 8, eabn7627, \dodoi{10.1126/sciadv.abn7627}

% type= article
\bibitem[{L.~O. {Drury}(1983){Drury}}]{Drury1983}
{Drury}, L.~O. 1983, \bibinfo{title}{{REVIEW ARTICLE: An introduction to the theory of diffusive shock acceleration of energetic particles in tenuous plasmas},} Reports on Progress in Physics, 46, 973, \dodoi{10.1088/0034-4885/46/8/002}

% type= article
\bibitem[{B. {Dubrulle}(1994){Dubrulle}}]{dubrulle1994}
{Dubrulle}, B. 1994, \bibinfo{title}{{Intermittency in fully developed turbulence: Log-Poisson statistics and generalized scale covariance},} \prl, 73, 959, \dodoi{10.1103/PhysRevLett.73.959}

% type= article
\bibitem[{B.~G. {Elmegreen} \& J. {Scalo}(2004){Elmegreen} \& {Scalo}}]{Elmegreen2004}
{Elmegreen}, B.~G., \& {Scalo}, J. 2004, \bibinfo{title}{{Interstellar Turbulence I: Observations and Processes},} \araa, 42, 211, \dodoi{10.1146/annurev.astro.41.011802.094859}

% type= article
\bibitem[{E. {Fermi}(1949){Fermi}}]{fermi1949}
{Fermi}, E. 1949, \bibinfo{title}{{On the Origin of the Cosmic Radiation},} Physical Review, 75, 1169, \dodoi{10.1103/PhysRev.75.1169}

% type= article
\bibitem[{D.~F.~G. {Fiorillo} {et~al.}(2025){Fiorillo}, {Comisso}, {Peretti}, {Petropoulou}, \& {Sironi}}]{Fiorillo2025}
{Fiorillo}, D. F.~G., {Comisso}, L., {Peretti}, E., {Petropoulou}, M., \& {Sironi}, L. 2025, \bibinfo{title}{{The contribution of turbulent AGN coronae to the diffuse neutrino flux},} arXiv e-prints, arXiv:2504.06336, \dodoi{10.48550/arXiv.2504.06336}

% type= article
\bibitem[{O. {French} {et~al.}(2023){French}, {Guo}, {Zhang}, \& {Uzdensky}}]{French2023}
{French}, O., {Guo}, F., {Zhang}, Q., \& {Uzdensky}, D.~A. 2023, \bibinfo{title}{{Particle Injection and Nonthermal Particle Acceleration in Relativistic Magnetic Reconnection},} \apj, 948, 19, \dodoi{10.3847/1538-4357/acb7dd}

% type= article
\bibitem[{D. {Giannios}(2013){Giannios}}]{giannios2013}
{Giannios}, D. 2013, \bibinfo{title}{{Reconnection-driven plasmoids in blazars: fast flares on a slow envelope},} \mnras, 431, 355, \dodoi{10.1093/mnras/stt167}

% type= article
\bibitem[{P. {Goldreich} \& S. {Sridhar}(1995){Goldreich} \& {Sridhar}}]{goldreich1995}
{Goldreich}, P., \& {Sridhar}, S. 1995, \bibinfo{title}{{Toward a Theory of Interstellar Turbulence. II. Strong Alfvenic Turbulence},} \apj, 438, 763, \dodoi{10.1086/175121}

% type= article
\bibitem[{K. {Gootkin} {et~al.}(2025){Gootkin}, {Haggerty}, {Caprioli}, \& {Davis}}]{Gootkin2025}
{Gootkin}, K., {Haggerty}, C., {Caprioli}, D., \& {Davis}, Z. 2025, \bibinfo{title}{{Efficient Particle Acceleration in 2.5-Dimensional, Hybrid-Kinetic Simulations of Decaying, Supersonic, Plasma Turbulence},} arXiv e-prints, arXiv:2509.18374.
\newblock \doarXiv{2509.18374}

% type= article
\bibitem[{T. {Ha} {et~al.}(2025){Ha}, {N{\"a}ttil{\"a}}, {Davelaar}, \& {Sironi}}]{Ha2025}
{Ha}, T., {N{\"a}ttil{\"a}}, J., {Davelaar}, J., \& {Sironi}, L. 2025, \bibinfo{title}{{Machine Learning Characterization of Intermittency in Relativistic Pair Plasma Turbulence: Single and Double Sheet Structures},} \apjl, 985, L31, \dodoi{10.3847/2041-8213/add47b}

% type= article
\bibitem[{C.~C. {Haggerty} \& D. {Caprioli}(2020){Haggerty} \& {Caprioli}}]{Haggerty2020}
{Haggerty}, C.~C., \& {Caprioli}, D. 2020, \bibinfo{title}{{Kinetic Simulations of Cosmic-Ray-modified Shocks. I. Hydrodynamics},} \apj, 905, 1, \dodoi{10.3847/1538-4357/abbe06}

% type= article
\bibitem[{C.~C. {Haggerty} {et~al.}(2025){Haggerty}, {Sikorski}, {Shay}, {Phan}, {Cassak}, {Murtas}, \& {Pyakurel}}]{Haggerty2025}
{Haggerty}, C.~C., {Sikorski}, D., {Shay}, M.~A., {et~al.} 2025, \bibinfo{title}{{The Enhancement of Ion Heating in Kinetic, Anti-Parallel Reconnection in the Presence of a Flow Shear},} arXiv e-prints, arXiv:2508.09424.
\newblock \doarXiv{2508.09424}

% type= article
\bibitem[{P.~S. {Iroshnikov}(1964){Iroshnikov}}]{iroshnikov1964}
{Iroshnikov}, P.~S. 1964, \bibinfo{title}{{Turbulence of a Conducting Fluid in a Strong Magnetic Field},} \sovast, 7, 566

% type= article
\bibitem[{C.~F. {Kennel} \& F.~V. {Coroniti}(1984){Kennel} \& {Coroniti}}]{Kennel1984}
{Kennel}, C.~F., \& {Coroniti}, F.~V. 1984, \bibinfo{title}{{Confinement of the Crab pulsar's wind by its supernova remnant.},} \apj, 283, 694, \dodoi{10.1086/162356}

% type= article
\bibitem[{T. Kohonen(1982)Kohonen}]{Kohonen1982}
Kohonen, T. 1982, \bibinfo{title}{Self-organized formation of topologically correct feature maps,} Biological Cybernetics, 43, 59, \dodoi{10.1007/BF00337288}

% type= article
\bibitem[{A. {Kolmogorov}(1941){Kolmogorov}}]{kolmogorov1941}
{Kolmogorov}, A. 1941, \bibinfo{title}{{The Local Structure of Turbulence in Incompressible Viscous Fluid for Very Large Reynolds' Numbers},} Akademiia Nauk SSSR Doklady, 30, 301

% type= article
\bibitem[{A. {Lazarian} {et~al.}(2012){Lazarian}, {Vlahos}, {Kowal}, {Yan}, {Beresnyak}, \& {de Gouveia Dal Pino}}]{lazarian2012}
{Lazarian}, A., {Vlahos}, L., {Kowal}, G., {et~al.} 2012, \bibinfo{title}{{Turbulence, Magnetic Reconnection in Turbulent Fluids and Energetic Particle Acceleration},} \ssr, 173, 557, \dodoi{10.1007/s11214-012-9936-7}

% type= article
\bibitem[{M. {Lemoine}(2021){Lemoine}}]{lemoine2021}
{Lemoine}, M. 2021, \bibinfo{title}{{Particle acceleration in strong MHD turbulence},} \prd, 104, 063020, \dodoi{10.1103/PhysRevD.104.063020}

% type= article
\bibitem[{Y.-H. Liu {et~al.}(2017)Liu, Hesse, Guo, Daughton, Li, Cassak, \& Shay}]{Liu2017}
Liu, Y.-H., Hesse, M., Guo, F., {et~al.} 2017, \bibinfo{title}{Why does Steady-State Magnetic Reconnection have a Maximum Local Rate of Order 0.1?} Phys. Rev. Lett., 118, 085101, \dodoi{10.1103/PhysRevLett.118.085101}

% type= article
\bibitem[{M. Lyutikov {et~al.}(2019)Lyutikov, Temim, Komissarov, Slane, Sironi, \& Comisso}]{Lyutikov2019}
Lyutikov, M., Temim, T., Komissarov, S., {et~al.} 2019, \bibinfo{title}{Interpreting Crab Nebula’s synchrotron spectrum: two acceleration mechanisms,} Monthly Notices of the Royal Astronomical Society, 489, 2403, \dodoi{10.1093/mnras/stz2023}

% type= article
\bibitem[{A.~P. {Marscher}(2014){Marscher}}]{marscher2014}
{Marscher}, A.~P. 2014, \bibinfo{title}{{Turbulent, Extreme Multi-zone Model for Simulating Flux and Polarization Variability in Blazars},} \apj, 780, 87, \dodoi{10.1088/0004-637X/780/1/87}

% type= article
\bibitem[{W.~H. {Matthaeus}(1982){Matthaeus}}]{Matthaeus1982}
{Matthaeus}, W.~H. 1982, \bibinfo{title}{{Reconnection in two dimensions: Localization of vorticity and current near magnetic X-points},} \grl, 9, 660, \dodoi{10.1029/GL009i006p00660}

% type= article
\bibitem[{R. Mbarek {et~al.}(2022)Mbarek, Haggerty, Sironi, Shay, \& Caprioli}]{Mbarek2022}
Mbarek, R., Haggerty, C., Sironi, L., Shay, M., \& Caprioli, D. 2022, \bibinfo{title}{Relativistic Asymmetric Magnetic Reconnection,} Phys. Rev. Lett., 128, 145101, \dodoi{10.1103/PhysRevLett.128.145101}

% type= article
\bibitem[{J.~M. {Mehlhaff} {et~al.}(2025){Mehlhaff}, {Zhou}, \& {Zhdankin}}]{Mehlhaff2025}
{Mehlhaff}, J.~M., {Zhou}, M., \& {Zhdankin}, V. 2025, \bibinfo{title}{{Radiative Relativistic Turbulence as an In Situ Pair-plasma Source in Blazar Jets},} \apj, 987, 159, \dodoi{10.3847/1538-4357/addb47}

% type= article
\bibitem[{W.-C. {M{\"u}ller} {et~al.}(2003){M{\"u}ller}, {Biskamp}, \& {Grappin}}]{muller2003}
{M{\"u}ller}, W.-C., {Biskamp}, D., \& {Grappin}, R. 2003, \bibinfo{title}{{Statistical anisotropy of magnetohydrodynamic turbulence},} \pre, 67, 066302, \dodoi{10.1103/PhysRevE.67.066302}

% type= article
\bibitem[{J. {N{\"a}ttil{\"a}} \& A.~M. {Beloborodov}(2021){N{\"a}ttil{\"a}} \& {Beloborodov}}]{Joonas2021}
{N{\"a}ttil{\"a}}, J., \& {Beloborodov}, A.~M. 2021, \bibinfo{title}{{Radiative Turbulent Flares in Magnetically Dominated Plasmas},} \apj, 921, 87, \dodoi{10.3847/1538-4357/ac1c76}

% type= article
\bibitem[{J. {N{\"a}ttil{\"a}} \& A.~M. {Beloborodov}(2022){N{\"a}ttil{\"a}} \& {Beloborodov}}]{joonas2022}
{N{\"a}ttil{\"a}}, J., \& {Beloborodov}, A.~M. 2022, \bibinfo{title}{{Heating of Magnetically Dominated Plasma by Alfv{\'e}n-Wave Turbulence},} \prl, 128, 075101, \dodoi{10.1103/PhysRevLett.128.075101}

% type= article
\bibitem[{J. Nunez-Iglesias {et~al.}(2018)Nunez-Iglesias, Blanch, Looker, Dixon, \& Tilley}]{skan-library}
Nunez-Iglesias, J., Blanch, A.~J., Looker, O., Dixon, M. W.~A., \& Tilley, L. 2018, \bibinfo{title}{A new Python library to analyse skeleton images confirms malaria parasite remodelling of the red blood cell membrane skeleton,} PeerJ, 6.
\newblock \url{https://api.semanticscholar.org/CorpusID:3590543}

% type= article
\bibitem[{T.~N. {Parashar} \& W.~H. {Matthaeus}(2016){Parashar} \& {Matthaeus}}]{Parashar2016}
{Parashar}, T.~N., \& {Matthaeus}, W.~H. 2016, \bibinfo{title}{{Propinquity of Current and Vortex Structures: Effects on Collisionless Plasma Heating},} \apj, 832, 57, \dodoi{10.3847/0004-637X/832/1/57}

% type= article
\bibitem[{M.~J. {Rees} \& J.~E. {Gunn}(1974){Rees} \& {Gunn}}]{Rees1974}
{Rees}, M.~J., \& {Gunn}, J.~E. 1974, \bibinfo{title}{{The origin of the magnetic field and relativistic particles in the Crab Nebula},} \mnras, 167, 1, \dodoi{10.1093/mnras/167.1.1}

% type= article
\bibitem[{R.~F. {Serrano} {et~al.}(2024){Serrano}, {N{\"a}ttil{\"a}}, \& {Zhdankin}}]{serrano2024}
{Serrano}, R.~F., {N{\"a}ttil{\"a}}, J., \& {Zhdankin}, V. 2024, \bibinfo{title}{{Scale Statistics of Current Sheets in Relativistic Collisionless Plasma Turbulence},} arXiv e-prints, arXiv:2408.12511, \dodoi{10.48550/arXiv.2408.12511}

% type= article
\bibitem[{Z.-S. {She} \& E. {Leveque}(1994){She} \& {Leveque}}]{she1994}
{She}, Z.-S., \& {Leveque}, E. 1994, \bibinfo{title}{{Universal scaling laws in fully developed turbulence},} \prl, 72, 336, \dodoi{10.1103/PhysRevLett.72.336}

% type=
\bibitem[{W. Silversmith(2021)Silversmith}]{cc3d}
Silversmith, W. 2021, {cc3d: Connected components on multilabel 3D \& 2D images.}, 3.2.1 \dodoi{https://zenodo.org/record/5535251}

% type= article
\bibitem[{L. {Sironi} \& A. {Spitkovsky}(2014){Sironi} \& {Spitkovsky}}]{Sironi2014}
{Sironi}, L., \& {Spitkovsky}, A. 2014, \bibinfo{title}{{Relativistic Reconnection: An Efficient Source of Non-thermal Particles},} \apjl, 783, L21, \dodoi{10.1088/2041-8205/783/1/L21}

% type= article
\bibitem[{E. {Sobacchi} {et~al.}(2023){Sobacchi}, {Piran}, \& {Comisso}}]{sobacchi2023}
{Sobacchi}, E., {Piran}, T., \& {Comisso}, L. 2023, \bibinfo{title}{{Ultrafast Variability in AGN Jets: Intermittency and Lighthouse Effect},} \apjl, 946, L51, \dodoi{10.3847/2041-8213/acc84d}

% type= article
\bibitem[{B.~U.~{\"O}. {Sonnerup} \& M. {Scheible}(1998){Sonnerup} \& {Scheible}}]{Sonnerup1998}
{Sonnerup}, B. U.~{\"O}., \& {Scheible}, M. 1998, \bibinfo{title}{{Minimum and Maximum Variance Analysis},} ISSI Scientific Reports Series, 1, 185

% type= article
\bibitem[{W.~J. {Sun} {et~al.}(2019){Sun}, {Slavin}, {Tian}, {Bai}, {Poh}, {Akhavan-Tafti}, {Lu}, {Yao}, {Le}, {Nakamura}, {Giles}, \& {Burch}}]{Sun2019}
{Sun}, W.~J., {Slavin}, J.~A., {Tian}, A.~M., {et~al.} 2019, \bibinfo{title}{{MMS Study of the Structure of Ion-Scale Flux Ropes in the Earth's Cross-Tail Current Sheet},} \grl, 46, 6168, \dodoi{10.1029/2019GL083301}

% type= article
\bibitem[{S. {van der Walt} {et~al.}(2014){van der Walt}, {Sch{\"o}nberger}, {Nunez-Iglesias}, {Boulogne}, {Warner}, {Yager}, {Gouillart}, {Yu}, \& {scikit-image Contributors}}]{scikit-image}
{van der Walt}, S., {Sch{\"o}nberger}, J.~L., {Nunez-Iglesias}, J., {et~al.} 2014, \bibinfo{title}{{scikit-image: Image processing in Python},} PeerJ, 2, e453, \dodoi{10.7717/peerj.453}

% type= article
\bibitem[{C. {Vega} {et~al.}(2022){Vega}, {Boldyrev}, {Roytershteyn}, \& {Medvedev}}]{vega2022}
{Vega}, C., {Boldyrev}, S., {Roytershteyn}, V., \& {Medvedev}, M. 2022, \bibinfo{title}{{Turbulence and Particle Acceleration in a Relativistic Plasma},} \apjl, 924, L19, \dodoi{10.3847/2041-8213/ac441e}

% type= article
\bibitem[{A. {Vincent} \& M. {Meneguzzi}(1991){Vincent} \& {Meneguzzi}}]{Vincent1991}
{Vincent}, A., \& {Meneguzzi}, M. 1991, \bibinfo{title}{{The spatial structure and statistical properties of homogeneous turbulence},} Journal of Fluid Mechanics, 225, 1, \dodoi{10.1017/S0022112091001957}

% type= article
\bibitem[{P. Virtanen {et~al.}(2020)Virtanen, Gommers, Oliphant, Haberland, Reddy, Cournapeau, Burovski, Peterson, Weckesser, Bright, {van der Walt}, Brett, Wilson, Millman, Mayorov, Nelson, Jones, Kern, Larson, Carey, Polat, Feng, Moore, {VanderPlas}, Laxalde, Perktold, Cimrman, Henriksen, Quintero, Harris, Archibald, Ribeiro, Pedregosa, {van Mulbregt}, \& {SciPy 1.0 Contributors}}]{2020SciPy}
Virtanen, P., Gommers, R., Oliphant, T.~E., {et~al.} 2020, \bibinfo{title}{{{SciPy} 1.0: Fundamental Algorithms for Scientific Computing in Python},} Nature Methods, 17, 261, \dodoi{10.1038/s41592-019-0686-2}

% type= article
\bibitem[{M. {Wan} {et~al.}(2016){Wan}, {Matthaeus}, {Roytershteyn}, {Parashar}, {Wu}, \& {Karimabadi}}]{wan2016}
{Wan}, M., {Matthaeus}, W.~H., {Roytershteyn}, V., {et~al.} 2016, \bibinfo{title}{{Intermittency, coherent structures and dissipation in plasma turbulence},} Physics of Plasmas, 23, 042307, \dodoi{10.1063/1.4945631}

% type= article
\bibitem[{Y. {Yang} {et~al.}(2019){Yang}, {Wan}, {Matthaeus}, {Shi}, {Parashar}, {Lu}, \& {Chen}}]{yang2019curv}
{Yang}, Y., {Wan}, M., {Matthaeus}, W.~H., {et~al.} 2019, \bibinfo{title}{{Role of magnetic field curvature in magnetohydrodynamic turbulence},} Physics of Plasmas, 26, 072306, \dodoi{10.1063/1.5099360}

% type= article
\bibitem[{Y. {Yang} {et~al.}(2017{\natexlab{a}}){Yang}, {Matthaeus}, {Parashar}, {Haggerty}, {Roytershteyn}, {Daughton}, {Wan}, {Shi}, \& {Chen}}]{Yang2017}
{Yang}, Y., {Matthaeus}, W.~H., {Parashar}, T.~N., {et~al.} 2017{\natexlab{a}}, \bibinfo{title}{{Energy transfer, pressure tensor, and heating of kinetic plasma},} Physics of Plasmas, 24, 072306, \dodoi{10.1063/1.4990421}

% type= article
\bibitem[{Y. {Yang} {et~al.}(2017{\natexlab{b}}){Yang}, {Matthaeus}, {Parashar}, {Wu}, {Wan}, {Shi}, {Chen}, {Roytershteyn}, \& {Daughton}}]{Yang2017NC}
{Yang}, Y., {Matthaeus}, W.~H., {Parashar}, T.~N., {et~al.} 2017{\natexlab{b}}, \bibinfo{title}{{Energy transfer channels and turbulence cascade in Vlasov-Maxwell turbulence},} \pre, 95, 061201, \dodoi{10.1103/PhysRevE.95.061201}

% type= article
\bibitem[{K. {Yoshimatsu} {et~al.}(2009){Yoshimatsu}, {Kondo}, {Schneider}, {Okamoto}, {Hagiwara}, \& {Farge}}]{yoshimatsu2009}
{Yoshimatsu}, K., {Kondo}, Y., {Schneider}, K., {et~al.} 2009, \bibinfo{title}{{Wavelet-based coherent vorticity sheet and current sheet extraction from three-dimensional homogeneous magnetohydrodynamic turbulence},} Physics of Plasmas, 16, 082306, \dodoi{10.1063/1.3195066}

% type= article
\bibitem[{K.~H. {Yuen} \& A. {Lazarian}(2020){Yuen} \& {Lazarian}}]{yuen2020}
{Yuen}, K.~H., \& {Lazarian}, A. 2020, \bibinfo{title}{{Curvature of Magnetic Field Lines in Compressible Magnetized Turbulence: Statistics, Magnetization Predictions, Gradient Curvature, Modes, and Self-gravitating Media},} \apj, 898, 66, \dodoi{10.3847/1538-4357/ab9360}

% type= article
\bibitem[{S. {Zenitani} {et~al.}(2011){Zenitani}, {Hesse}, {Klimas}, \& {Kuznetsova}}]{zenitani2011}
{Zenitani}, S., {Hesse}, M., {Klimas}, A., \& {Kuznetsova}, M. 2011, \bibinfo{title}{{New Measure of the Dissipation Region in Collisionless Magnetic Reconnection},} \prl, 106, 195003, \dodoi{10.1103/PhysRevLett.106.195003}

% type= article
\bibitem[{S. {Zenitani} \& M. {Hoshino}(2001){Zenitani} \& {Hoshino}}]{Zenitani2001}
{Zenitani}, S., \& {Hoshino}, M. 2001, \bibinfo{title}{{The Generation of Nonthermal Particles in the Relativistic Magnetic Reconnection of Pair Plasmas},} \apjl, 562, L63, \dodoi{10.1086/337972}

% type= article
\bibitem[{H. {Zhang} {et~al.}(2023){Zhang}, {Marscher}, {Guo}, {Giannios}, {Li}, \& {Negro}}]{Zhang2023}
{Zhang}, H., {Marscher}, A.~P., {Guo}, F., {et~al.} 2023, \bibinfo{title}{{First-principles-integrated Study of Blazar Synchrotron Radiation and Polarization Signatures from Magnetic Turbulence},} \apj, 949, 71, \dodoi{10.3847/1538-4357/acc657}

% type= article
\bibitem[{V. {Zhdankin} {et~al.}(2016){Zhdankin}, {Boldyrev}, \& {Uzdensky}}]{zhdankin2016}
{Zhdankin}, V., {Boldyrev}, S., \& {Uzdensky}, D.~A. 2016, \bibinfo{title}{{Scalings of intermittent structures in magnetohydrodynamic turbulence},} Physics of Plasmas, 23, 055705, \dodoi{10.1063/1.4944820}

% type= article
\bibitem[{V. {Zhdankin} {et~al.}(2013){Zhdankin}, {Uzdensky}, {Perez}, \& {Boldyrev}}]{Zhdankin2013}
{Zhdankin}, V., {Uzdensky}, D.~A., {Perez}, J.~C., \& {Boldyrev}, S. 2013, \bibinfo{title}{{Statistical Analysis of Current Sheets in Three-dimensional Magnetohydrodynamic Turbulence},} \apj, 771, 124, \dodoi{10.1088/0004-637X/771/2/124}

% type= article
\bibitem[{V. {Zhdankin} {et~al.}(2020){Zhdankin}, {Uzdensky}, {Werner}, \& {Begelman}}]{zhdankin2020}
{Zhdankin}, V., {Uzdensky}, D.~A., {Werner}, G.~R., \& {Begelman}, M.~C. 2020, \bibinfo{title}{{Kinetic turbulence in shining pair plasma: intermittent beaming and thermalization by radiative cooling},} \mnras, 493, 603, \dodoi{10.1093/mnras/staa284}

% type= article
\bibitem[{J. {Zrake} \& A.~I. {MacFadyen}(2012){Zrake} \& {MacFadyen}}]{zrake2012}
{Zrake}, J., \& {MacFadyen}, A.~I. 2012, \bibinfo{title}{{Numerical Simulations of Driven Relativistic Magnetohydrodynamic Turbulence},} \apj, 744, 32, \dodoi{10.1088/0004-637X/744/1/32}

\end{thebibliography}
\bibliographystyle{aasjournalv7}

\end{document}